\documentclass[namedreferences,hyperref,optionalrh,solaromanenum]{spr-sola}

\usepackage{graphicx}                    
\usepackage{amssymb}                    
\usepackage{amsmath}
\usepackage{color}                       
\usepackage{wrapfig}
\usepackage{gensymb}
\usepackage{hyperref}
\usepackage[normalem]{ulem}

\newcommand{\kms}{{\mathrm{\, km \,s^{-1}}}}

\newcommand{\mins}{{\mathrm{\, minutes}}}
\newcommand{\gcm}{{\mathrm{\, g \,cm^{-3}}}}

\chardef\us=`\_

\begin{document}

\begin{frontmatter}

\title{Numerical Modeling of Prominences and Coronal Rain with the \texttt{MPI-AMRVAC} Code}

%
 \author[addressref={aff1,aff2},corref,email={valeriia.liakh@astro.uio.no}]{\inits{V.}~\fnm{Valeriia}~\snm{Liakh}\orcid{0000-0002-9570-8145}}

 \author[addressref={aff3},corref,email={}]{\inits{J. M.}~\fnm{Jack}~\snm{Jenkins}\orcid{}}

 \address[id=aff1]{Institute of Theoretical Astrophysics, University of Oslo, P.O. Box 1029, Blindern N-0315, Oslo, Norway}

\address[id=aff2]{Rosseland Centre for Solar Physics, University of Oslo, P.O. Box 1029, Blindern N-0315, Oslo, Norway}

\address[id=aff3]{European Space Agency (ESA), European Space Astronomy Centre (ESAC), Camino Bajo del Castillo s/n, 28692 Villanueva de la Ca\~nada, Madrid, Spain}

%
\runningauthor{Liakh and Jenkins}
\runningtitle{Numerical modeling of prominences and coronal rain with the \texttt{MPI-AMRVAC} code}


\begin{abstract}
This review surveys recent advances in the numerical modeling of solar prominences and coronal rain achieved with the fully open-source adaptive-grid, parallelized Adaptive Mesh Refinement Versatile Advection Code (\texttt{MPI-AMRVAC}). We examine how these models have contributed to our understanding of the formation and evolution of cool plasma structures in the solar corona. We first discuss prominence models that focus on prominence formation and their dynamic behavior. We then turn to coronal rain, highlighting its connection to thermal instability and its role in the exchange of mass and energy between the corona and chromosphere. Particular attention is given to the growing efforts to connect simulations with observations through synthetic emission and spectral diagnostics.
\end{abstract}

%

\end{frontmatter}

%
 \section{Introduction}\label{sec:intro} 


The solar atmosphere contains a wealth of plasma dynamics, the study of which continues to raise more questions year on year.
Many authors focus on the particularly energetic evolutions, be that to understand the precursors to the explosive flares or coronal mass ejections, or the ubiquitous background coronal heating to which a solution continues to evade us, more than 100 years since its discovery.
The past $\approx$~15 years have seen a large uptick in interest for less energetic but equally as enigmatic phenomena, too, given advancements in observational and computational capabilities alike.
In this review, we will focus in particular on the discrete coronal condensations that are coronal rain and solar filaments/prominences.

In the early 2000s, it was already clear that understanding discrete condensations in the solar corona—whether in the form of large‐scale filaments and prominences or as smaller coronal rain blobs—required a unified picture of magnetic support, thermal instability / nonequilibrium, and dynamic plasma flows \citep[][]{Antiochos:1999bapj}. 
Observationally, prominences were known to reside in magnetic dips whose depth and curvature must be sufficient to balance the weight of 10$^4$~K plasma against gravity \citep[e.g.][]{Aulanier:1998aaap, Aulanier:1998baap,Aulanier:1999aap}. 
Yet, how these dips form and evolve in an inherently three‐dimensional (3D), time‐dependent corona remained an open question.
One burning question at the time was whether the magnetic support was provided by the `sheared-arcade' or `flux rope' topologies, for example, since this had broad consequences on our understanding of the eventual eruption of these features \citep[e.g.][]{Rust:1996apj,Antiochos:1999aapj,Amari:2000apj}.
Within the review, we won't cover such eruptive behaviour, and instead focus on the preceding formation and development of the cool coronal condensations and their hosting magnetic field.

High‐resolution imaging and spectroscopy during the mid‐2000s then began to flesh out the fine structure of both prominences and coronal rain. 
In a series of seminal papers, \citet{Lin:2005solphys,Lin:2007solphys,Lin:2009apj} used observations taken by the Swedish Solar Telescope (SST) hydrogen H$\alpha$ filtergraphs to show that quiescent and active‐region filament threads have widths and lengths of only a few hundred to a few thousand kilometers, respectively, with preferential horizontal (active‐region) or vertical (quiescent) alignment relative to the spine \cite[see also][]{Kucera:2003solphys,Okamoto:2007sci,Berger:2008apj,Karpen:2015assl}.
At the time, it was widely considered that both the fundamental primitive and observational characteristics of prominence plasma were governed by the hosting magnetic field, mediated by the region in which they were located, that is, internal, external, internal/external, and diffuse bipolar regions \cite[][]{Mackay:2008solphys}.
It would be a few years before authors would begin to consider the influence of mass despite the low plasma-$\beta$ (ratio of gas pressure to magnetic pressure) assumptions.

Around the same time, high‐resolution coronal rain studies with the same instrument revealed that these condensations have similar properties to filament condensations with widths $\approx$300 km, slightly shorter lengths $\approx$700 km, comparable temperatures $<$~7000~K, larger dynamic velocities  $\leq$ 70 km s$^{-1}$, and accelerations well below free‐fall \citep[][]{Antolin:2012apj}.
This strongly suggested that condensations trace the multistranded coronal magnetic field flux tubes, whose \textit{finite} lateral extent and accompanying magnetic pressure gradient forces the leading thermodynamic pressure (compression) to act against the falling material so as to impose a lower terminal velocity \citep[][]{Schrijver:2001solphys}.
Complementary extreme ultraviolet (EUV) studies with Transition Region And Coronal Explorer (TRACE) and EUV Imaging Telescope aboard the Solar and Heliospheric Observatory (SOHO/EIT) had earlier identified rapid loop‐cooling events, in which hot loops (T $\approx$ 1\,--\,2 MK) abruptly transition through 0.1\,--\,0.2 MK passbands before fading, indicative of catastrophic cooling at the apex and subsequent drainage as rain \citep[][]{Schrijver:2002solphys,DeGroof:2004aap}.
It is in light of such observations that numerical modelling efforts have sought to establish agreement and offer explanatory insight. 
However, it was not evident whether the observations themselves adequately resolved the underlying dimensionality or physics. 
During this period, limitations in computational resources, as well as the relative immaturity of numerical methods and the theoretical frameworks underpinning them, meant that simulations were generally unable to match, let alone critically assess or extend, the conclusions drawn from observations.

On the theoretical side, numerous prominence formation models were under consideration, in the form of direct plasma injection or cold plasma `levitation' \citep[e.g.][]{Litvinenko:1999solphys}.
It was unclear whether coronal rain and prominences were related at this time, in particular as the sporadic nature of these prominence processes struggled to reconcile the `out-of-nowhere' observational characteristics of coronal rain where the formation was shown to be clearly in-situ within the corona \citep[][]{Schrijver:2001solphys}.
One‐dimensional hydrodynamic (HD) loop models were commonplace during this period, and first to convincingly demonstrate that catastrophic cooling (thermal non‐equilibrium/instability) naturally arises when their heating is concentrated near the footpoints \citep{Mok:1990apj,Antiochos:1991apj,Klimchuk:2006solphys}.
\citet{Muller:2003aap, Muller:2004aap} showed that reducing the heating scale‐height below a critical value triggers runaway radiative losses near the loop apex, forming condensations that subsequently drain as coronal rain with properties matching TRACE and EIT observations.
It was simultaneously shown that the distinction in behaviour between prominence condensations and that of coronal rain lay solely in the magnetic topology; where concave-up topology is present, a condensation remains suspended within the corona instead of draining completely to the surface \citep[e.g.][]{Karpen:2005apj}.
Early models like these quantified how the onset of instability depends sensitively on loop length, heating asymmetry, and the balance between conduction and radiation losses--laying the groundwork for later multi‐dimensional studies \citep[e.g.][]{Mendoza-Briceno:2005apj}.
Although different regimes of thermal non-equilibrium/instability had been identified, this pointed to a universal but mediated process responsible for condensation formation.

At the closing of the decade, reviews by \citet{Mackay:2010ssr} and \citet{Labrosse:2010ssr} summarised the outstanding questions in the field of solar prominences such as the filament‐channel formation processes, the fine‐scale thread morphology, and the global force balance in low plasma-$\beta$ conditions, as well as outlining the ambitious endeavours of solving the transfer of radiation through optically thick and thin mediums.
For the coronal rain phenomenon, no unified review was available, but a large body of work was forming \citep[][]{Schrijver:2001solphys, Winebarger:2002apj, Muller:2003aap, Bradshaw:2003aap, Muller:2004aap, DeGroof:2004aap, Cargill:2004apj, Klimchuk:2006solphys}.
These authors collectively stated a clear need for a comprehensive mapping of the parameter space associated with thermal-non-equilibrium, given the wide ranging conditions within the solar atmosphere in terms of heating characteristics (magnitude, asymmetry) and loop parameters (length, expansion, topology).

In what follows, we will highlight the advances in understanding of the coronal rain and solar prominence/filament phenomena that have been facilitated by the \texttt{MPI-AMRVAC} simulation suite. This broadly covers the period between 2010 and the publishing date of this article. In Section \ref{sec:methods} we will provide a quick overview of the \texttt{MPI-AMRVAC} capabilities, in Section \ref{sec:prom_field_plasma} we outline the different magnetic field approximations that have been adopted by authors, after which we explore conclusions drawn on the primitive behaviour of plasma formation within these magnetic fields. In Section \ref{sec:prom_dyn} we present the study of the dynamics found within prominence plasma on both macroscopic and microscopic scales, still within the magnetohydrodynamic (MHD) approximation, before zooming in on the discrete condensations of coronal rain in Section \ref{sec:rain}. Finally, modern methods to diagnose the primitive plasma through their observational counterparts are presented in Section \ref{sec:synthetic} before we outline our view for future research focus in Section \ref{sec:future}.

Two other independent reviews that also consider prominence and coronal rain modeling are under review at the time of our writing: \citet{Zhou:2025rmpp} stresses the various formation pathways that are available to solar prominences, emphasising that the real solar atmosphere likely contains a mix of these processes at any given time as backed up by observations, and \citet{Keppens:2025lrsp} zooms in on the theoretical, linear `MHD spectroscopic' development of the instabilities behind these observed and modeled processes, before presenting general and detailed statements as to the current, developing, and future state-of-the-art.

\section{\texttt{MPI-AMRVAC} Code for Prominence and Coronal Rain Modeling}\label{sec:methods} 


\texttt{MPI-AMRVAC} is a versatile, block-adaptive simulation framework designed to integrate general conservation laws of the form,
\begin{equation}
  \partial_t \textbf{U} + \nabla \cdot \textbf{F}(\textbf{U}) = \textbf{S}_\mathrm{Phys}(\textbf{U},\partial_\mathrm{i}\textbf{U}, \partial_\mathrm{i}\partial_\mathrm{j}\textbf{U},\textbf{x},t).
\end{equation}
Here, \(\mathbf{U}\) collects the conserved variables, \(\textbf{F}(\mathbf{U})\) their fluxes, and \(\textbf{S}_{\rm Phys}\) all additional source terms—ranging from gravity, background heating, radiative cooling, viscosity, or thermal conduction—that may depend on the fields themselves, their spatial derivatives, $\partial_\mathrm{i}\textbf{U}$ and  $\partial_\mathrm{i}\partial_\mathrm{j}\textbf{U}$, position, $\textbf{x}$, and time, $t$.
We will go into this later.
Within this simulation framework, users can select from a suite of pre-configured modules to tackle different physical regimes. In the context of solar applications, three modules prove especially germane:

\begin{itemize}
    \item HD: Considers the pure fluid equations of mass, momentum, and energy, treating magnetic forces as a fixed background. This lightweight setup excels at studying thermal structures and flows along prescribed field geometries.
    \item MHD: Couples the HD equations to the induction equation, thereby allowing magnetic field lines and plasma to evolve in unison. Full MHD is necessary for studies aiming to capture the interplay between the magnetic field and plasma.
    \item Magneto-frictional (MF): For long term studies of magnetic and plasma equilibria that are not concerned with dynamics and instabilities. MF approaches are useful for estimates of global stability of prominence hosting magnetic configurations, e.g., active regions that transit the solar disk.
\end{itemize}

Our review concentrates on how each of these methodologies has been applied—within \texttt{MPI-AMRVAC}—to probe the formation and evolution of solar prominences and coronal rain. For an in-depth description of the latest \texttt{MPI-AMRVAC} release, including its expanded library of physics modules (e.g., different equation systems, non-ideal effects) and high-order numerical schemes, see \cite{Keppens:2023aap}.

\subsection{HD Approximation}\label{subsec:hydrodynamics}

The solar atmosphere’s magnetic field is both highly structured and continuously evolving across a wide range of spatial and temporal scales. To render numerical studies of coronal plasma dynamics computationally feasible, it is common to reduce the problem dimensionality—often to two dimensions (2D) or even to a one‐dimensional (1D) loop geometry—while retaining the essential thermodynamic physics. In the 1D limit, the prominence plasma is assumed to flow and heat along a rigid magnetic flux tube of prescribed shape, and one solves only the hydrodynamic equations:
\begin{align}
    \partial_t \rho + \nabla \cdot (\textbf{v}\rho) = 0, &:\mathrm{Continuity~equation}\\
    \partial_t (\rho\textbf{v}) + \nabla \cdot (\textbf{v}\rho\textbf{v}) + \nabla p = 0, &: \mathrm{Momentum~equation}\\
    \partial_t e + \nabla \cdot (\textbf{v}e + \textbf{v}p) = 0, &: \mathrm{Energy~equation}
\end{align}
closed by the ideal‐gas law for a fully ionized hydrogen + helium (10:1 ratio) plasma,
\[
  p = 2.3\,n_H\,k_{\mathrm B}\,T,
  \quad
  e = \frac{p}{\gamma - 1}
  + \tfrac12\,\rho\,|\mathbf v|^2.
\]
Here, $\rho$ is the mass density, $\textbf{v}$ the plasma velocity, $p$ the thermal pressure, 
$e$ the total energy density, $n_H$ the hydrogen number density, $k_{\mathrm B}$ the Boltzmann constant, 
$T$ the temperature, and $\gamma$ the adiabatic index (typical value is $5/3$). 

By omitting the induction equation, this framework implicitly assumes that the evolving plasma neither modifies nor is modified by the magnetic field geometry. Such an approximation is often justified for condensations embedded in strong, nearly static field regions—typical of active regions or their peripheries—where magnetic forces dominate and plasma motions remain confined along preexisting loops.

\subsection{MHD Approximation}\label{subsec:magnetohydrodynamics}

The magnetic field in the solar atmosphere is continually reshaped by convective motions at photospheric heights—where the plasma-$\beta$ often exceeds unity—yet in the overlying corona $\beta\ll1$, so that magnetic pressure dominates over gas pressure. In this low-$\beta$ regime, a static‐field HD model breaks down as soon as mass loading or footpoint motions perturb the field lines. This is especially true for solar prominences, where cool, dense condensations exert significant weight on their supporting magnetic dips.

To capture the two-way coupling between plasma motions and the evolving field, one must augment the HD system with the induction equation and include the magnetic field contribution within the momentum and energy equations,
\begin{align}
    \partial_t (\rho\mathbf v ) + \nabla \cdot (\mathbf v \rho\mathbf v  -\mathbf{BB}) + \nabla p_{tot} = 0, &: \mathrm{Momentum~equation}\\
    \partial_t e + \nabla \cdot (\textbf{v}e -\mathbf{BB}\cdot \mathbf v + \textbf{v}p_{tot}) = 0, &: \mathrm{Energy~equation}\\
    \partial_t \mathbf B - \nabla \times (\mathbf v \times \mathbf B) = 0. &: \mathrm{Induction~equation}
\end{align}
Mass continuity and the equation of state are unchanged, and total energy now takes the definition,
\[
  e = \frac{p}{\gamma - 1}
  + \tfrac12\,\rho\,|\mathbf v|^2
  + |\mathbf B|^2/2,
  \quad
  p_{tot}= p +  |\mathbf B|^2/2.
\]
This is the full system of MHD equations, expressed in terms of the conservative variables: mass density $\rho$, momentum density $\mathbf{m}=\rho\mathbf{v}$, total energy density $e$, and magnetic field $\mathbf{B}$. The magnetic field is given in normalized units with the magnetic permeability set to unity.

Inclusion of this equation completes the `Ideal MHD equation set' and allows field lines to be modified by plasma flows, enabling heavy mass to sag field lines into deeper dips or to trigger instabilities when magnetic tension can no longer sustain the load.
This significantly increases the computational load, with the additional need to constrain the induction equation with explicit methods to maintain the solenoidal condition $\nabla \cdot \mathbf B=0$, \texttt{MPI-AMRVAC} has multiple ways to do this, which can be tailored to the specific use case.

\subsection{MF Approximation}\label{subsec:magneto-frictional}
In the magneto–frictional approach, the usual momentum equation is replaced by an artificial velocity field proportional to the Lorentz force,
\begin{equation}
  \mathbf{v} \;=\;\frac{1}{\nu}\,\frac{(\nabla\times\mathbf{B})\times\mathbf{B}}{B^2},  
\end{equation}
where \(\nu\) is a prescribed friction coefficient. The magnetic evolution is then governed by the induction equation alone so that the field relaxes quasi‐statically toward a force‐free or magnetohydrostatic equilibrium. By damping out fast MHD waves, this method efficiently follows the slow accumulation of currents and field‐line twisting that underlie plasma support, without the timestep constraints imposed by Alfvénic or acoustic dynamics. Its drawback, however, is the deliberate omission of true plasma inertia and wave phenomena, which means it cannot capture dynamic instabilities or oscillatory behaviour.

\subsection{Optional Source Terms}\label{subsec:source_terms}

In realistic solar simulations, authors often augment the ideal MHD equations with additional source terms that capture gravity, background heating, radiative losses, thermal conduction, viscosity, and resistivity. 

\begin{itemize}
  \item \textbf{Gravity:} $\rho\,\mathbf{g}$ in the momentum equation, and the work term $\rho\,\mathbf{g}\cdot\mathbf{v}$ in the energy equation, where $\mathbf{g}$ is the solar gravitational acceleration.
  
  \item \textbf{Background heating:} $Q_{bg}$ is tuned to balance radiative losses and thermal conduction in the initial state, often taking the form of an exponential following hydrostatic equilibrium.
  Since the true coronal heating mechanism remains uncertain, a spatially or density-dependent, often time-independent background heating term is introduced to maintain a stable equilibrium before any subsequent evolutions are induced.

  \item \textbf{Radiative losses:} $Q_{\rm rad} = -n_e\,n_H\,\Lambda(T)\,$ where $n_e$ and $n_H$ are electron and hydrogen number densities and $\Lambda(T)$ is the optically thin loss function. 
  This term governs cooling under the optically thin approximation by removing energy from the system proportional to density-squared.

  \item \textbf{Spitzer (parabolic) thermal conduction:} $\nabla \cdot q \hat{\mathbf b}$, where $q = \kappa_{\parallel} T^{5/2} \bigl(\hat{\mathbf b}\cdot\nabla T\bigr)$ is the conductive heat flux, $\hat{\mathbf b}=\mathbf{B}/|\mathbf{B}|$ is the unit vector along the magnetic field, and $\kappa_{\parallel}$ is the Spitzer conductivity along the field assuming that $\kappa_{\perp}\ll\kappa_{\parallel}$.
  Heat is transported efficiently along magnetic field lines, maintaining coronal temperature gradients while cross-field conduction is suppressed.

  \item \textbf{Viscosity:} $\nabla\!\cdot\!\boldsymbol{\Pi}$ in the momentum equation and $\nabla\!\cdot\!(\mathbf{v}\!\cdot\boldsymbol{\Pi})$ in the energy equation, where $\Pi_{ij} = \mu \Bigl(\partial_i v_j + \partial_j v_i - \tfrac{2}{3}\,\delta_{ij}\,\nabla\!\cdot\!\mathbf v\Bigr)$ is the viscous stress tensor, $\mu$ is the dynamic viscosity.
  Viscous stresses dissipate shear flows, convert kinetic energy into heat, and help stabilise small-scale velocity gradients.
  The magnitude of this process is not well understood within the solar corona, nor is it clear whether our current numerical frontiers properly resolve the process.
  Hence, this is commonly neglected.

  \item \textbf{Resistivity:} -$\nabla\times\bigl(\eta\,\nabla\times\mathbf B\bigr)$ in the induction equation; $Q_{\rm Joule} = \eta\,\bigl|\nabla\times\mathbf B\bigr|^2\,$ as Joule heating in the energy equation with $\eta$ the resistivity. This term enables magnetic reconnection and diffusion of current sheets, providing a pathway for magnetic energy to be converted into thermal energy. The actual value of $\eta$ is understood to be of order 10$^{-9}$, requiring spatial resolutions orders of magnitude beyond current capabilities of several kms. To avoid unphysical numerical oscillations, however, this is often set just above the gridscale, $\approx$~10$^{-4}$ in dimensionless units ($\approx$~10$^{12}\mathrm{\, cm^{2}\, s^{-1}}$).
\end{itemize}


  \section{Modeling of Prominence Magnetic Field and Plasma}\label{sec:prom_field_plasma} 

  In this section, we discuss recent advances in modeling the filament channel and prominence plasma using the \texttt{MPI-AMRVAC} code. Several key questions remain central to understanding prominence formation in both quiet-Sun and active regions: What is the structure of the magnetic field supporting prominences, and how does it form? How does prominence plasma form and accumulate within this magnetic field?

  \subsection{Modeling of Magnetic Field}\label{sec:field}
   
   
   In the numerical modeling of solar prominences, it is often assumed that the magnetic field is represented by a flux rope \citep[][]{vanBallegooijen:1989apj,Priest:1989apj,Rust:1994solphys,Aulanier:1998aaap,Gibson:2006jgr}, or a sheared arcade \citep[][]{Antiochos:1994apjl,DeVore:2000apj,Aulanier:2002aap,DeVore:2005apj} that lies horizontally above the polarity-inversion line. The flux rope can be defined analytically \citep[e.g.][]{Titov:1999aap} or formed through footpoint motions starting from the magnetic arcade structure following \citet{vanBallegooijen:1989apj} scenario. According to this scenario, the initial magnetic field is an ordinary potential arcade. In order to form the flux rope or sheared arcade, the motions of the footpoints of the magnetic field are applied. This can be shearing, converging, or vortex motions, or a combination therein. A comprehensive overview of earlier observational and theoretical studies on prominence magnetic structures is provided in \citet{Mackay:2010ssr}.
  
\begin{figure*}[!h]
    \centering
    \includegraphics[width=0.9\textwidth]{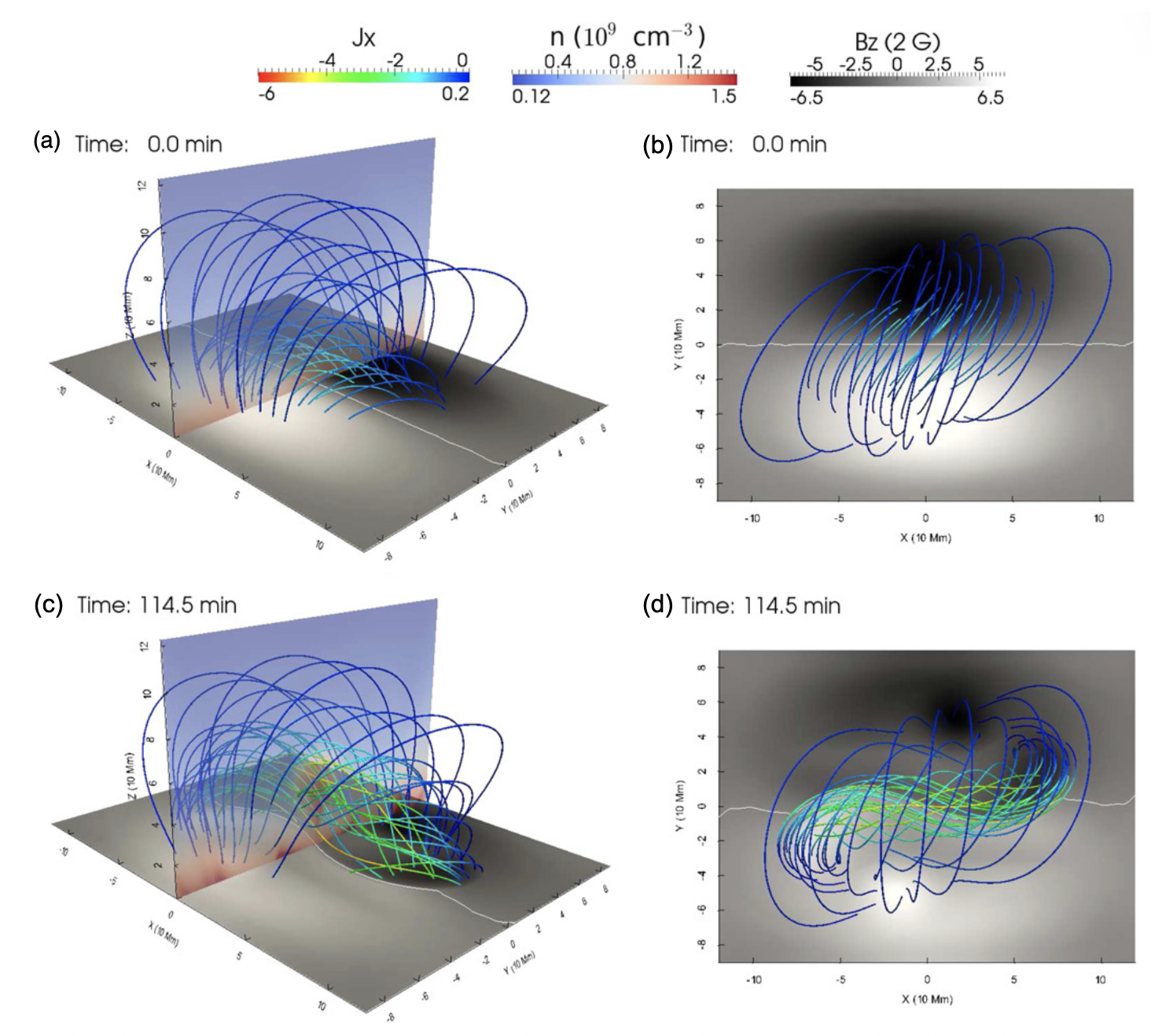}
    \caption{Side (left) and top (right) views of the flux rope at time $0.0$ and $114.5$ minutes. The bottom magnetograms are shown in gray with the polarity-inversion line plotted in white. Magnetic field lines are colored by the local current density $J_x$ in the rainbow color table. The vertical planes are colored by number density in the blue-red color table. Adapted from \citet{Xia:2014apj}. \textcopyright\ AAS. Reproduced with permission.}
    \label{fig:Xia_2014_1}
\end{figure*}

\begin{figure*}[!b]
    \centering
    \includegraphics[width=0.9\textwidth]{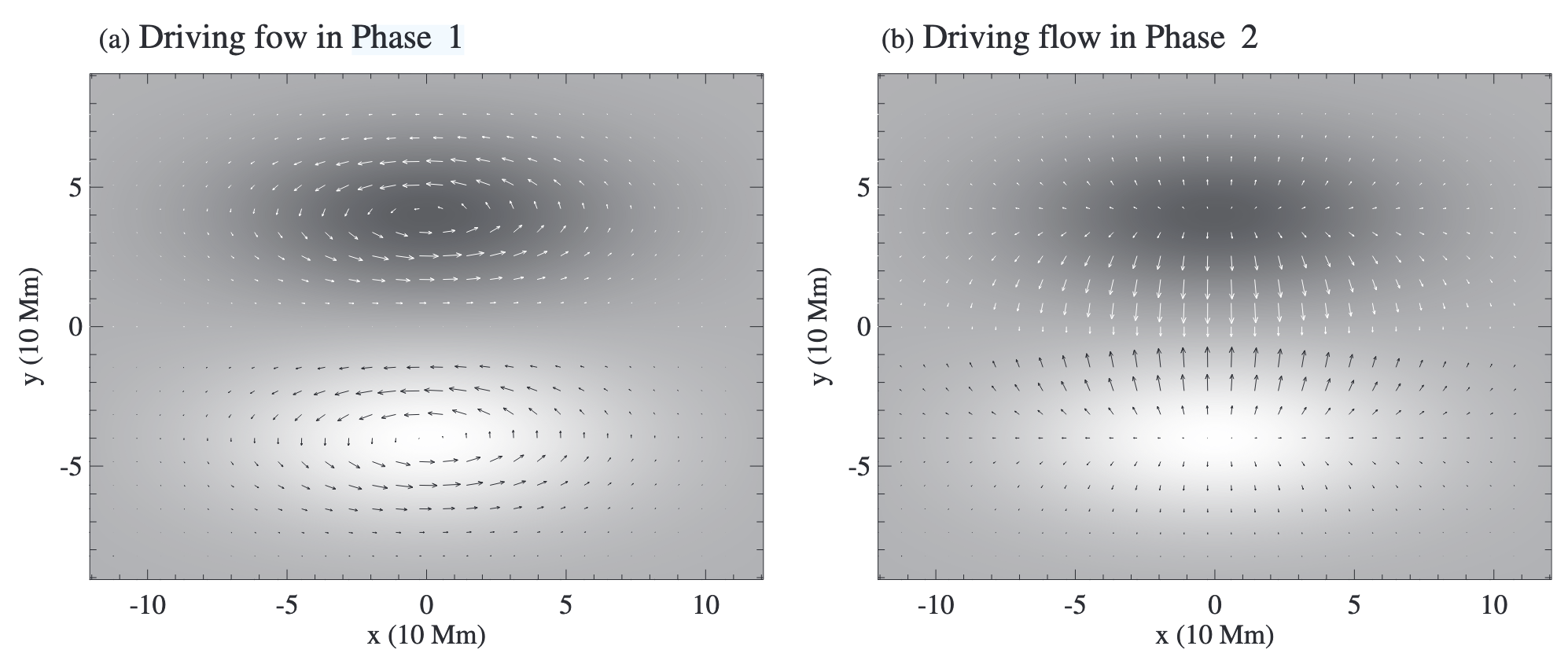}
    
    \caption{(a) Twisting and (b) converging velocity fields (indicated by arrows) applied at the footpoints, overlaid on the bottom boundary magnetogram shown in grayscale. Adapted from \citet{Xia:2014apj}. \textcopyright\ AAS. Reproduced with permission.}
    \label{fig:Xia_2014_2}
\end{figure*}

   The first attempt to model 3D flux rope formation through the footpoint vortex and converging motions using the \texttt{MPI-AMRVAC} code was made by \citet{Xia:2014apj}. The numerical experiment begins with an isothermal, gravitationally stratified corona. This atmosphere is permeated by the magnetic field linear force-free extrapolated from an analytic bipolar magnetogram (Figure \ref{fig:Xia_2014_1}, top). To form the flux rope, the authors prescribed boundary conditions that assume twisting motions around the two main polarities of the magnetic field towards the polarity-inversion line (Figure \ref{fig:Xia_2014_2}, left). The vertical component of the magnetic field is assumed to be preserved. After viscous relaxation, converging motions are imposed at the footpoints to drive them toward the polarity inversion line (Figure \ref{fig:Xia_2014_2}, right), initiating magnetic reconnection and leading to the formation of a flux rope (Figure \ref{fig:Xia_2014_1}, bottom). The authors identified that a significant portion of the formed flux rope consists of dipped field lines, and therefore, it is suitable for supporting prominences. In this study, the authors did not investigate the possible origins of these footpoint motions, such as associating them with the differential rotation of the Sun, granular and supergranular motions, and other factors. In numerical studies with the \texttt{MPI-AMRVAC} code, footpoint motions have often been applied to form flux ropes in two-and-a-half-dimensional (2.5D) and 3D studies \citep{Xia:2014apjl, Zhao:2017apj, Jenkins:2021aap, Jenkins:2022natas, Brughmans:2022aap, Liakh:2023apjl, Donne:2024apj, Liakh:2025aap} starting from the magnetic arcade field. 

  \begin{figure*}[!h]
    \centering
    \includegraphics[width=0.9\textwidth]{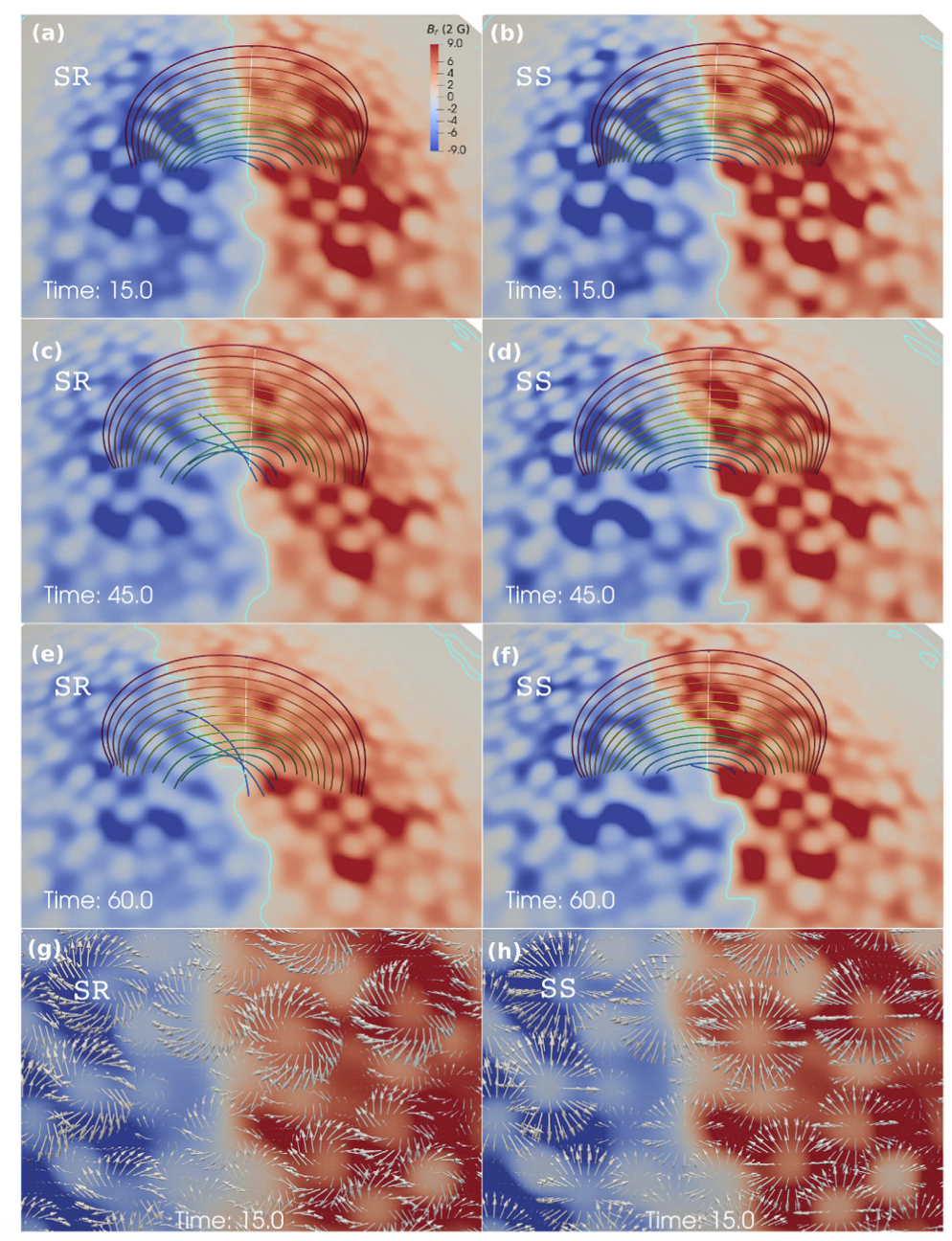}
    
    \caption{Two models applying the supergranular motions and with or without Coriolis force at times 15, 45, and 60. In panels a–f, magnetic field lines are shown in rainbow colors, and the polarity-inversion line in cyan. The photosphere is colored by a radial magnetic field saturated at $\pm20$ G. Zoom-in views are shown in panels g and h, where gray arrows present the photospheric horizontal velocity field. Adapted from \citet{Liu:2022apjl}.}
    \label{fig:Liu_2022_2}
\end{figure*} 

    
  
 The potential of supergranular motions to form a flux rope from the dipolar magnetic field has been studied by \citet{Liu:2022apjl}. In this study, the authors employed potential field extrapolations in spherical coordinates to obtain the dipolar initial magnetic field structure. This magnetic field has then been subjected to supergranular diverging motions, assuming that the supergranular cell area varies in time (Figure \ref{fig:Liu_2022_2}, right). Additionally, they considered a model that also includes vortical motions caused by the Coriolis force in addition to the diverging motions of granular/magnetic diffusion (Figure \ref{fig:Liu_2022_2}, left). The comparison of the two models reveals significantly different resulting magnetic structures, as shown in Figure \ref{fig:Liu_2022_2}e, f. The model, including the counterclockwise vortical motions, shows the injection of the negative magnetic helicity, which is accumulated along the polarity-inversion line and finally leads to the formation of the dextral filament channel in the northern hemisphere. The conclusion is that the Coriolis force is a crucial component in understanding the formation of the filament channel through helicity injection, as the first model without the Coriolis force does not produce a flux rope.
 On the other hand, previous simulations on helicity condensation \citep[][]{Knizhnik:2015apj,Zhao:2015} did not produce flux ropes due to the absence of supergranular diverging flows, which are essential to bring the opposite polarities and initiate reconnection.
 
 This model has also been used to explain the polar crown prominences \citet{Chen:2024apj}. The central puzzle regarding this type of prominence was related to the fact that at those latitudes, no significant flux emergence is observed, on which models that use converging and shearing motions rely. This model incorporated various types of photospheric motions, including differential rotation, meridional circulation, and supergranulation. In addition, the Coriolis force has been applied, which forms vortices that lead to magnetic helicity condensation into the corona consistent with \citet{Antiochos:2013apj}. This model allowed the reproduction of the helicity-chirality pattern of these structures. For instance, in the standard model, when the polarity-inversion line is oriented east-west, helicity-chirality is reproduced as negative-dextral in the northern hemisphere, and positive-sinistral in the southern hemisphere. This type of modeling of filament channel formation opens the way for further study of the evolution of quiescent prominences, including the formation of prominence plasma structures, their dynamics and flows, and eruptions.

  Data-constrained and data-driven simulations have been widely employed to reproduce flux rope eruptions and to investigate their triggering mechanisms and dynamics. However, in this part of the review, we focus specifically on studies that examine the formation, structure, and morphology of solar prominences. For a comprehensive overview of recent advances in modeling prominence eruptions and coronal mass ejections (CMEs) using the aforementioned approaches, we refer the reader to the recent reviews by \citet{Schmieder:2024rvmpp} and \citet{Guo:2024rvmpp}, along with the references therein.

\begin{figure*}
    \centering
    \includegraphics[trim= 0 20 0 0 ,clip,width=0.9\textwidth]{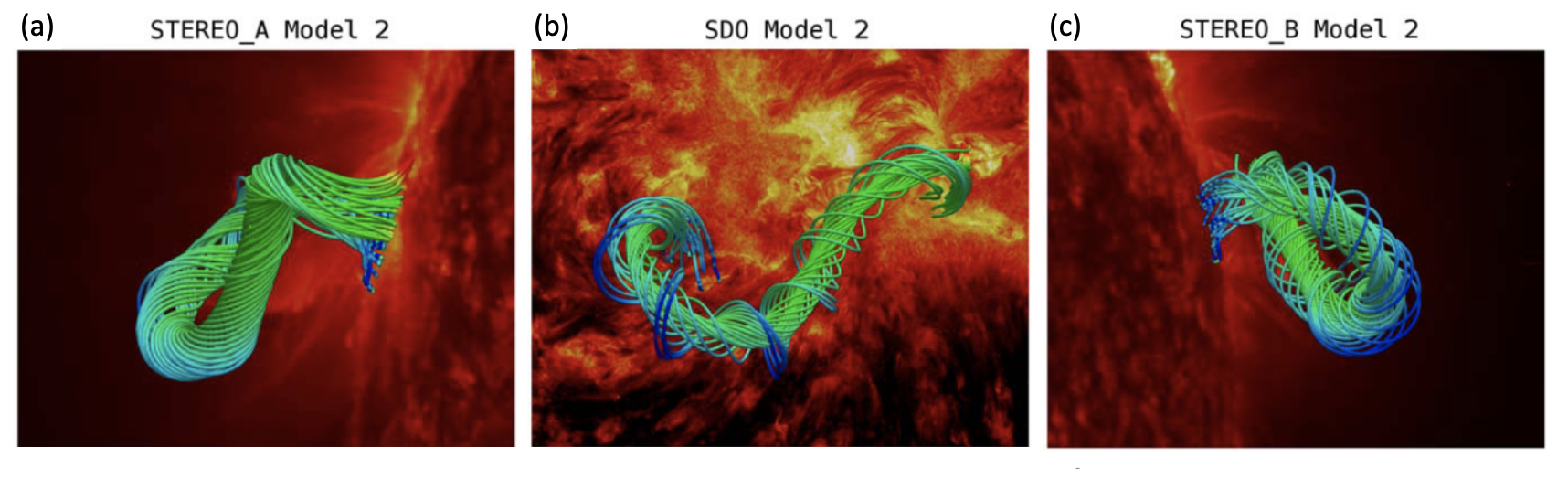}
    
    \caption{(a) Magnetic flux rope (MFR) constructed by the MFR embedding method, further relaxed by the MF method, overlaid on the 304 \AA image, which was observed by STEREO-A/EUVI at 01:11 UT on 2011 June 21. The viewing angle is from STEREO-A. (b) Same as (a) but viewed from SDO. (c) Same as (a) but viewed from STEREO-B. Adapted from \citet{Guo:2019apjl}. \textcopyright\ AAS. Reproduced with permission.}
    \label{fig:Guo_2019_1}
\end{figure*}  

\begin{figure*}[!b]
    \centering
    \includegraphics[width=0.9\textwidth]{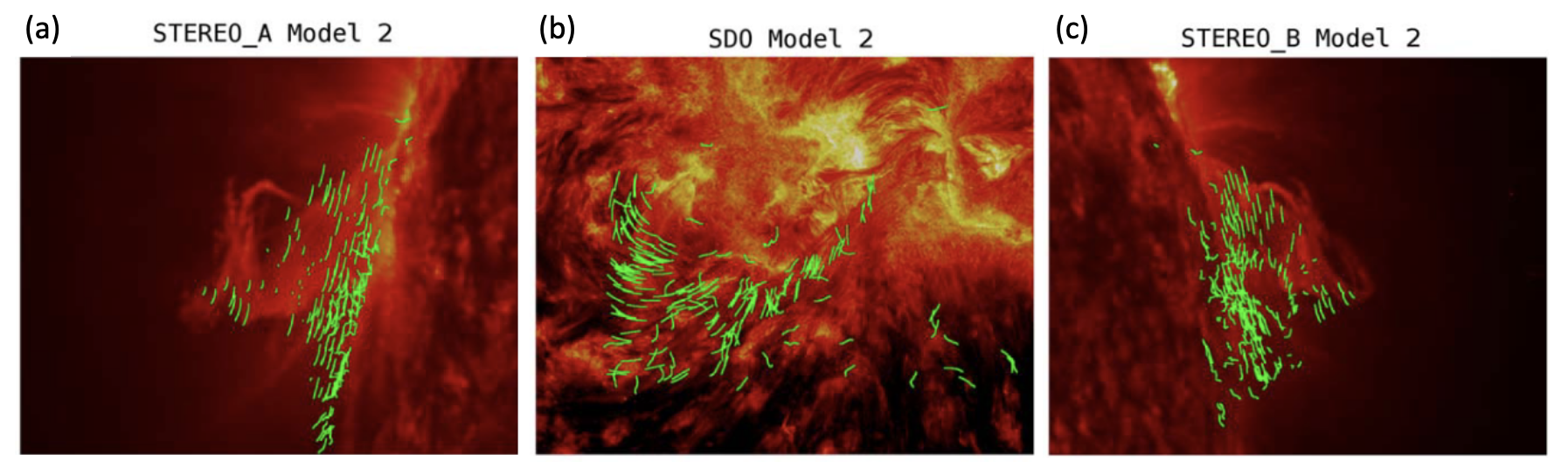}
    
    \caption{(a) Magnetic dips computed from the MFR constructed by the MFR embedding method, further relaxed by the MF method, overlaid on the 304 \AA image, which was observed by STEREO-A/EUVI at 01:11 UT on 2011 June 21. The viewing angle is from STEREO-A. (b) Same as (a) but viewed from SDO. (c) Same as (a) but viewed from STEREO-B. Adapted from \citet{Guo:2019apjl}. \textcopyright\ AAS. Reproduced with permission.}
    \label{fig:Guo_2019_2}
\end{figure*} 

  A class of studies constructs the prominence magnetic field according to the flux rope embedding method \citep{Titov:2014apj} and the regularized Biot–Savart laws \citep{Titov:2018apjl}. An example of the application of this method has been shown by \citet{Guo:2019apjl}. In this work, the authors obtained the axis of the flux rope from multipoint 304 \AA\ observations taken by the Extreme Ultraviolet Imager (EUVI) on board Solar Terrestrial Relations Observatory (STEREO) and the Atmospheric Imaging Assembly (AIA) on board the Solar Dynamics Observatory (SDO). This axis is then used to construct the flux rope according to the regularized Biot–Savart laws. The flux rope is then embedded in the potential field obtained from the Helioseismic and Magnetic Imager on board SDO. This initial state is relaxed using the MF approach \citep{Guo:2016apj1,Guo:2016apj2}. The result of this relaxation is shown in the left panels of Figure \ref{fig:Guo_2019_1}. In Figure \ref{fig:Guo_2019_2}, the authors highlighted the location of magnetic dips within the resulting magnetic flux rope that bear resemblance to the prominence threads observed in 304 \AA\ images from both STEREO/EUVI and SDO/AIA.

  \begin{figure*}
    \centering
    \includegraphics[width=0.7\textwidth]{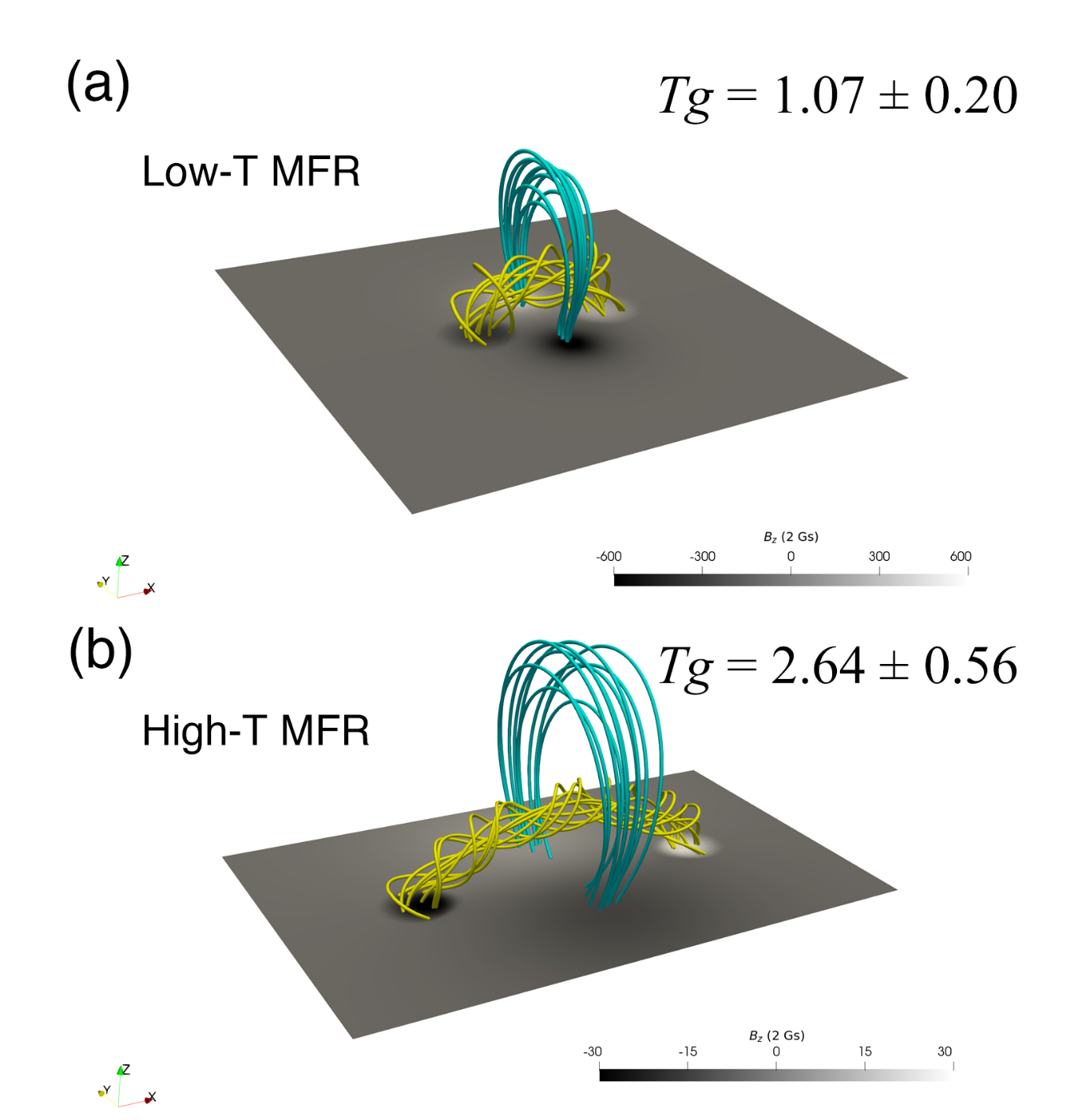}
    
    \caption{Magnetic field models with low and high twists. Yellow lines denote the core flux-rope field, and cyan lines denote the background potential field. Adapted from \citet{Guo:2022aap}.}
    \label{fig:Guo_2022_1}
\end{figure*}  

\begin{figure*}[!h]
    \centering
    \includegraphics[width=0.9\textwidth]{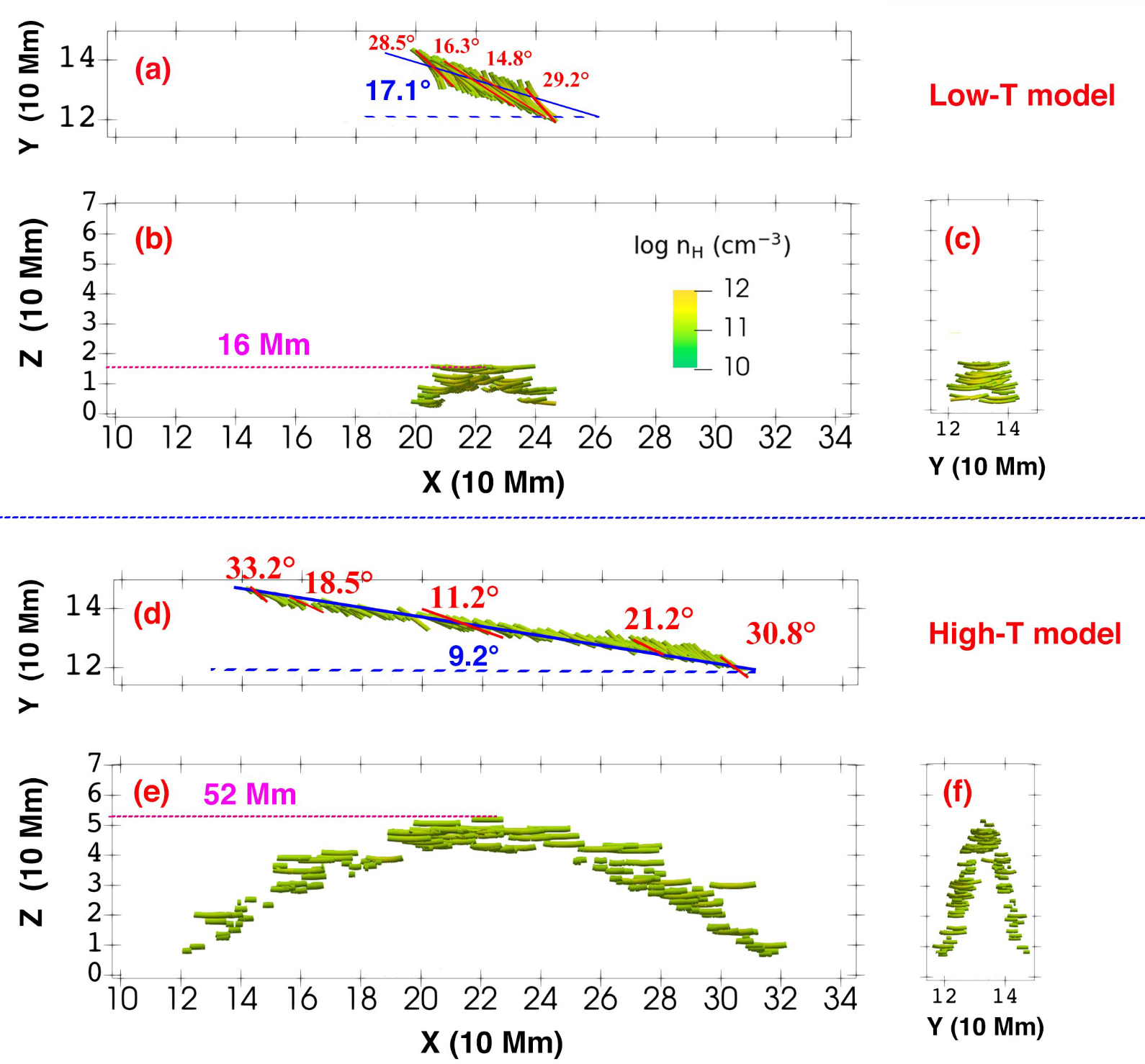}
    
    \caption{Distribution of prominence threads from different viewing angles. Panels a–c show the top, side, and end views of the low-twist model, while panels d–f present the corresponding views for the high-twist model. In panels a and d, blue solid lines indicate the filament spines. Blue angles denote the orientation between the flux rope axes and the filament spines, whereas red angles indicate the orientation between individual threads and the filament spines. Adapted from \citet{Guo:2022aap}.}
    \label{fig:Guo_2022_2}
\end{figure*}

  \citet{Guo:2021apj1} performed a parametric survey of the Titov and Demoulin modified flux rope using the regularized Bio-Savart laws. The authors found that the twist of the magnetic flux rope is proportional to the ratio of the flux rope's length to its minor radius and is independent of the background magnetic field. This is an important conclusion for interpreting the observations, suggesting that quiescent prominences have a larger twist compared to active-region prominences. This is of particular interest as this suggests that long quiescent filaments may be supported by flux ropes with twist numbers exceeding the commonly cited kink instability threshold \citep{Hood:1981}, yet these structures remain stable for extended periods of time. However, the critical twist number can depend on other factors such as the external magnetic field \citep{Torok:2005apjl}, flux rope configuration \citep{Baty:2001aap}, plasma motions \citep{Zaqarashvili:2010aap}, plasma-$\beta$ \citep{Hood:1979solphys}. Gravity can also play an important role in the stabilization of flux ropes \citep{Bi:2014apj,Jenkins:2018solphys,Fan:2020apj}.
  
  In a related study, \citet{Guo:2022aap} has investigated the possible threads configuration according to the different twists of the Titov and Demoulin flux rope (Figure \ref{fig:Guo_2022_1}). Using this magnetic configuration, the authors performed pseudo-3D simulations with the \texttt{MPI-AMRVAC} code, a method briefly described in the previous section. To form the prominence threads, they included non-adiabatic terms in the energy equation, such as thermal conduction, radiative cooling, and background heating, and also applied localized heating. Overall, thread formations are performed via an evaporation-condensation scenario, which we discuss in more detail in Section \ref{sec:plasma}. The resulting prominence threads are shown in Figure \ref{fig:Guo_2022_2}. The important conclusion of this study is that the low-twist flux ropes host more transient but longer threads, while the high-twist flux ropes, on average, host quasi-stationary and shorter threads, respectively. In this way, short, vertically stacked threads can account for the observed morphology of quiescent prominences, while long, low-lying threads correspond to active region prominences. Such endeavours could be extremely useful in ascertaining the impact of the coronal heating mechanism on prominence threads moving forward.

  Based on the regularized Bio-Savart laws, \citet{Kang:2023mnras} also studied the morphology of an observed prominence. Using H$\alpha$ from the ground-based Optical and Near infrared Solar Eruption Tracer (ONSET) and 304 \AA\ observations from SDO/AIA and STEREO/EUVI, the authors consider the overlap in the positions of the dips of their relaxed magnetic field with that of the cool absorbing filament plasma to indicate agreement between the real and simulated configurations. As evident from the H$\alpha$ observations in particular, more attention is needed on extending the modelling to consider multiple possible footpoints, and so barbs, in lieu of a continuous flux rope between its extremities.

   Combining observations from New Vacuum Solar Telescope (NVST) and SDO with a data-driven model from \citet{Guo:2024apj}, itself based on a time-dependent MF approach, \citet{Li:2025apjl} studied the formation of the complex filament magnetic field. In the early stage of the evolution of this magnetic field, the authors obtained a set of magnetic arcades located almost perpendicular to the polarity-inversion line, suggesting a potential field. However, in the later stages, the authors obtained the magnetic field with the two developed flux ropes connected with a region of sheared arcades. Later on, this magnetic structure evolved into one flux rope, became unstable, and erupted. The converging motions in the photosphere were stated as the main reason for the formation of the multi-component magnetic field structure. In this study, the authors confirmed that observationally derived shearing motions can serve to deform the potential arcade field, forming an elongated sheared arcade or a flux rope if reconnection occurs. The converging motions are responsible for driving reconnection at the polarity-inversion line and formation of the shorter flux rope, as observed in their study. Moreover, the authors suggest that the reconnection between two flux ropes could possibly develop into the double-decker filament configuration as explored in \citet{Shen:2024apj}.

  The recent modeling using the \texttt{MPI-AMRVAC} code has contributed to our understanding of the formation and structure of prominence-supporting magnetic fields. On the one hand, 3D simulations have demonstrated the formation of magnetic flux ropes driven by various types of footpoint motions, highlighting the roles of supergranular flows, differential rotation, and vortical motions induced by the Coriolis force. The motions caused by the Coriolis force appear to be key ingredients in the formation of polar crown prominences. On the other hand, reconstructions of the prominence magnetic field, combined with comparisons to observations, have confirmed that cold plasma resides in magnetic dips. These studies have also shed light on the morphological differences between quiet-Sun and active region prominences, which are likely related to differences in magnetic twist, lower twist in quiescent prominences, and higher twist in active region ones.

    \subsection{Modeling of Prominence Plasma}\label{sec:plasma}
    
   In this section, we first review prominence‐formation models driven by the evaporation–condensation mechanism, in which steady or varying heating at the magnetic footpoints induces plasma evaporation. In the latter half, we discuss levitation-condensation models, where preexisting or forming magnetic dips lift plasma that eventually forms condensation. Earlier prominence formation models and key results obtained before the advent of high-resolution numerical tools such as \texttt{MPI-AMRVAC} are comprehensively reviewed in \citet{Mackay:2010ssr}.
    
   The evaporation–condensation scenario explains prominence formation as a two-step thermal process. Initially, heating at the footpoints of magnetic field lines anchored in the chromosphere drives hot plasma upward into the corona \citep{Aschwanden:2001apj,Winebarger:2002apj}. The physical relevance of this approach derives from the understanding that discrete heating events within the low solar atmosphere in the form of nanoflares deposit their energy within the upper layers of the chromosphere \citep[][]{Priest:2002apj}. In all of the cases that follow, the physics of the nanoflares are not explicitly considered, only the influence that their energy deposition has, that is, direct heating \citep[][]{Klimchuk:2006solphys}. Subsequently, radiative losses and the onset of thermal instability cause the evaporated plasma to cool and condense into dense, cool material.

  \citet{Xia:2011apj} performed a non-adiabatic 1D HD study employing the evaporation-condensation scenario with the \texttt{MPI-AMRVAC} code. The energy equation included the non-adiabatic terms of thermal conduction, radiative losses, and background heating. The initial atmosphere of the 1D loop is a gravitationally stratified chromosphere, transition region, and corona. The radiative losses term is included in the energy equation as described in Section \ref{subsec:source_terms}. The energy equation also includes a fixed background heating term that decreases exponentially with distance away from the nearest footpoint along the loop and remains constant over time. It is designed to initially compensate for the radiative loss profile derived from the hydrostatic equilibrium. The prescribed atmosphere is in a state of force equilibrium but not in exact thermal equilibrium due to the non-zero initial thermal conduction introduced by the transition region. Therefore, a brief relaxation stage is needed.
  
  After the relaxation, the localized heating term in the energy equation was activated to form a prominence thread as done in previous 1D simulations \citep[see, e.g.][]{Antiochos:1991apj, Karpen:2001apjl}. The authors distinguish three stages in their experiment: a) evaporation of the plasma into the corona, thus, the radiative losses show an increase, and a persistent temperature decrease; b) both plasma pressure and temperature drop rapidly; c) strong inflow is produced from the footpoints due to the large pressure gradient. The authors also investigated the possibility of symmetric and asymmetric heating at the footpoints. Thus, in the symmetric case, one condensation forms rapidly in agreement with previous studies \citep[see, e.g.][]{Antiochos:1999bapj} and continues to grow, even when the localized heating is switched off due to the subsequent siphoning of plasma from the chromosphere. When the localized heating is asymmetric, one condensation forms, but its location is shifted from the midpoint towards the more weakly heated region. Finally, when the heating is strongly asymmetric and the heating scale length is short, two threads form at the shoulders of the tube. A similar approach to studying the formation and dynamics using one-dip or two-dip structure has been employed in many numerical studies with the \texttt{MPI-AMRVAC} code \citep{Zhang:2012aap, Zhang:2013aap, Zhou:2017apj, Zhang:2020aap, Huang:2021apjl, Ni:2022aap}.
  
  \citet{Huang:2021apjl} performed a 1D parametric study that allowed for the unifying of the evaporation-condensation and injection models. The injection model suggests that prominence forms as a result of the direct injection of plasma into the magnetic dips as a result of the magnetic reconnection within the chromosphere \citep{Wang:1999apjl,Chae:2001apj}. Their experiment suggests that if localized heating is located in the lower chromosphere, the enhanced gas pressure pushes plasma in the upper chromosphere into the corona, similar to the injection model. In contrast, if localized heating occurs in the upper chromosphere, the local plasma is heated and subsequently evaporates and condenses later due to thermal instability following the evaporation-condensation scenario.
  
\begin{figure*}
    \centering
    \includegraphics[width=0.9\textwidth]{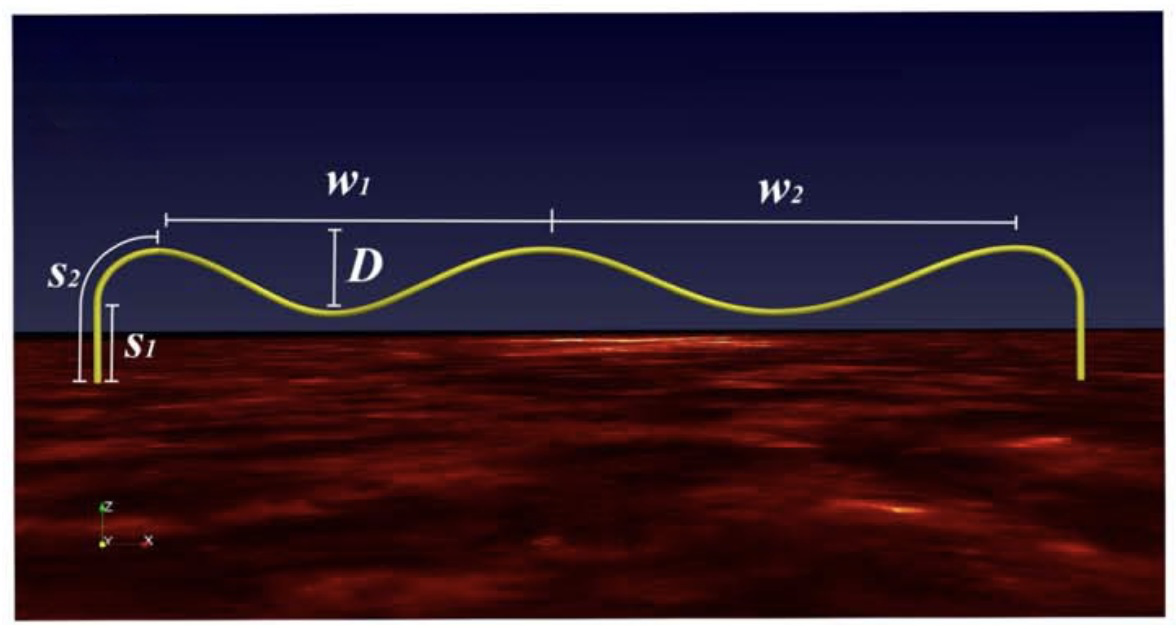}
    
    \caption{Geometry of a helical flux tube used for 1D HD simulations of filament threads. Adapted from \citet{Guo:2021apj2}.}
    \label{fig:Guo_2021b_1}
\end{figure*}

\begin{figure*}[!h]
    \centering
    \includegraphics[width=0.9\textwidth]{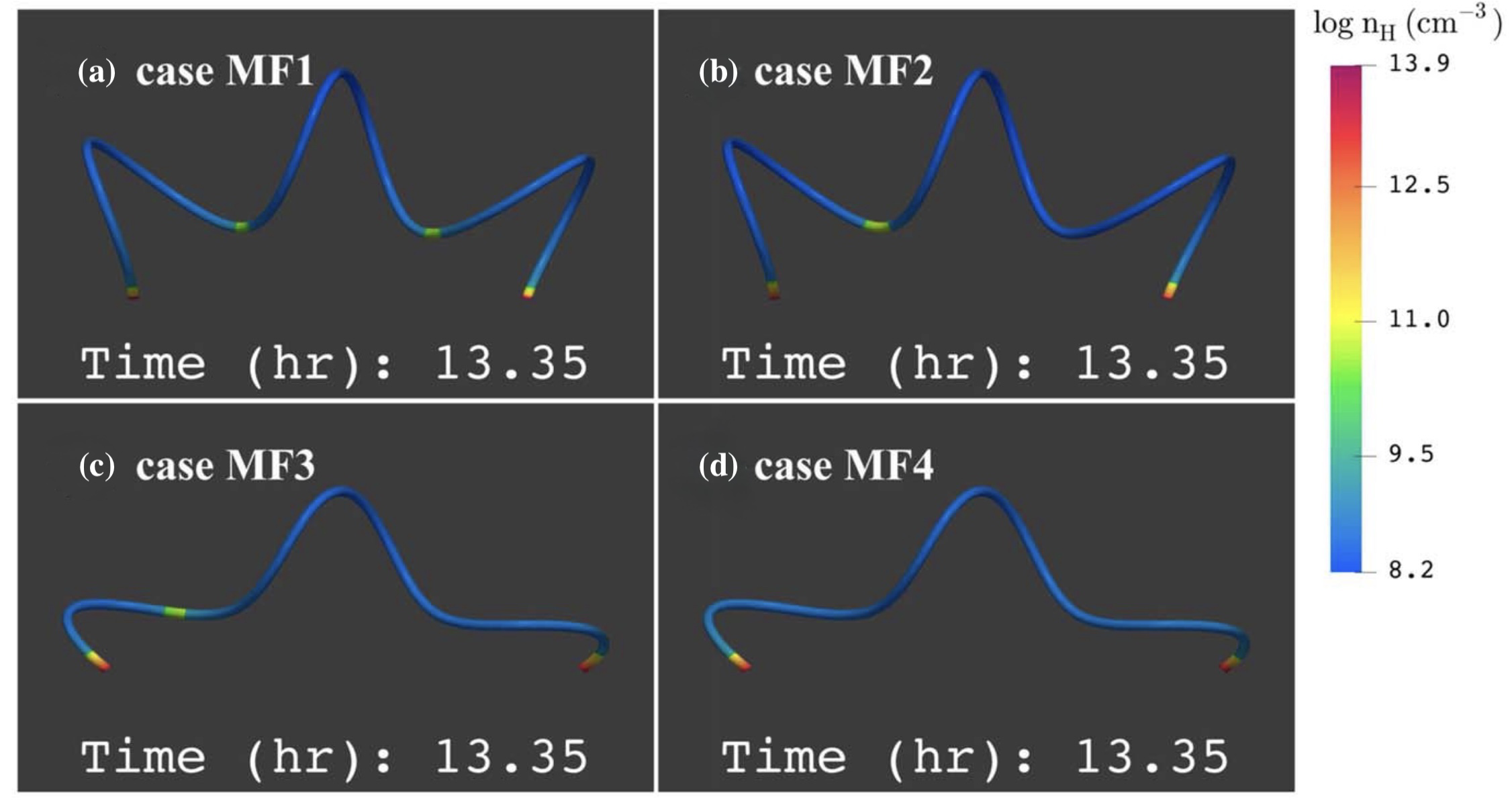}
    
    \caption{The number density distribution along the magnetic field line in the different localized heating models at the end of the runs, respectively. MF1: twist number 2.64, heating scale 10 Mm, two stable threads formed. MF2: twist number 2.64, heating scale 20 Mm, one stable thread formed. MF3: twist number 1.88, heating scale 10 Mm, one stable thread formed, another thread drains. MF4: twist number 1.88, heating scale 20 Mm, only one draining thread forms. Adapted from \citet{Guo:2021apj2}.}
    \label{fig:Guo_2021b_2}
\end{figure*}

 \citet{Guo:2021apj2} performed a 1D study of the formation of prominence threads according to the evaporation-condensation scenario in the flux tube that contains two dips (Figure \ref{fig:Guo_2021b_1}). They found that only a specific combination of the localized heating parameters and tube geometry can lead to the formation of the two persistent threads in the dips. The results from the different models are shown in Figure \ref{fig:Guo_2021b_2}. In this figure, only panel a shows the formation of two persistent threads in both dips. In this model, localized heating is highly concentrated at the footpoints, and the flux tube should contain deep dips. This two-dip flux tube is more typical of quiet-Sun prominences than those associated with active regions. This is because the short filaments in active regions are typically associated with sheared magnetic fields. In contrast, long quiescent filaments are more often linked to twisted magnetic fields, sometimes with more than two full turns \citep{Su:2015apj,Mackay:2020aap}.

  Two main classes of models explain the formation of magnetic dips that support prominence plasma. The first class involves the formation of dips due to the deformation of magnetic fields by the weight of the dense plasma, as in magneto-hydrostatic models based on the configuration of \citet{Kippenhahn:1957zap}. The second class assumes that magnetic dips pre-exist in nearly force-free fields, flux ropes or sheared arcades, where plasma accumulates along already dipped field lines without significantly altering the magnetic structure. An example of the first class was developed by \citet{Xia:2012apjl}, who used the \texttt{MPI-AMRVAC} code in a 2.5D setup to simulate prominence formation via the evaporation–condensation scenario in a magnetic arcade. Localized heating at the footpoints drives chromospheric evaporation, increasing plasma density along the field lines. This leads to runaway radiative cooling and subsequent condensation near the apex of the arcade, forming a vertically extended prominence structure. The initial magnetic configuration did not contain dips; however, they developed dynamically due to the gravitational influence of the heavy condensed plasma, consistent with the mechanism proposed by \citet{Kippenhahn:1957zap}. This dip formation is facilitated by the relatively weak magnetic field in the model, which decreases from 6.8 G at the base to 3.7 G at the top.

  A subsequent 3D study was carried out by \citet{Moschou:2015adspr}. In this case, the magnetic field structure was a superposition of the sheared arcades that included magnetic dips in the central region. In comparison to the previous 2.5D study, no solid prominence body formed in the 3D model; the majority of the condensations drained back to the lower atmosphere, forming coronal rain, and the remaining plasma showed indications of the Rayleigh-Taylor instability. These falling condensations collected into `falling fingers' and moved largely perpendicular to the local magnetic field, following the direction of gravity. Such behaviour was largely short-lived, with the plasma eventually following the magnetic field towards the local dips. There have been studies on the Rayleigh-Taylor instability in prominences with \texttt{MPI-AMRVAC}. We discuss this in more detail in Section \ref{sec:prom_dyn}.
  
\begin{figure*}
    \centering
    \includegraphics[width=0.9\textwidth]{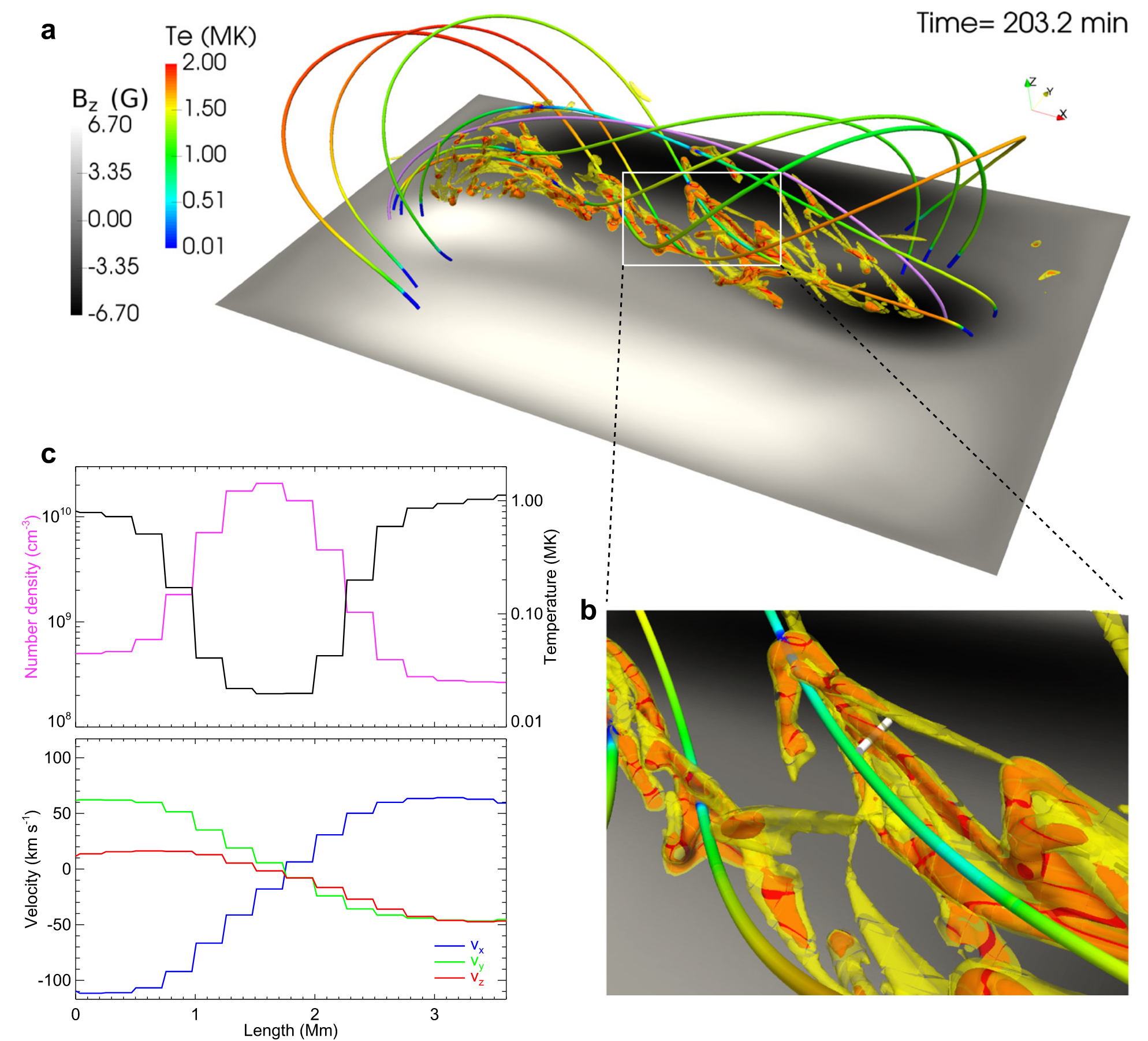}
    
    \caption{Magnetic field lines and prominence threads at 203.2 minutes. Panel a: a global view of the prominence with the density contours, magnetic field lines colored by temperature Te. A purple line represents the axis of the flux rope. Panel b: Zoomed-in view of the region in the white rectangular box in panel a. Panel c: The number density (pink curve), temperature (black curve), and velocities (blue curve $V_x$, green curve $V_y$, red curve $V_z$) along the cutting white line displayed in (b). Adapted from \citet{Xia:2016apj}. \textcopyright\ AAS. Reproduced with permission.}
    \label{fig:Xia_2016_1}
\end{figure*}

  \citet{Xia:2016apj} then considered the behaviour of condensed plasma within the 3D flux rope model of \citet{Xia:2014apj} using the following strategy: the initial condition of a flux rope preformed in the isothermal corona (in the absence of the energy equation) is modified to include a chromospheric layer (and localised heating in the now-introduced energy equation). The evaporated plasma condenses due to runaway cooling and forms the prominence material, which consists of multiple dynamic threads and blobs, as shown by the temperature isocontours in Figure \ref{fig:Xia_2016_1}. The authors studied the full cycle of the prominence lifetime, starting from evaporation and condensation and proceeding through drainage back to the lower atmosphere. In order to increase comparability with the real SDO/AIA observations, the authors constructed synthetic representations of their simulation in the SDO/AIA 211, 171, 304 \AA\ channels based on the \texttt{Chianti} database. Further details on synthetic images are presented in Section \ref{sec:synthetic}.
 
 Another class of models assumes that the magnetic field rises into the corona as a result of magnetic reconnection \citep{Rust:1994solphys}. As it rises, it can lift the denser plasma from the lower layers of the solar atmosphere. There are two distinct models: levitation and levitation-condensation. While the first type relies only on lifting chromospheric plasma into magnetic dips in the corona to form a prominence, the second type assumes that the lifted material is not yet condensed but originates from the lower, denser corona. Because the lifted plasma is denser than the surrounding corona, it experiences enhanced radiative losses, which in turn trigger thermal instability and lead to the formation of condensations. Both types have been modeled using the \texttt{MPI-AMRVAC} code. 

\begin{figure*}[!h]
    \centering
    \includegraphics[width=0.9\textwidth]{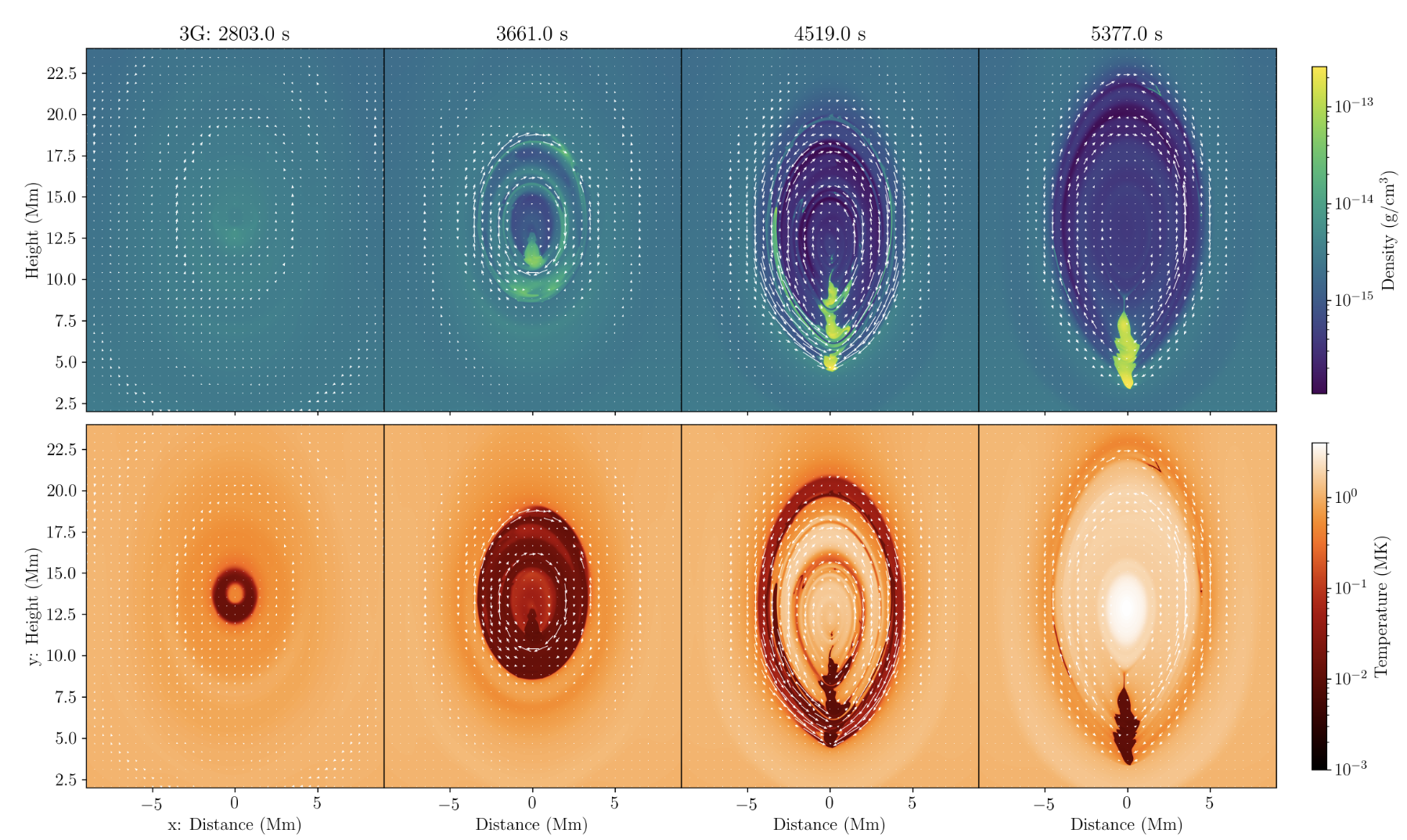}
    
    \caption{Evolution of the condensing material within the $3$ G formed flux rope using grid resolution $5.7$ km. First row: evolution of the density, second: temperature. The velocity field of relative magnitude is overplotted with the white arrows. Adapted from \citet{Jenkins:2021aap}. \textcopyright\ ESO. Reproduced with permission.}
    \label{fig:Jenkins_2021_3}
\end{figure*}

 \citet{Jenkins:2021aap} studied the 2.5D flux rope formation beginning with a periodic sheared arcade and introducing the levitation-condensation scenario. These authors targeted the small-scale properties, applying several levels of AMR reaching an at-the-time extreme spatial resolution of $5.7$ km. In this case, the initial atmosphere included only a gravitationally stratified corona. They also considered two cases of the $3$ and $10$ G. The evolution of the flux rope and plasma inside it in the high-resolution case with a weaker magnetic field is shown in Figure \ref{fig:Jenkins_2021_3}. The first two columns show that the plasma inside the flux rope starts to cool down until the first condensation appears in the density distribution. The next panel shows that the flux rope gradually becomes depleted of plasma. In the last panels, it is evident that the prominence occupies a small region at the bottom of the flux rope, and the flux rope cavity recovers its MK temperature. The authors distinguish several important stages: a) the lifted coronal plasma is slowly redistributed according to convective continuum instability before condensation formation \citep{Blokland:2011aap}; b) the increased radiative losses of the lifted plasma lead to the thermal instability and condensation and the field-projected Brunt-Väisälä frequency confirms that the condensations are forming out of the pressure balance; c) condensations flow towards the center of the dipped region of the flux rope; d) under the influence of the heavy plasma, the magnetic field lines are compressed and the gradient of the current density increases, leading to an increase of the resistive term in the induction equation. This leads to local magnetic dissipation (akin to reconnection) and the motion of plasma across the magnetic field, even under frozen-field conditions, known as mass slippage \citep[][]{Low:2012apj}. 
 
\begin{figure*}[!b]
    \centering
    \includegraphics[width=0.7\textwidth]{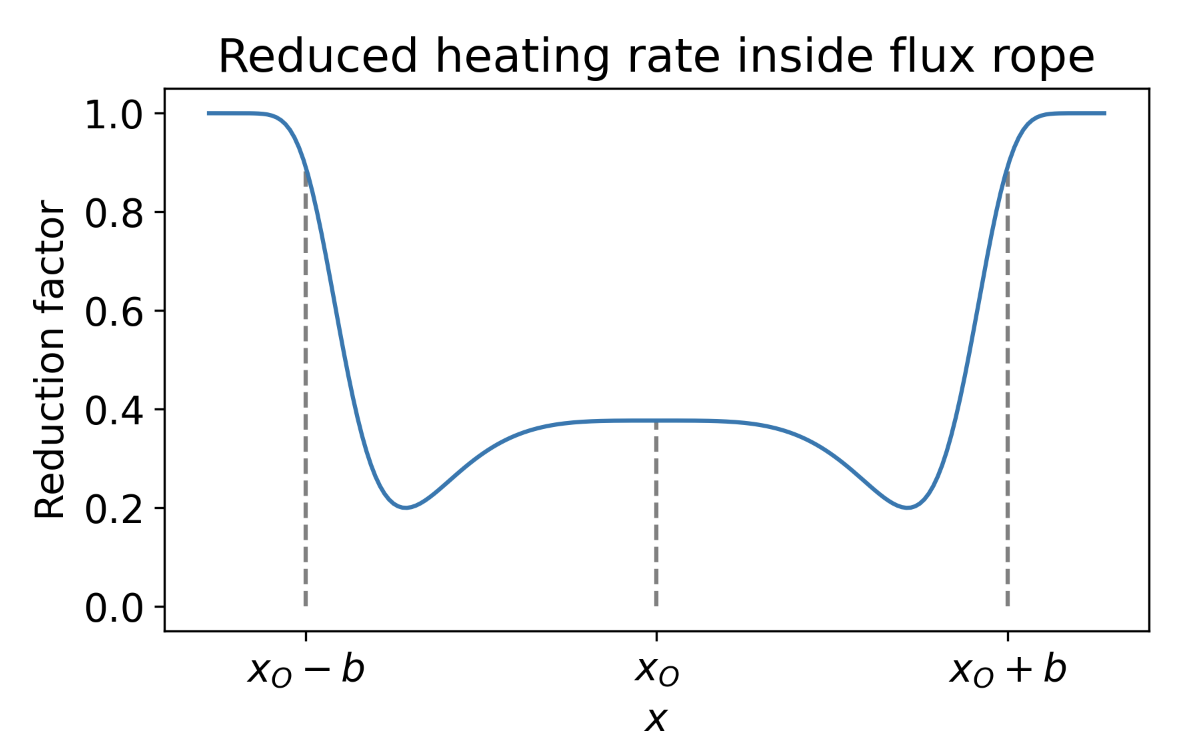}
    
    \caption{Reduction heating factor along the horizontal flux rope axis. Heating is reduced most at the flux rope edges since the field lines are longer. Outside the flux rope, there is no reduction. Adapted from \citet{Brughmans:2022aap}.}
    \label{fig:Brughmans_2022_1}
\end{figure*}

 \citet{Brughmans:2022aap} extended this study, exploring the influence of the different background heating prescriptions on thermal instability development. The authors explored two heating models in addition to the common exponential heating model. The first used the mixed heating model, which depends on some combination of the local magnetic field strength and density \citep{Mok:2008apjl}. The second is the reduced heating model that employs a dynamic detection of the flux rope; a mask within which the heating is reduced to account for the 3D effects of a flux rope, assuming that the heating is applied with some characteristic length scale from the lower atmosphere where the flux rope is anchored. Such a reduced heating profile across the flux rope is shown in Figure \ref{fig:Brughmans_2022_1}. The authors also tested the case of shearing or antishearing footpoints motions in combination with converging. The results of this study showed that the combination of shearing motions with the mixed models hinders the production of condensation. The reason is the temperature increase in the flux rope center caused by the Ohmic heating. However, in the reduced mixed heating model, this problem is solved, and the reduced heating leads to large condensations by eliminating the residual flux rope heating. This creates a cooler flux rope and prominence with almost uniform density and pressure. This study demonstrated that the choice of background heating has a significant impact on the formation, evolution, and morphology of condensations in the 2.5D models. It is clear from the wildly varying results that more work is needed in this area to constrain the parameter space.

\begin{figure*}
    \centering
    \includegraphics[width=0.9\textwidth]{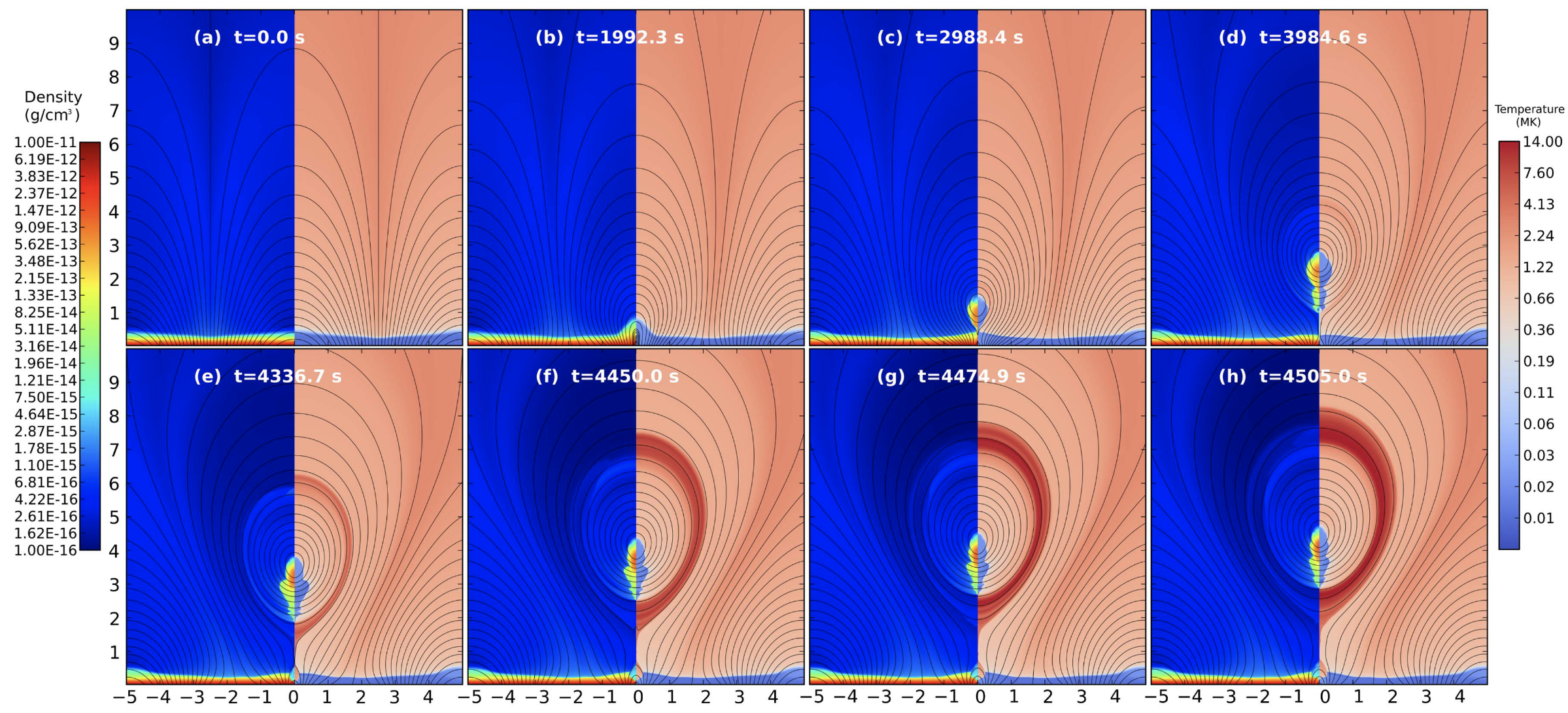}
    
    \caption{The evolution of density and temperature with magnetic field lines overlaid (in black). Adapted from \citet{Zhao:2017apj}. \textcopyright\ AAS. Reproduced with permission.}
    \label{fig:Zhao_2017_1}
\end{figure*}

 \citet{Zhao:2017apj} and \citet{Zhao:2022apj} studied the scenarios in which the prominence forms not during the phase where the flux rope is stable but during its eruption. In this review, we do not focus on the eruption itself, but rather on the formation of prominence plasma under these conditions. \citet{Zhao:2017apj} performed the numerical experiment using 2.5D periodic sheared arcades and formed the flux rope via footpoints shearing and converging motions, including the lower solar atmosphere (i.e., chromosphere and transition region). The evolution of the density, temperature, and magnetic field is shown in Figure \ref{fig:Zhao_2017_1}. As more field lines reconnect, the flux rope rises higher. From this figure, the rising flux rope lifts a significant amount of the cold and dense material directly from the chromosphere, where the bottom of the flux rope is initially located. Assuming a very conservative prominence length of only $20$ Mm, the authors estimated the total mass, $3.2\times10^{14}$ g in agreement with observation by \citet{Gilbert:2005apj}. However, the authors note that plasma drainage to the footpoints is not possible in this case due to the limitations of the 2.5D model and the flux rope’s main axis being oriented along the invariant direction. This means the amount of plasma lifted in this model could be overestimated. 
 
 \citet{Zhao:2022apj} employed a slightly different approach in 2.5D modeling, assuming that the flux rope is represented by the cylindrical current located in the corona, with the main flux rope axis aligned with the invariant direction. A mirror current is placed below the bottom boundary to produce the upward `hoop' force. The quadrupolar magnetic field was chosen as the background field to provide the downward force. The parameters of the cylindrical current field and the quadrupolar magnetic field were chosen so that the resulting configuration was not in equilibrium, with a net positive (upwards) hoop force. When the flux rope then rises, reconnection outflows inject dense, hot plasma from the low corona into the flux rope, where it accumulates. The flux rope lifts this plasma even higher, where this plasma cools down and condenses due to the thermal instability. Another interesting aspect of this model is the chromospheric plasma supply by the plasmoids formed at the lower heights in the current sheet. In this way, the plasmoids transport chromospheric plasma, making a significant contribution to the mass budget of flux ropes and the formation of prominences \citep[][]{Zhang:2014apj,Zhang:2021aap}. 
 
\begin{figure*}
    \centering
    \includegraphics[width=0.7\textwidth]{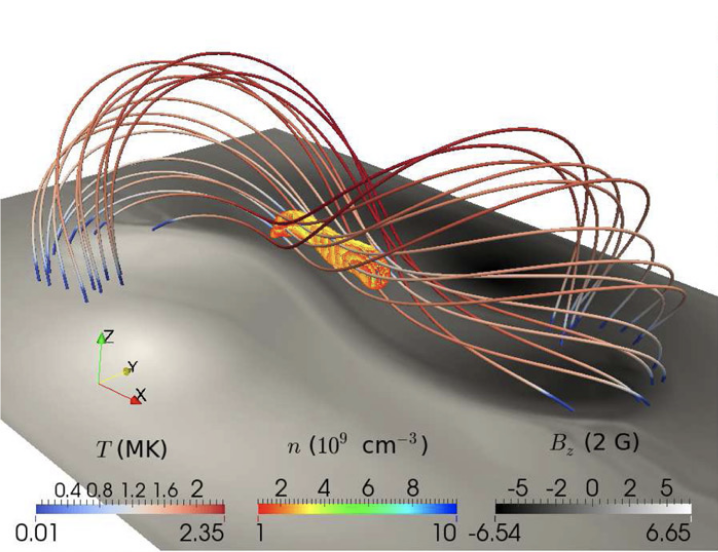}
    
    \caption{Flux rope with a prominence formed in-situ at $150\mins$. The magnetic field lines are colored by temperature in blue to red, the prominence is colored by density in a rainbow of colors, and the bottom magnetogram is in gray. Adapted from \citet{Xia:2014apjl}. \textcopyright\ AAS. Reproduced with permission.}
    \label{fig:Xia_2014b_1}
\end{figure*}

 Moving to 3D, \citet{Xia:2014apjl} used footpoint motions to form a flux rope from an initial periodically sheared arcade embedded in a non-adiabatic, gravitationally stratified corona (Figure \ref{fig:Xia_2014b_1}). After the formation of a stable flux rope, plasma lifted into the corona within magnetic dips condenses and forms a prominence, while other plasma drains to the footpoints along the helical magnetic field lines. This can be considered a 3D example of a merged levitation-condensation (given the lifted flux rope during relaxation) and evaporation-condensation (given the subsequent pressure gradient siphoning) model, albeit not falling exactly within either definition. The slow drainage of the material forms a feature that resembles a right-bearing prominence barb. The authors obtained synthetic images in the different SDO/AIA channels and confirmed the appearance of the dark cavity, prominence, and horn configuration in agreement with the observations \citep{Fuller:2009apj,Schmit:2013apj}.    
 
\begin{figure*}[!h]
    \centering
    \includegraphics[width=0.9\textwidth]{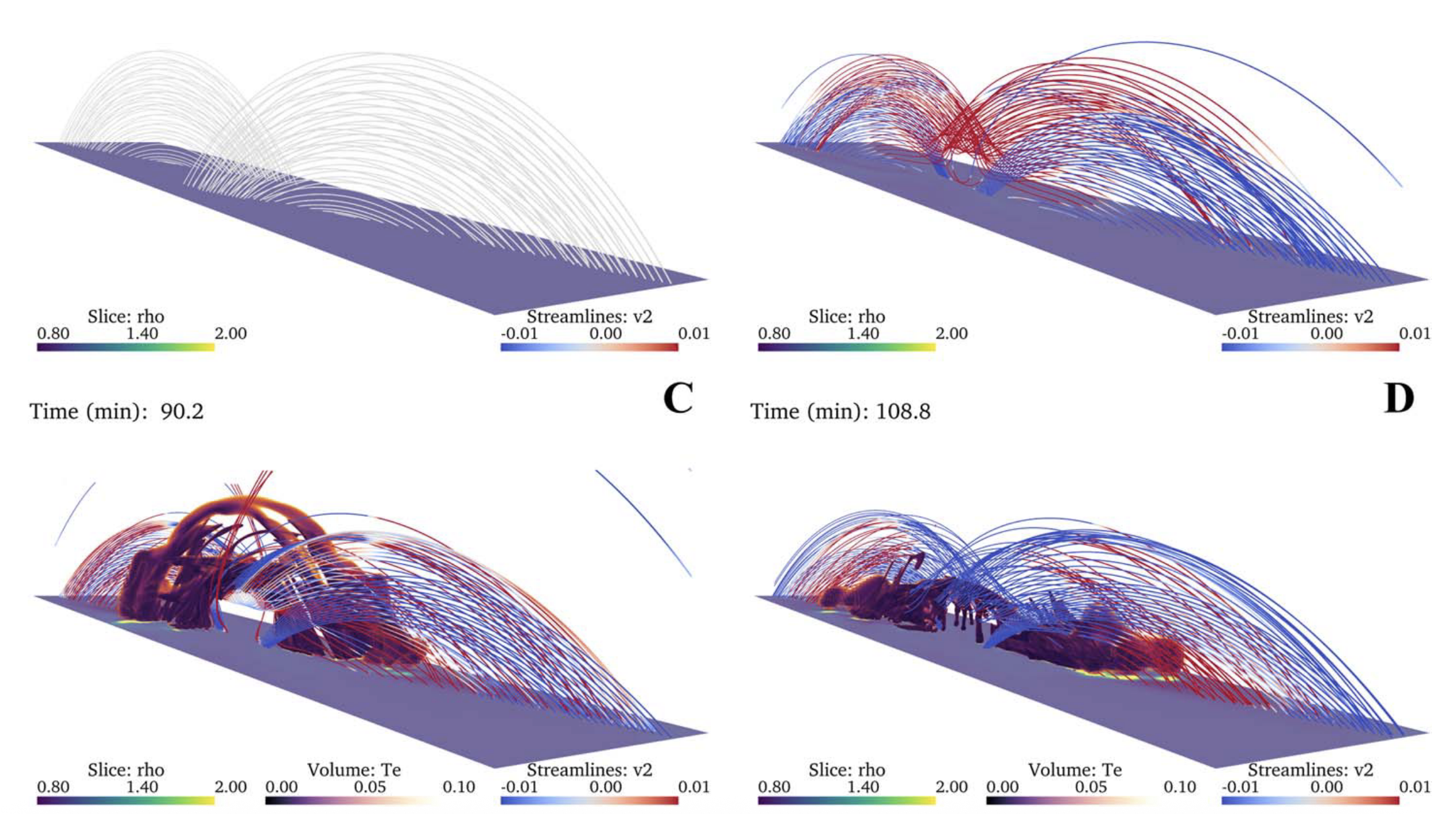}
    
    \caption{Temporal evolution of the magnetic field and prominences/coronal rain formation. The bottom plane shows the density (in units of $2.3\times 10^{-15}\gcm$), while the magnetic field lines are colored by the vertical velocity of the plasma $V_y$ (in units of $116\kms$). Panel a: The initial magnetic arcades. Panel b: The reconnection of the magnetic field lines and altering the magnetic topology into a flux rope. Panel c: First condensations of prominence and coronal rain occurrence. Panel d: The Rayleigh–Taylor instability signatures in the central part of the filament. Adapted from \citet{Donne:2024apj}.}
    \label{fig:Donne_2024_1}
\end{figure*}

\begin{figure*}[!b]
    \centering
    \includegraphics[width=0.9\textwidth]{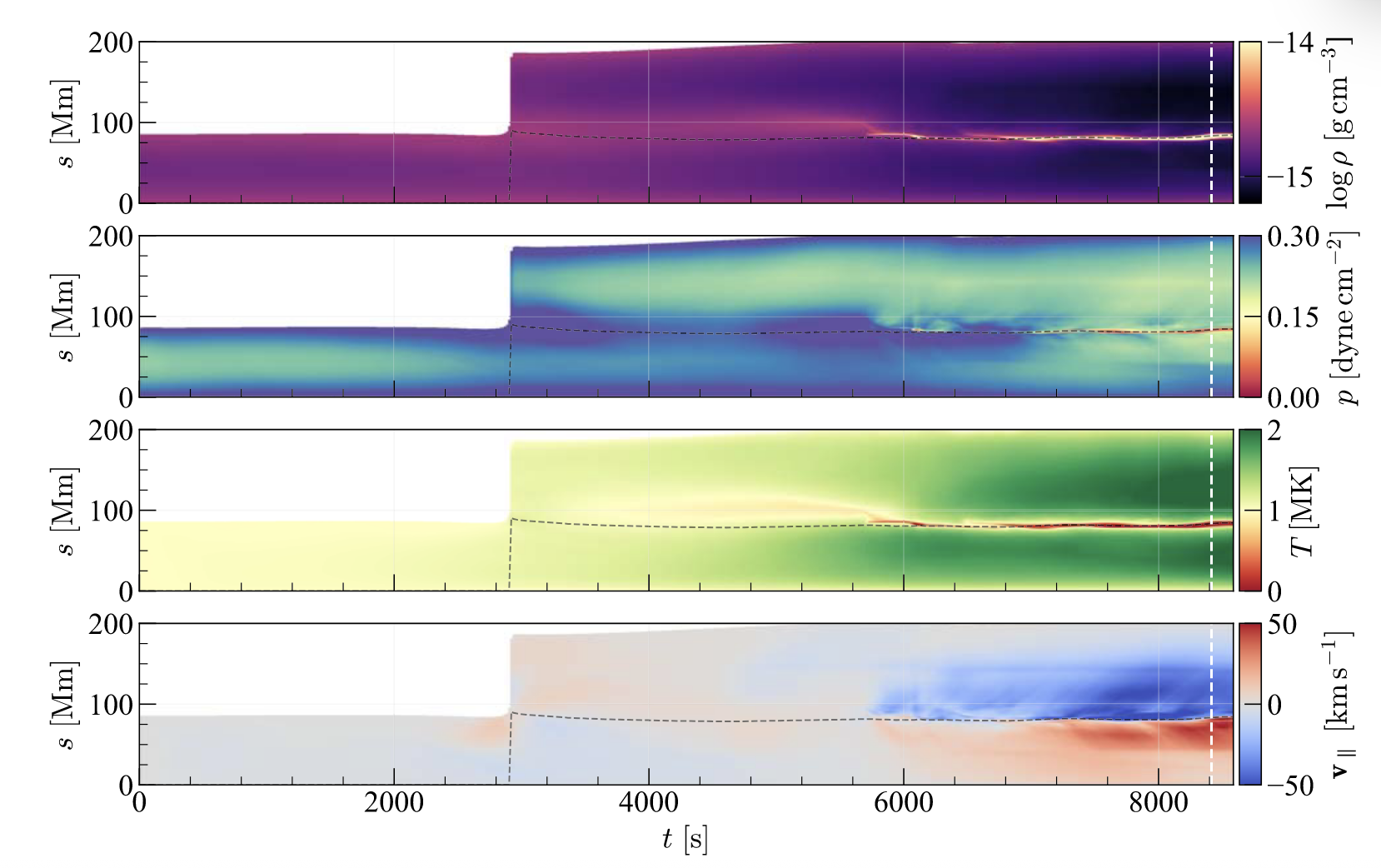}
    
    \caption{Temporal evolution of the logarithmic density, the pressure, the temperature, and the tangent velocity along a magnetic field line. The coordinate $s$ is the coordinate along the magnetic field line. The black dashed line indicates the position of the dip within the magnetic field line. The noncolored region in the left corner indicates the stage before the reconnection of the magnetic field line. Adapted from \citet{Donne:2024apj}.}
    \label{fig:Donne_2024_2}
\end{figure*}

 In a recent study, \citet{Donne:2024apj} demonstrated in situ prominence formation within a coronal flux rope using 3D modeling at a very high spatial resolution of $20.8$ km. This study represents the currently highest resolution simulation of prominence formation that is non-periodic and includes a magnetic connection to the bottom of the domain, so far without the inclusion of a chromosphere. The evolution of the magnetic field lines and condensations is shown in Figure \ref{fig:Donne_2024_1}. The modeling is more directly related to the levitation-condensation mechanism as the results show that enhanced radiative losses within the lifted plasma lead to a local pressure drop within the flux rope, which not only triggers the formation of condensations but also drives siphon flows from the footpoints along the field lines due to the resulting pressure gradient (once again akin to features found in the evaporation-condensation model). This moment is clearly shown in Figure \ref{fig:Donne_2024_2}, where the main parameters of density, gas pressure, temperature, and longitudinal velocity are traced along one magnetic line of the flux rope. Starting from approximately $6000$~s, there is an inflow of plasma from the boundary towards the center of the flux rope.

 \begin{figure*}[!h]
    \centering
    \includegraphics[width=0.7\textwidth]{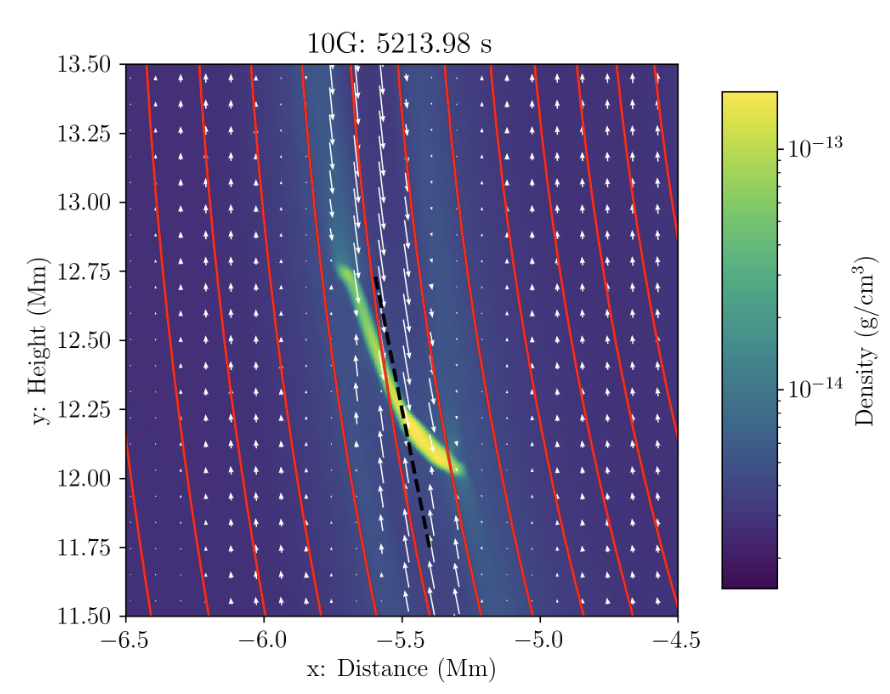}
    
    \caption{Distribution of density within the forming filamentary condensation. The velocity field of relative magnitude is denoted by white arrows. Magnetic field lines are overplotted as solid red lines. Adapted from \citet{Jenkins:2021aap}. \textcopyright\ ESO. Reproduced with permission.}
    \label{fig:Jenkins_2021_5}
\end{figure*}

  
 An important question regarding prominences is related to the orientation and alignment of their threads to the magnetic field and the flux rope's axis. In the observational case, it is often difficult to reconstruct the 3D magnetic field structure. Therefore, the observations rely on the plasma dynamics, highlighting the otherwise invisible magnetic field. As numerical simulations have direct access to such quantities and their interplay, they clearly present an important tool for diagnosing and constraining any interpretation of observations.
 
 \citet{Hermans:2021aap} performed a 2D experiment of the corona permeated by a diagonal magnetic field without gravity but with non-adiabatic terms for thermal conduction, radiative losses, and background heating. The idea was to investigate the triggering of the linear thermal instability by a slow magnetoacoustic wave. The authors found that the initial condensation shows a growth in the direction perpendicular to the magnetic field lines. Eventually, due to the ram pressure, these structures become fragmented, forming the threads that follow the magnetic field.
 
 Similarly to those studies, using extreme spatial resolution simulations within a 2.5D flux rope, \citet{Jenkins:2021aap} obtained that the initial condensation tends to form perpendicular to the hosting magnetic field (Figure \ref{fig:Jenkins_2021_5}). For the 2.5D dipped arcade structure, \citet{Zhou:2023aap} and \citet{Jercic:2024aap} found a similar formation of the condensation growing perpendicular to the magnetic field lines. In both cases, this formation was observed for symmetric heating at the footpoints, although in the case of \citet{Zhou:2023aap}, the heating varied over time. This is currently understood to be related to the influence of anisotropic thermal conduction, as explored by \citet{VanderLinden:1991solphysa,VanderLinden:1991solphysb}, that, even though it is very small in the coronal context (entirely neglected in these models but thus reduced to a numerical diffusive lengthscale), can lead to highly structured eigenfunctions. This requires more dedicated study, but currently lies far beyond our computational capabilities in multidimensional MHD.
 
 On a more global scale, 3D simulations continue to explore the acute angles between the filament spine, threads, and the hosting magnetic field axis. \citet{Xia:2014apjl} obtained the prominence formed in situ in the magnetic dips of the flux rope. They pointed out that the angle between the prominence axis and the magnetic field vector in the horizontal plane was around $18\degree$. Using a 3D prominence model and the synthetic images in 171 and 193 \AA, \citet{Jenkins:2022natas} has confirmed the result by \citet{Xia:2014apjl}, obtaining the same acute angle of the filament structure with respect to the flux rope axis in agreement with observations \citep{Bommier:1994solphys}. Any relationship herein is, of course, highly dependent on the specific magnetic topology adopted, but it at least confirms the observational conclusion of the presence of a sheared magnetic field.
 
 \citet{Guo:2022aap} studied the formation of the prominence threads in the highly and weakly twisted magnetic flux rope limits. The orientation of the threads with respect to the filament spine and flux rope axis is shown in Figure \ref{fig:Guo_2022_2}. For the low-twist model, they obtained an angle between the threads and the filament axis that varied from the center of the flux rope to the footpoint of the flux rope, ranging from $15\degree$ to $30\degree$ in agreement with observations \citep{Hanaoka:2017apj}. The angle between the polarity-inversion line and the prominence axis was around $17\degree$. The corresponding angles for the high-twist model were $33\degree$, $11\degree$, and $9\degree$, respectively. Importantly, they noticed that not only did the prominence body deviate from the flux rope axis, but that only the lower quarter of the radial extent of the flux rope hosted prominence plasma. This important result suggests that we must be cautious in interpreting the observational morphology of the prominence plasma, as it is not necessarily as representative of the prominence's magnetic field as initially proposed.
 
 In summary, the evaporation–condensation and levitation–condensation scenarios have been extensively explored using the \texttt{MPI-AMRVAC} code, showing the key role of thermal instability in prominence formation. These models demonstrate that the structure, dynamics, and thermodynamics of prominences are highly sensitive to the underlying magnetic topology, heating prescriptions, and spatial resolution. In particular, recent 3D simulations have proven essential for understanding the complete prominence mass cycle. Furthermore, the 3D simulations clarify the alignment of the prominence thread with the magnetic field and flux rope axis.
 
   \section{Modeling of Prominence Dynamics}\label{sec:prom_dyn} 
  Prominences are very dynamic structures showing a variety of different motions. There are downflows and upflows, counterstreaming motions, and oscillations, among other phenomena. Different dynamics, such as oscillations or downflows, are observed just before the global loss of equilibrium of the solar prominence, suggesting that these dynamics can be an early precursor to prominence eruptions \citep{Isobe:2006aap, Chen:2008aap, Bi:2014apj, Jenkins:2018solphys}. Using numerical simulations, we can reproduce the different types of motions and study their driving and damping mechanisms. In some experiments, we attempt to trigger a complex dynamic evolution, while in others, the dynamics are a natural consequence of prominence formation.

   
   \subsection{Prominence Oscillations}
   Prominence oscillations are repetitive motions along or transverse with respect to the magnetic field. These oscillations are classified according to their amplitude. The oscillations above $10\kms$ are called large amplitude oscillations (LAOs). Typically, LAOs are triggered by external perturbations such as Moreton and EIT (EUV) waves \citep{Eto:2002pasj,Okamoto:2004apj,Gilbert:2008apj,Asai:2012apjl,Shen:2014apj1,Xue:2014solpol,Takahashi:2015apj} or nearby activity such as jets, flares \citep{Jing:2003apjl,Jing:2006solphys,Vrsnak:2007aap}. A large portion of the filament oscillates, reflecting the global characteristics of the plasma and the magnetic field structure. Typically, these oscillations are present in a significant portion of the prominence and exhibit a more global character. The oscillations with amplitudes below $10\kms$ are called small amplitude oscillations (SAOs). In contrast, SAOs are, in general, not related to activity in the immediate area external to the prominence or any large scale energetic disturbance, and only a small part of the prominence is involved.
   
   The first catalog based on several months of the Global Oscillation Network Group (GONG) observations in H$\alpha$ by \citet{Luna:2018apjs} provided a statistical study of the oscillatory characteristics of these motions. The longitudinal oscillations, those being motions along the magnetic field, have periods of around one hour \citep{Jing:2003apjl, Jing:2006solphys, Vrsnak:2007aap, Zhang:2017apj}. Observations show that these oscillations are usually strongly damped, with the distribution of the damping time per period peaking at 1.25. For the vertical oscillations, those being radially aligned, the observed periods and damping times vary in the range 11-37 minutes and 25-99 minutes \citep{Eto:2002pasj, Okamoto:2004apj, Gilbert:2008apj, Gosain:2012apj, Liu:2012apj, RiuLiu:2013apj, Shen:2014apj1, Xue:2014solpol, Zhang:2018apj}, respectively. For the transverse oscillations, those being parallel to the solar surface, it was found that the period varies in the range of 30 - 80 minutes, and the damping time is equal to several oscillatory periods, also indicating strong damping.
   
   The main questions surrounding these phenomena are related to the restoring force, triggering factors, and damping mechanisms. More information on the classification of prominence oscillations and results from earlier studies can be found in the review by \citet{Arregui:2018lrsp}. In general, prominence oscillations represent a powerful tool for seismology as they enable the obtaining of properties of the local magnetic field and plasma from the observed periods and damping times. These oscillations can be reproduced in various magnetic geometries through numerical modeling, allowing for a parametric study.

\begin{figure*}
    \centering
    \includegraphics[width=0.7\textwidth]{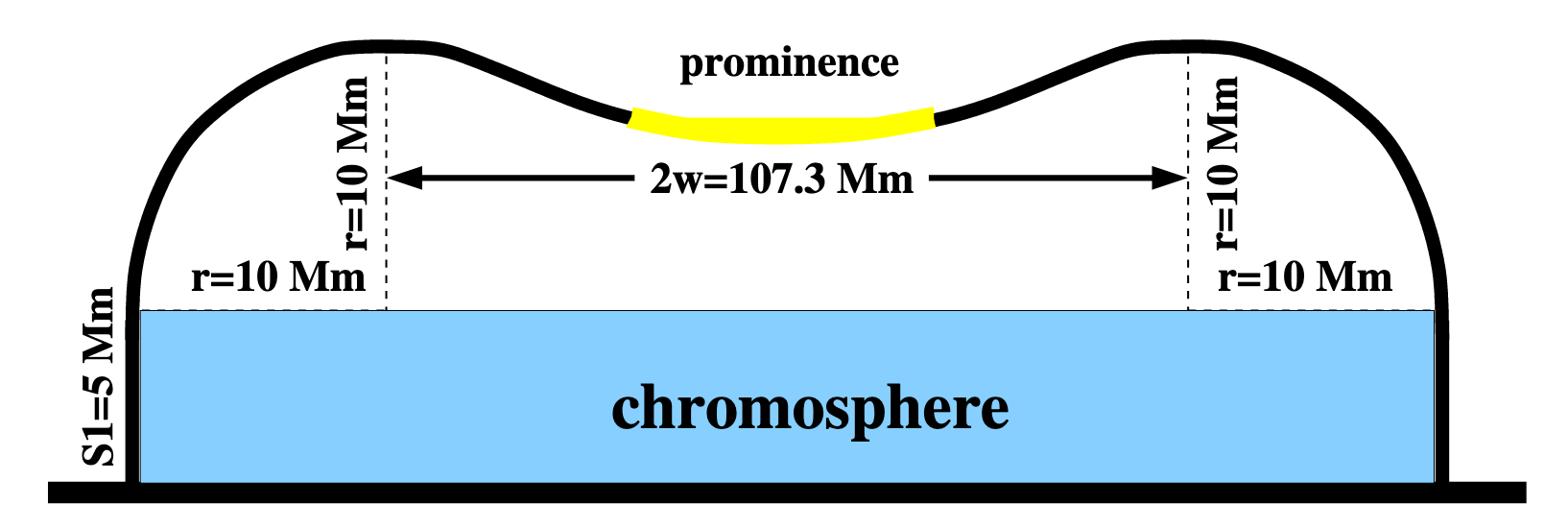}
    
    \caption{Flux tube configuration used for the 1D HD simulation of the prominence oscillation. Note that the horizontal and vertical sizes are not to scale. Adapted from \citet{Zhang:2012aap}. \textcopyright\ ESO. Reproduced with permission.}
    \label{fig:Zhang_2012_1}
\end{figure*}

\begin{figure*}[!h]
    \centering
    \includegraphics[width=0.5\textwidth]{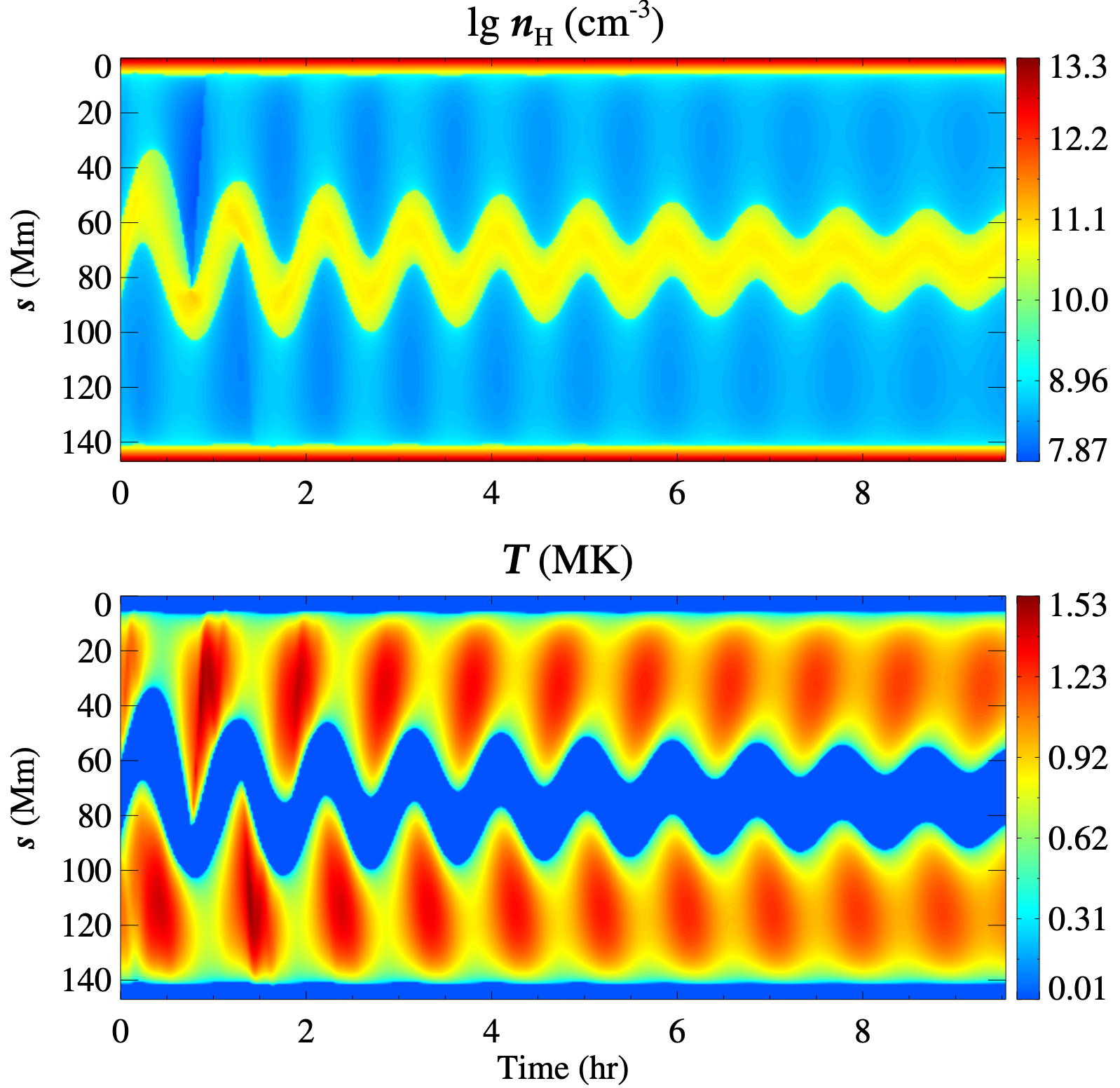}
    
    \caption{Time-distance diagrams of the density (top) and the temperature (bottom) along the flux tube, which indicate that the prominence experiences a damped oscillation. Adapted from \citet{Zhang:2012aap}. \textcopyright\ ESO. Reproduced with permission.}
    \label{fig:Zhang_2012_4}
\end{figure*}

   
   The first attempt to explain the observed prominence longitudinal oscillations using a 1D geometry in the \texttt{MPI-AMRVAC} code has been made by \citet{Zhang:2012aap}. The geometry of the tube hosting the prominence threads is shown in Figure \ref{fig:Zhang_2012_1}. In order to estimate the period and damping time from the observation or simulation, the authors fitted the trajectory of the oscillating prominence with a damped sinusoidal function as follows: $s = A \sin\left( \frac{2\pi}{P} t + \phi \right) e^{-t/\tau}+s_0$, where $s$ in the distance, $s_0$ is the initial position, $A$ is the initial amplitude, $P$ is period, $\tau$ is damping time, and $\phi$ is a phase shift. The observational period and damping time obtained using the artificial tracing slit were $52$ and $133$ minutes, respectively. To explain the observed characteristics of the prominence oscillations, the 1D thread was perturbed by introducing additional momentum to the domain, resulting in a velocity perturbation of approximately $40\kms$. The time-distance diagrams of the temperature and density demonstrating oscillations are shown in Figure \ref{fig:Zhang_2012_4}. The model reproduced the observed oscillation period, $56\mins$, but the damping time, $202$ minutes, exceeded the observed one. \citet{Zhang:2012aap} concluded that the damping mechanism of these oscillations cannot be explained only by the non-adiabatic effects, i.e., radiative losses and thermal conduction, but other mechanisms should be taken into account, such as wave leakage or mass accretion. 
    
    To explore this conclusion more concretely, \citet{Zhang:2013aap} used the model from \citet{Zhang:2012aap} to perform a systematic study, varying parameters such as prominence length and mass, the time duration of chromospheric heating and evaporation, prominence height, and the depth of the magnetic dip. These authors demonstrated that the period of oscillations varies weakly with the length and height of the prominence, as well as the perturbation velocity. In all cases, the period of the longitudinal oscillations agrees well with the pendulum model, where the main restoring force is the gravity projected onto the magnetic field. Hence, the period depends only on the curvature of the magnetic field and is defined as $P_{pendulum}=2\pi \sqrt{R/g}$, where $R$ is the radius curvature and $g$ is the solar gravitational acceleration \citep{Luna:2012apjl}. For the damping, it was found that if momentum is removed via mass drainage\,--\,that is, the overflowing of material past the shoulders of the magnetic field caused by particularly strong flows\,--\,significantly reduces the damping time compared to damping caused solely by non-adiabatic effects.
   
   
    
   
  \citet{Zhou:2017apj} extended previous 1D studies of prominence oscillations, changing the configuration of the flux tube and adding another dip. The aim is to investigate the potential energy transfer between the different threads within the same tube. For these purposes, they vary the geometries of the active (the one perturbed) and passive threads. When considering the identical dips and one thread is perturbed, the energy is efficiently transferred to the passive thread, and it starts oscillating with a delay due to the finite speed of a sound wave. However, the damping time of the passive thread is $2-2.5$ times longer than in the case of the single-dip configuration. Then, the oscillations in the case of non-identical dips have been studied. In this case, the period of the shorter dip does not depend on the thread-thread interaction. The shorter dip suggests a smaller radius of curvature. In this case, the main restoring force is still the projected gravity, as suggested by the pendulum model \citep{Luna:2012apj,Luna:2016apj}. On the contrary, the period inside the longer dip is significantly reduced when compared to the single-dip case. For the longer dips with a larger radius of curvature, the gas pressure gradient also plays a significant role in the restoring process \citep{Luna:2012apj,Terradas:2013apj}. Therefore, the sound waves excited by the active short dip act as an external driving force on the passive long dip. The damping times are significantly influenced by the thread-thread interaction, leading to a more complex behavior than in the single-dip case. For the active thread, the damping time decreases with increasing magnetic dip length, consistent with the behavior observed in isolated dips. In contrast, the damping time of the passive thread increases with dip length. These results are important to consider when interpreting periods and damping times derived from observations.

\begin{figure*}[!h]
    \centering
    \includegraphics[width=0.7\textwidth]{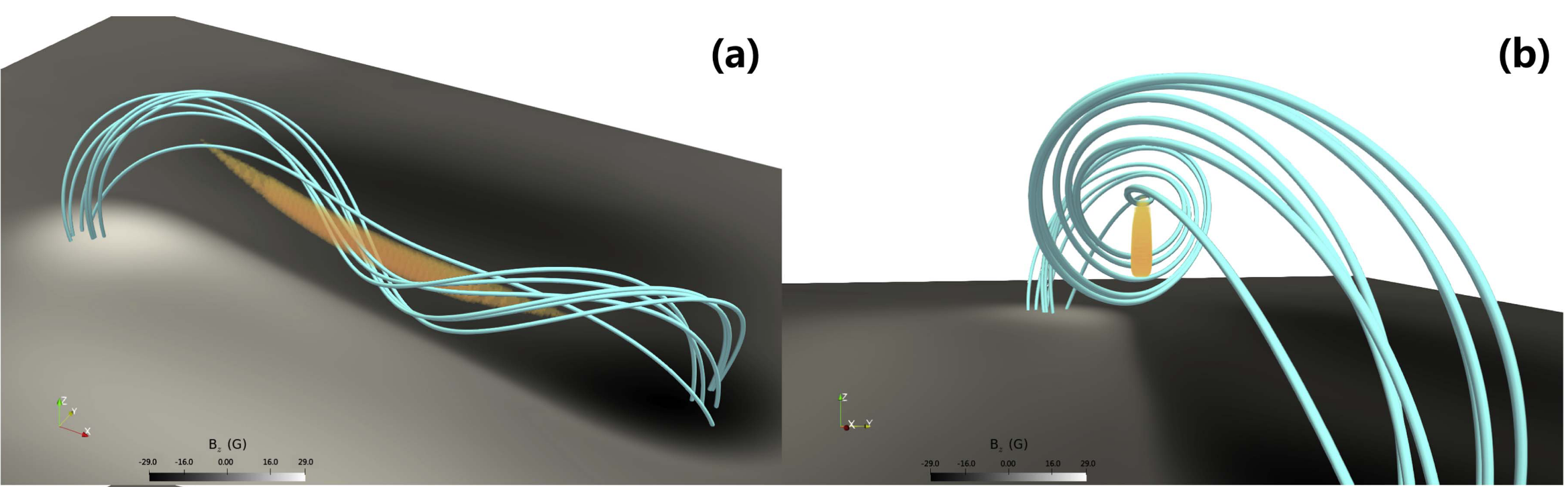}
    
    \caption{Two perspectives of the prominence hosted by the 3D force-free magnetic flux rope. In this figure, the yellow isosurface indicates prominence plasma with a density 20 times that of the background coronal density, and the light blue lines represent selected magnetic field lines. The grayscale in the bottom plane indicates the vertical component of the magnetic field. Adapted from \citet{Zhou:2018apj}. \textcopyright\ AAS. Reproduced with permission.}
    \label{fig:Zhou_2018_1}
\end{figure*}

   The 1D simulations provided us with important information regarding the period and damping times of the longitudinal oscillations, achieved with extremely high resolution. However, the 1D restriction does not allow us to study other types of oscillations, i.e., transverse to the magnetic field. Using the \texttt{MPI-AMRVAC} code, multiple studies have been conducted using 2D and 3D geometries. \citet{Zhou:2018apj} formed their 3D flux rope using converging and shearing motions in the gravitationally stratified corona. Then, following \citet{Xia:2016apj}, the isothermal corona was replaced with an idealized chromosphere, transition region, and corona. The radiative losses and thermal conduction have not been taken into account. Therefore, the formation due to thermal instability is not possible. After the flux rope forms and relaxes, a prominence is artificially inserted by increasing the density within the magnetic dips while maintaining an unchanged gas pressure. Then, the flux rope and prominence freely evolve until velocities in the domain are smaller than $2\kms$. The resulting magnetic field and prominence configuration are shown in Figure \ref{fig:Zhou_2018_1}. Then, the authors applied velocity perturbations in different directions relative to the magnetic field, namely longitudinal and transverse. The transverse perturbations included horizontal and vertical components. 
   
\begin{figure*}[!b]
    \centering
    \includegraphics[width=0.7\textwidth]{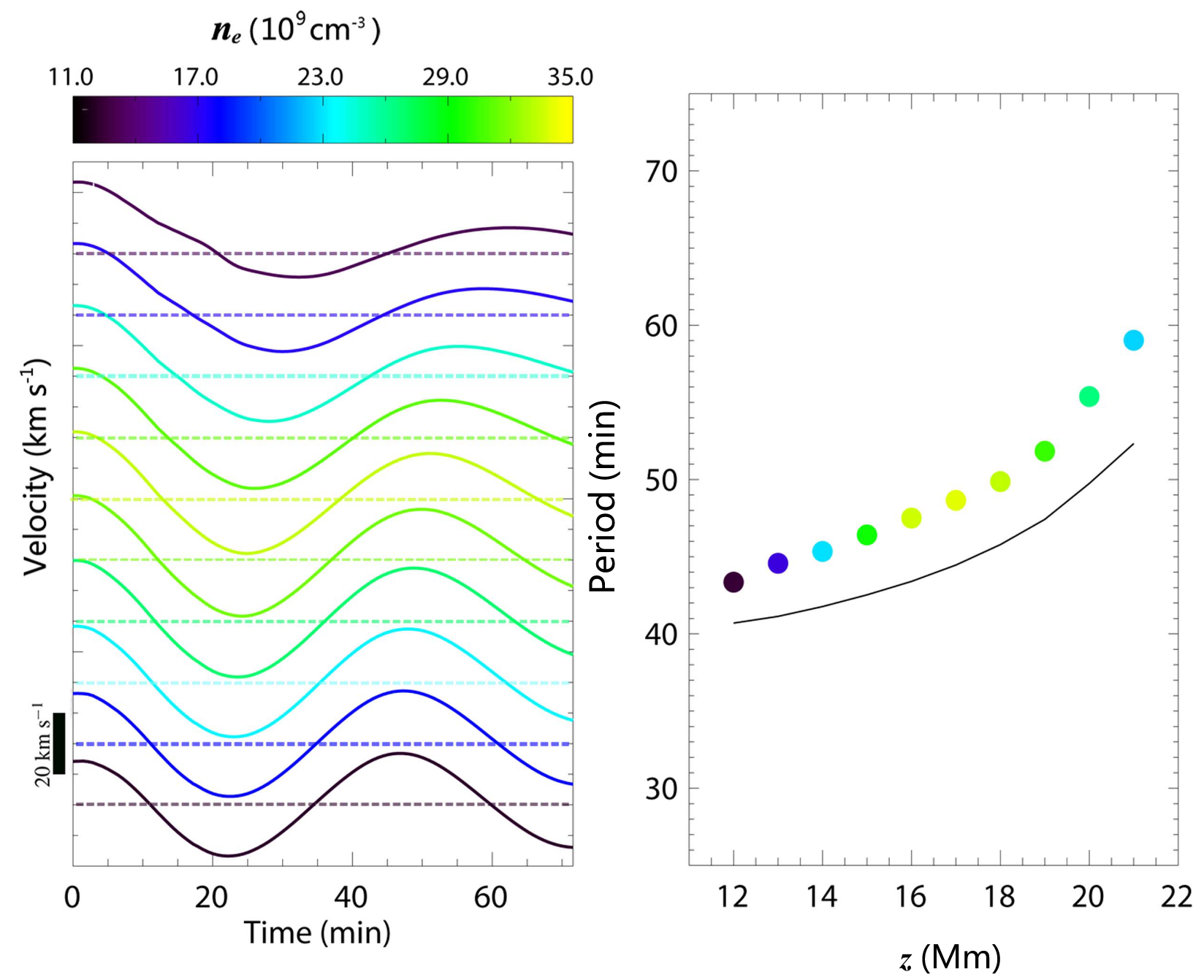}
    
    \caption{Left: Temporal evolution of the longitudinal velocity at the center of mass of 10 selected magnetic field lines at different heights. Colors indicate the initial number density at the center of mass of each field line, as shown in the color bar. The velocity scale is given in the lower-left corner. Right: Oscillation periods of the same field lines as a function of height. Solid circles represent simulation results, while the black solid line shows the theoretical pendulum periods. Adapted from \citet{Zhou:2018apj}. \textcopyright\ AAS. Reproduced with permission.}  \label{fig:Zhou_2018_2}
\end{figure*}
   
\begin{figure*}[!h]
    \centering
    \includegraphics[width=0.45\textwidth]{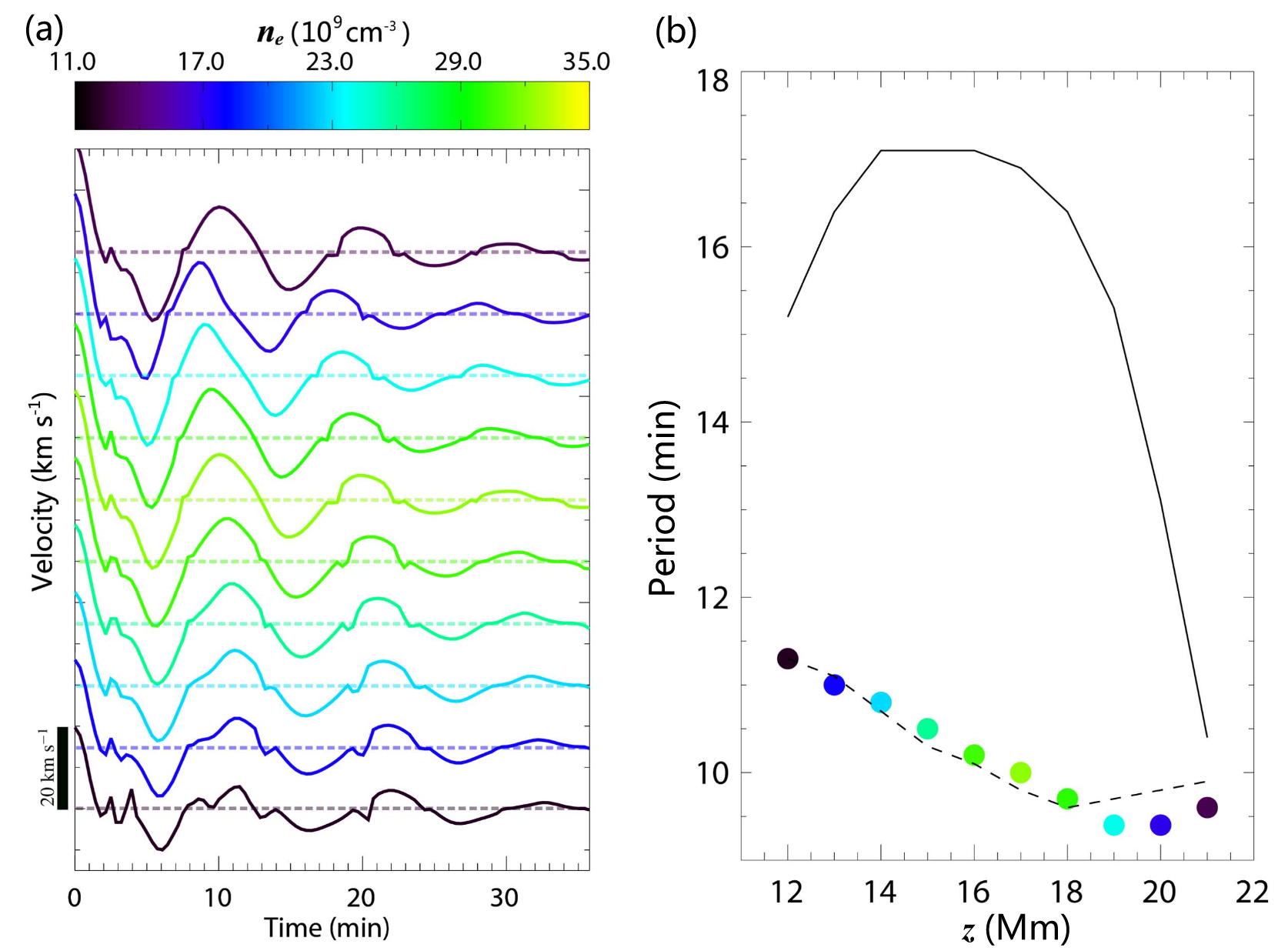}
        \includegraphics[width=0.45\textwidth]{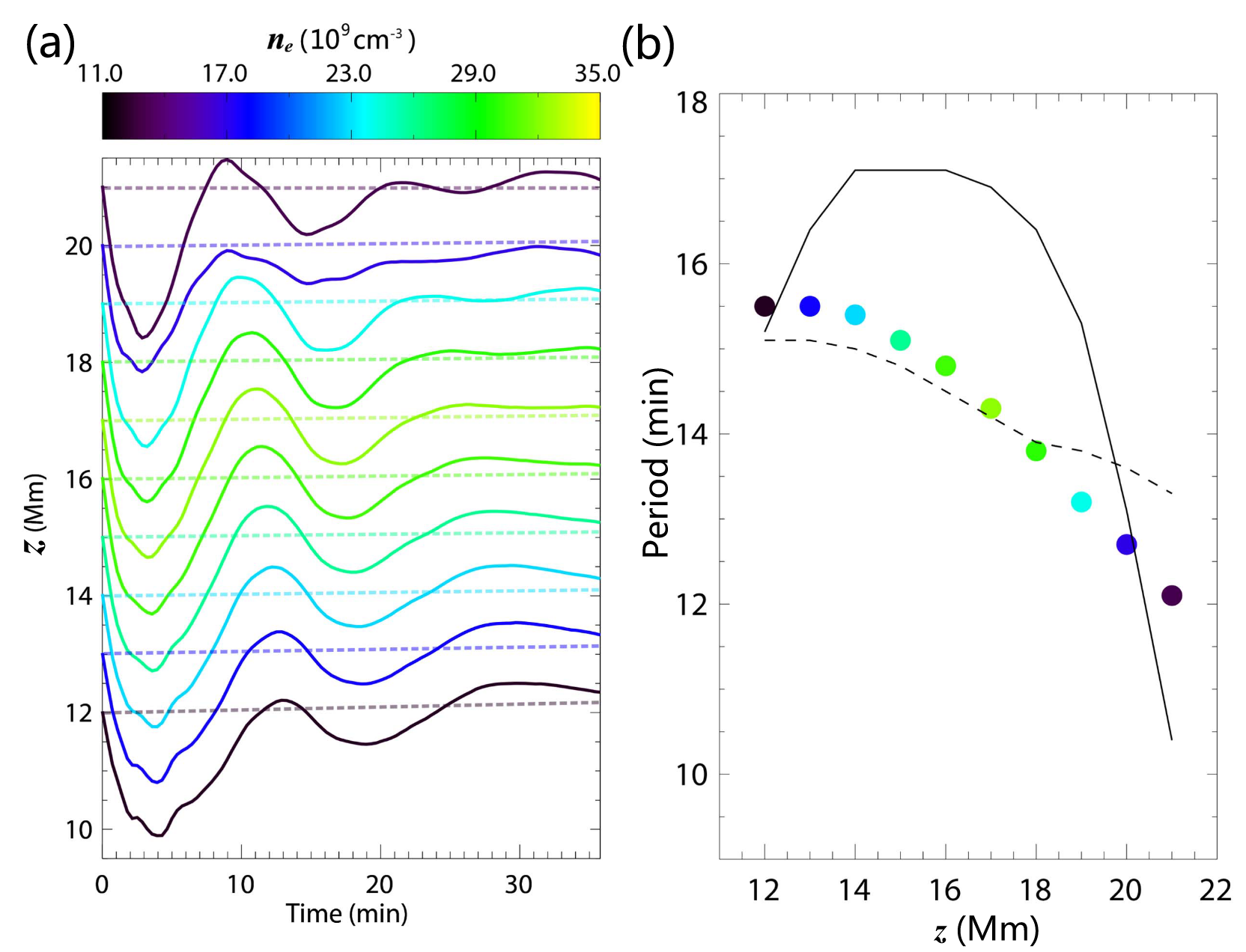}
    \caption{Same as Figure \ref{fig:Zhou_2018_2}, but showing the y-component of velocity for horizontal transverse oscillations (left panels a and b) and the z-component for vertical transverse oscillations (right panels a and b). The black solid line represents the theoretical period from the 1D string model \citep[][]{Joarder:1992aap2}, while the black dashed line corresponds to the period predicted by the 2D slab model \citep{Diaz:2001aap}. Adapted from \citet{Zhou:2018apj}. \textcopyright\ AAS. Reproduced with permission.}
    \label{fig:Zhou_2018_3}
\end{figure*}

 The period of the longitudinal oscillations roughly agrees with the pendulum model, increasing with height, where the radius of curvature of the field lines is larger. The authors noted that the periods are approximately 10 percent longer than the pendulum period, using the actual values of the field line curvature (Figure \ref{fig:Zhou_2018_2}, left). Their further consideration has shown that the magnetic field lines cannot be assumed to be rigid when the heavy prominence plasma moves along them. They defined the parameter, $\delta$, which estimates the ratio between local gravity and magnetic pressure force. If this parameter is close to unity, the motion of heavy prominence plasma deforms the magnetic dips, and the plasma follows a shallower trajectory than the actual field line dip. This results in a longer period, as per the pendulum model.
   
 The 3D experiment reveals that the periods of the transverse horizontal and vertical oscillations are shorter than those of the longitudinal oscillations. Transverse oscillations show only a slight decrease with height. Both periods have been compared to the analytical formulas of the oscillations of the 2D slab \citep{Diaz:2001aap}. The comparison shows a very good agreement (dashed lines in Figure \ref{fig:Zhou_2018_3}). The 1D string model developed by \citet{Joarder:1992aap2} fails to reproduce the periods of horizontal oscillations accurately and performs less accurately than the 2D slab model for vertical oscillations, as indicated by the solid lines in Figure \ref{fig:Zhou_2018_3}. It has been found that the main parameters defining the period of these oscillations are prominence width or height, respectively, and the main restoring force for these oscillations is magnetic tension.
   
 Using a 2D dipped arcade magnetic configuration, \citet{Zhang:2019apj} confirmed that the parameter $\delta$ plays a key role in determining the period of longitudinal oscillations. They found that this parameter can also be important for the damping mechanism. As the prominence plasma moves along the magnetic field lines, it deforms the field lines, inducing local magnetic pressure perturbations. These perturbations propagate as fast magnetoacoustic waves in a process known as wave leakage \citep{Cally:1986solphys}. Previous theoretical studies demonstrated that, in a 2D slab model, waves carry energy away from the system and lead to the damping of the oscillations \citep{Brady:2005aap, Diaz:2006aap, Verwichte:2006aap}. \citet{Zhang:2019apj} also compared the damping of the prominence oscillations in the adiabatic and non-adiabatic cases. They have shown that the inclusion of the radiative losses and thermal conduction significantly reduces the damping time from $211$ to $34$ minutes. The results suggest that non-adiabatic effects are the dominant mechanism for damping the longitudinal oscillations in the considered model. 

   
 Numerous studies have focused on identifying the drivers of different types of prominence oscillations. Observationally, it is often challenging to determine the exact triggering mechanism for a given type of oscillation. However, longitudinal oscillations are often associated with nearby activity, such as jets or solar flares that are magnetically connected to the prominence structure. In contrast, transverse oscillations are often associated with large-scale coronal waves generated by eruptive events or flares, which sometimes occur far from the perturbed prominence. Numerical modeling offers a valuable tool for reproducing and exploring these various triggering scenarios in detail.
   
\begin{figure*}
    \centering
    \includegraphics[width=0.7\textwidth]{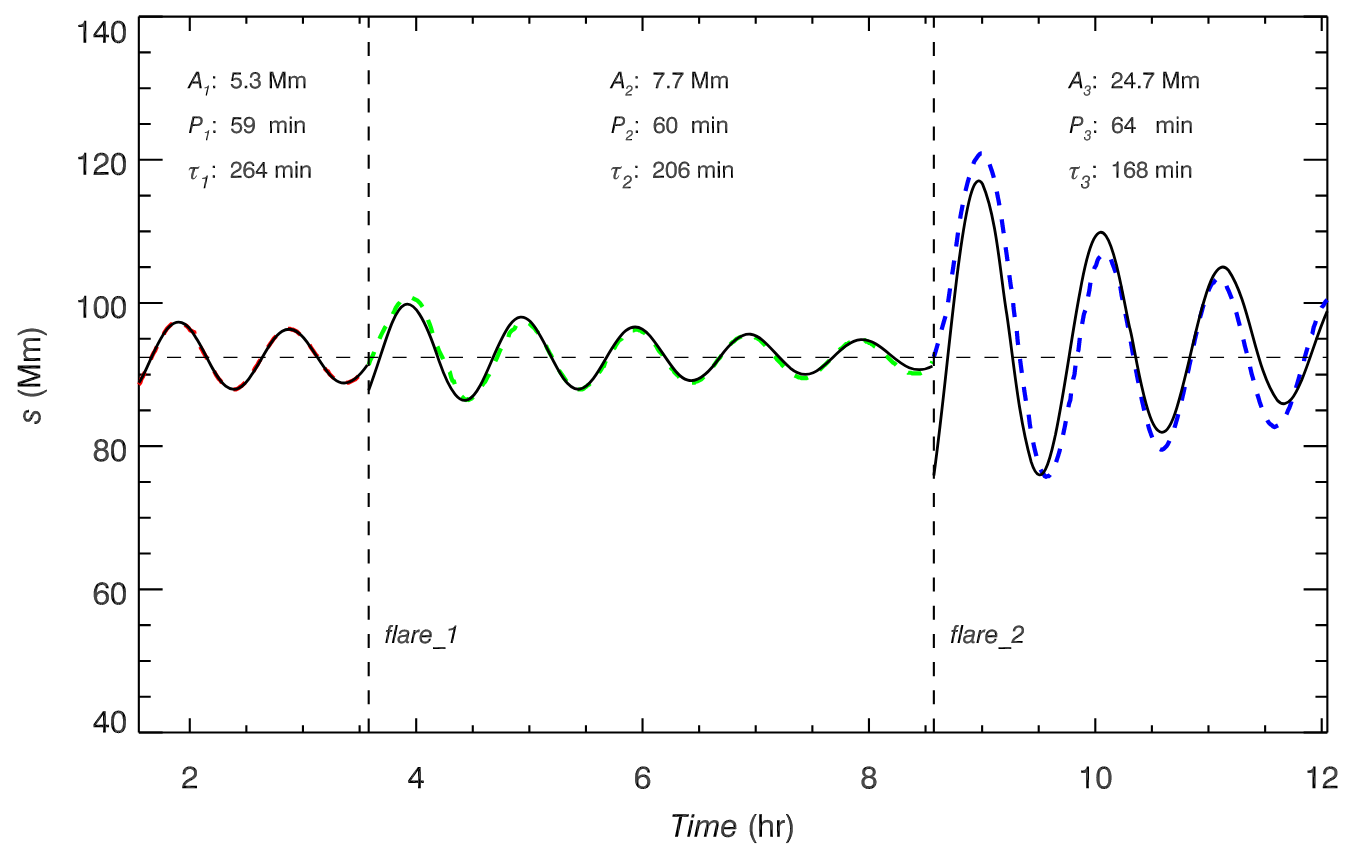}
    
    \caption{Temporal evolution of the center of mass of the filament. The black solid line represents the results of 1D simulations. The red, green, and blue lines represent the results of curve fittings by damped sine functions. The fitted parameters, including initial amplitudes, periods, and damping times, are labeled. Adapted from \citet{Zhang:2020aap}. \textcopyright\ ESO. Reproduced with permission.}
    \label{fig:Zhang_2020_1}
\end{figure*}

\begin{figure*}[!h]
    \centering
    \includegraphics[width=0.7\textwidth]{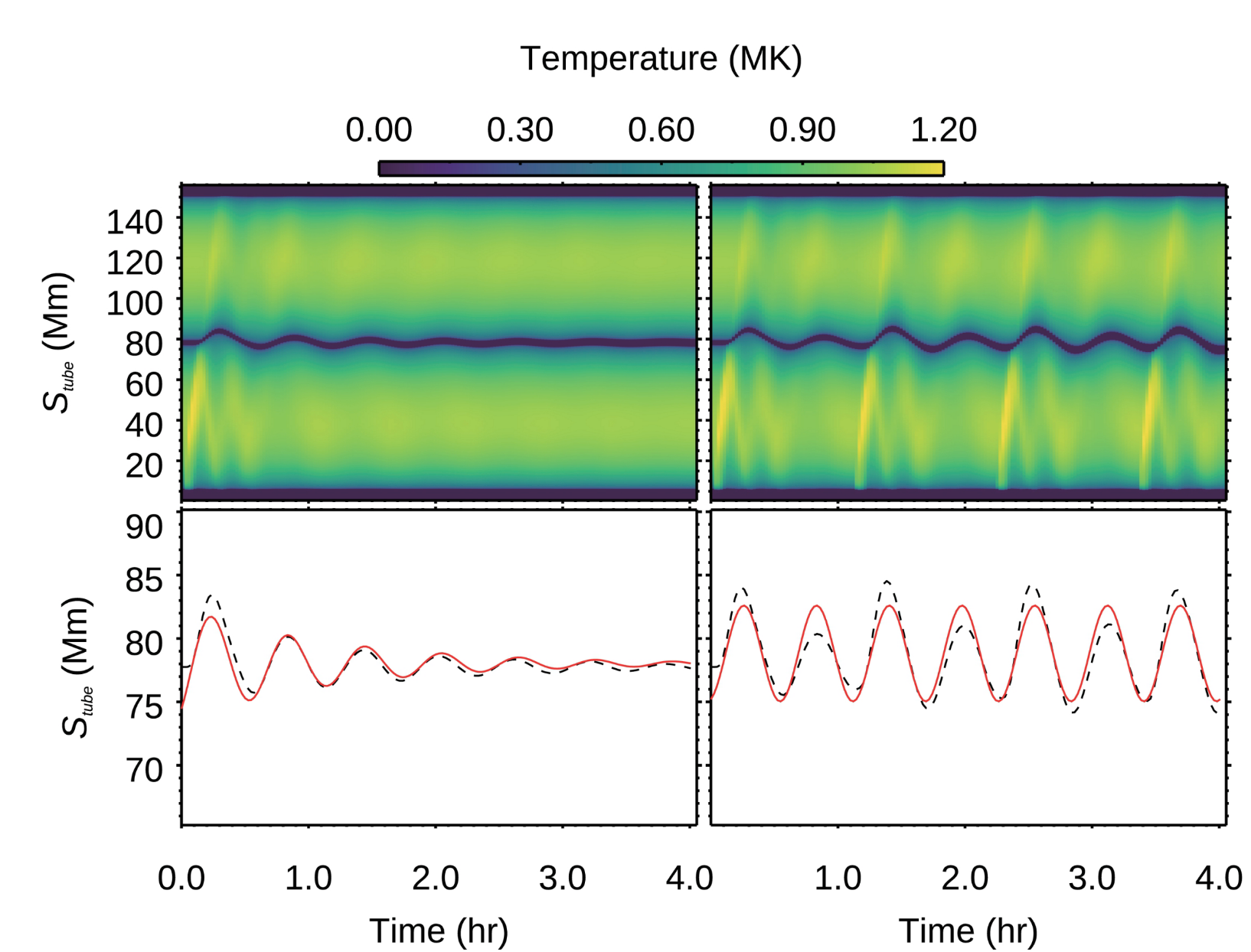}
    
    \caption{Top row: time-distance diagrams of the temperature along the flux tube in the one-pulse case (left) and in the multipulse case (right). Bottom row: temporal evolution of the displacement of the filament thread center in the one-pulse case (left) and in the multipulse case (right), where the dashed black lines correspond to the simulation results, and the solid red lines correspond to the fittings with damped sine functions. Adapted from \citet{Ni:2022aap}. \textcopyright\ ESO. Reproduced with permission.}
    \label{fig:Ni_2022_1}
\end{figure*}

 In the work by \citet{Zhang:2020aap}, the authors attempted to explain the characteristics of the observed prominence threads' oscillations triggered by two solar flares. Interestingly, the damping time of the oscillations decreases following the second flare in this observation. The 1D simulations, similar to one used in \citet{Zhang:2012aap} and \citet{Zhang:2013aap}, but applying impulsive heating at one footpoint to mimic the energy deposition associated with the observed flares. The temporal evolution of the center of mass of the filament is shown in Figure \ref{fig:Zhang_2020_1}. The model reproduced the observed periods; however, the damping time after the second flare could not be explained solely by non-adiabatic effects. Using a similar model, \citet{Ni:2022aap} produced the oscillations in a 1D experiment to explain the observation of the oscillations triggered by quasiperiodic jets (Figure \ref{fig:Ni_2022_1}). In this case, the period deviates from the pendulum period as they are being actively driven by the periodicity of the jets. In the absence of the quasiperiodic jets, i.e., in the initial and final stages of the experiment, the periods agree once again with the pendulum period.

\begin{figure*}
    \centering
    \includegraphics[width=0.4\textwidth]{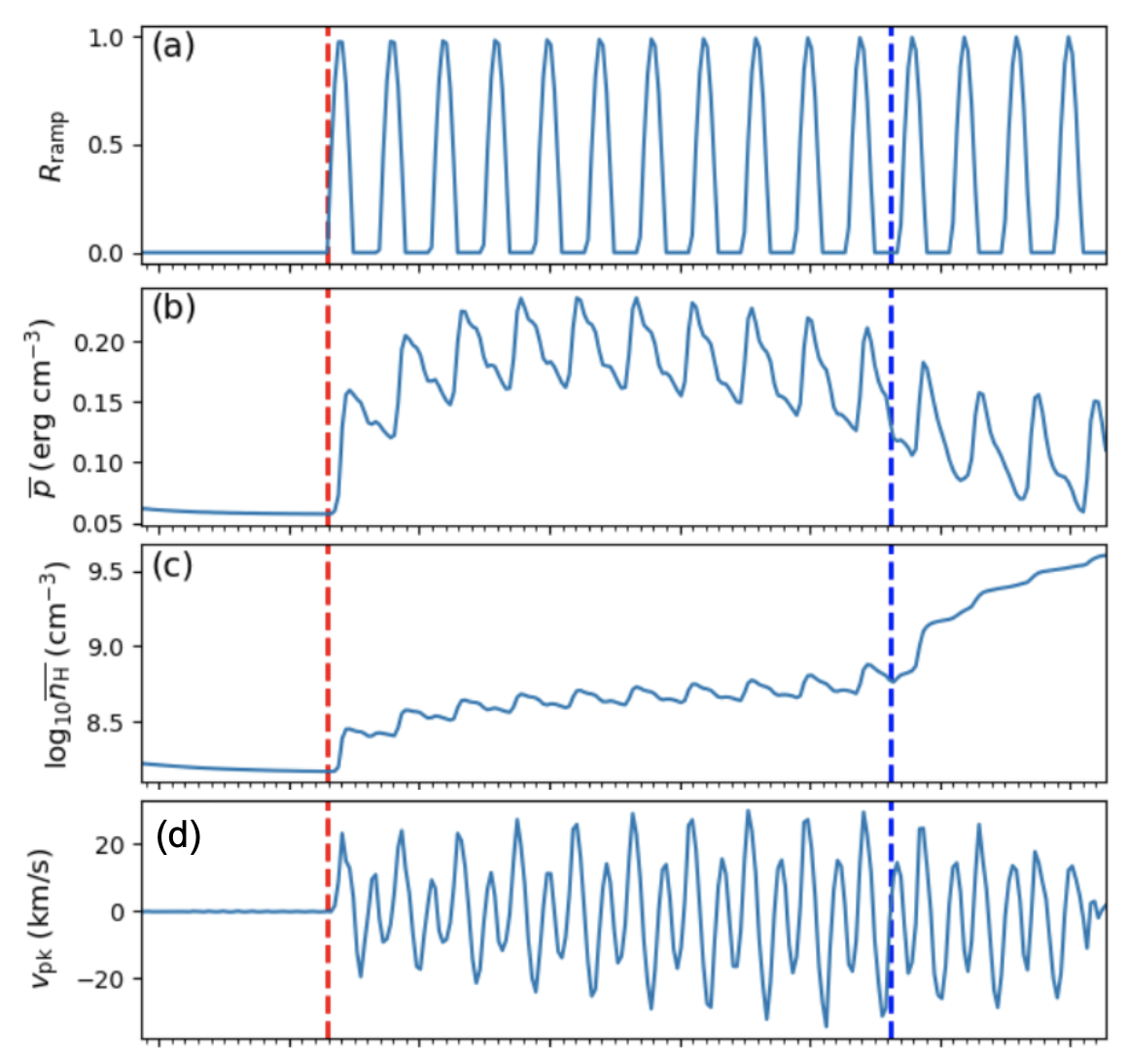}
    \includegraphics[width=0.55\textwidth]{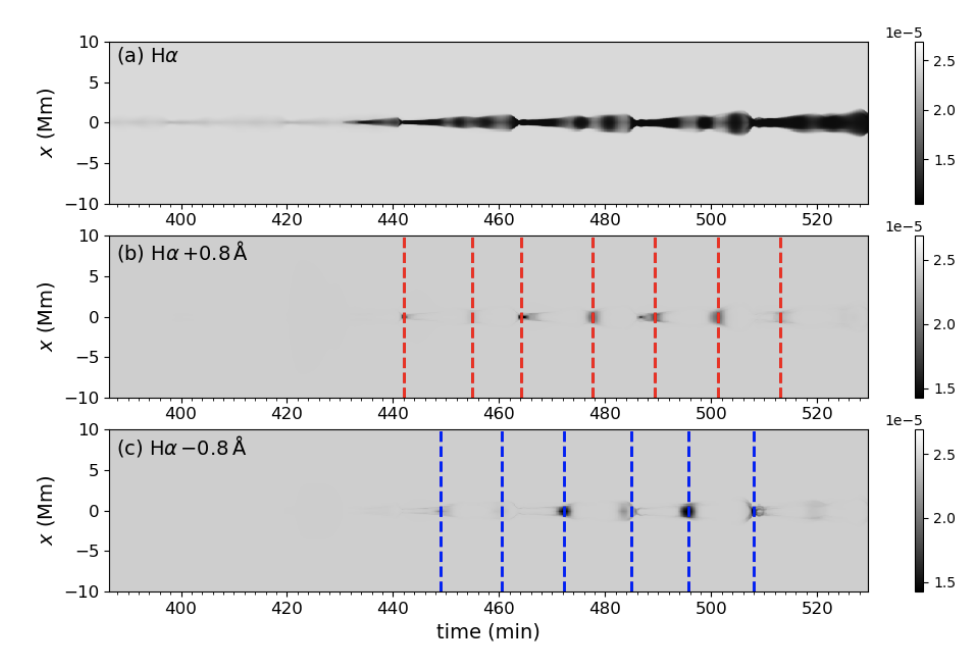}
    \caption{Left: Time evolution of (a) the modulating ramp function, indicating the gradual increase and decrease of the localized heating; (b) average pressure; (c) average number density; and (d) vertical velocity at the apex of the selected magnetic field line.
    Right: Time evolution of the synthetic H$\alpha$ (a) line center, (b) red wing, and (c) blue wing emission from the filament (top-down view). Vertical dashed lines are included to identify periodicities. Adapted from \citet{Zhou:2023aap}.}
    \label{fig:Zhou_2023_2}
\end{figure*}

 Similarly to \citet{Ni:2022aap}, \citet{Zhou:2023aap} applied periodic heating at both footpoints of the 2D dipped arcade, using a time-dependent function with a period of 20 minutes. As shown by \citet{Zhou:2023aap}, applying periodic heating at both footpoints of the 2D dipped arcade led to the formation of a vertical prominence structure. The resulting heavy prominence mass stretched the magnetic field, causing vertical oscillations (Figure \ref{fig:Zhou_2023_2}, left). The period of these oscillations is determined by the interplay between the intrinsic kink mode period and the period of the external driver. The synthetic images in the right panel of Figure \ref{fig:Zhou_2023_2} show the alternate disappearance and reappearance of the prominence in the accompanying H$\alpha$ line core and wings syntheses, caused by significant Doppler shifts along the line of sight (which, in this case, is aligned with the vertical direction). In this way, the observed phenomena of a bulk winking filament \citep{Ramsey:1965aj,Ramsey:1966aj,Hyder:1966zap,Shen:2014apj1} have been reproduced in numerical simulations \citep[cf.][]{MartinezGomez:2022aap}.
   
   The triggering of the oscillations by the energetic wave is an interesting aspect that raises several important questions. How can prominence structures be perturbed to have such large amplitudes as measured in observations? This would suggest that the wave should possess a large amount of energy. Another question is how these waves are produced by eruptive events or solar flares and how they can travel over such large distances in the strongly non-uniform magnetic field of the solar corona. This necessarily requires that the wave experiences reflection, refraction, and transmission, which can decrease its energy. Finally, when this energetic wave reaches prominence, how does it interact with the prominence plasma magnetic field, represented by the dipped arcade or flux rope, and what type of oscillations are triggered in this case? 
   
   \citet{Jercic:2022aap} have studied wave-driven oscillations of the threads located in the identical magnetic dips of a 2D dipped arcade structure anchored in the gravitationally stratified chromosphere, transition region, and corona. The non-adiabatic effects were not taken into account. The threads are formed artificially as individual regions of enhanced density and decreased temperature. After the relaxation phase, the threads are perturbed by a wave generated through a source term in the energy equation, approximating the influence of a solar flare at one footpoint of the dipped arcade. The produced wave triggers the oscillations of the threads, but the authors found that their results strongly disagreed with the pendulum model. Because the magnetic dips are identical, the pendulum period is by definition constant for each dip. However, the periods measured by the authors varied from thread to thread. Moreover, their period values are much smaller than predicted by the pendulum model, with the difference varying in the range of $10-55$ minutes. The authors explained this discrepancy by noting that the shock wave that propagates through the domain creates compressional waves that propagate through the thread, reflecting internally off the prominence-coronal transition region (PCTR) gradients and interfering with each other. These waves also travel through the coronal plasma in between the threads, thereby causing additional compression and providing an external driving force. This interplay aspect emphasises the importance of numerical modeling for interpreting observational periods that are ordinarily analysed by means of simplified prominence seismology approximations (rigid field, constant radius of curvature, etc.). Another interesting aspect of this study is that some threads show amplification rather than damping. This result confirms the study by \citet{Liakh:2021aap}, which showed that energy and momentum are transferred between the threads that belong to the different magnetic field lines. In this way, the oscillations are significantly damped in some prominence threads, while in others, they are amplified. 
   
   
   
\begin{figure*}
    \centering
    \includegraphics[width=0.51\textwidth]{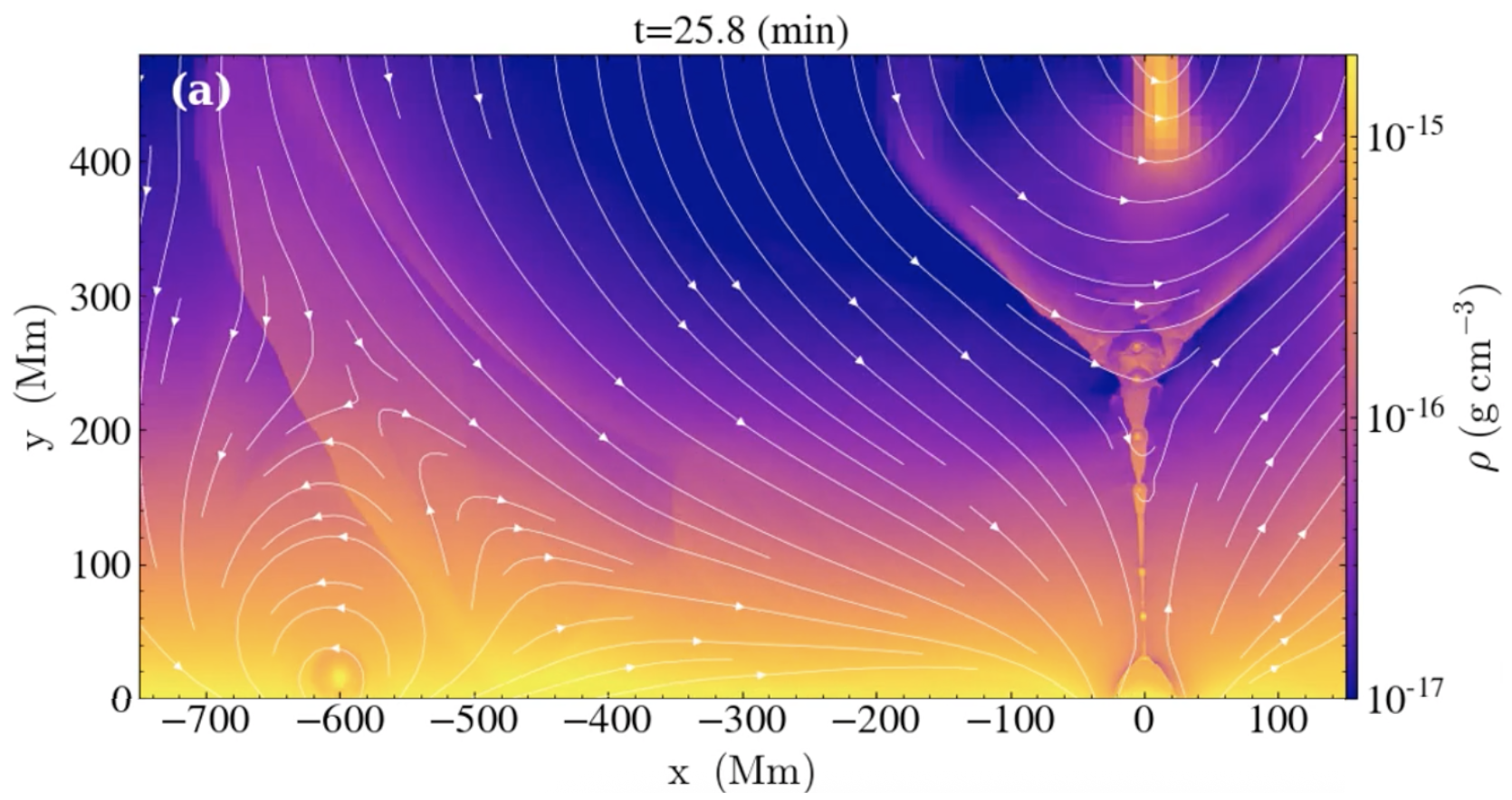}
    \includegraphics[width=0.41\textwidth]{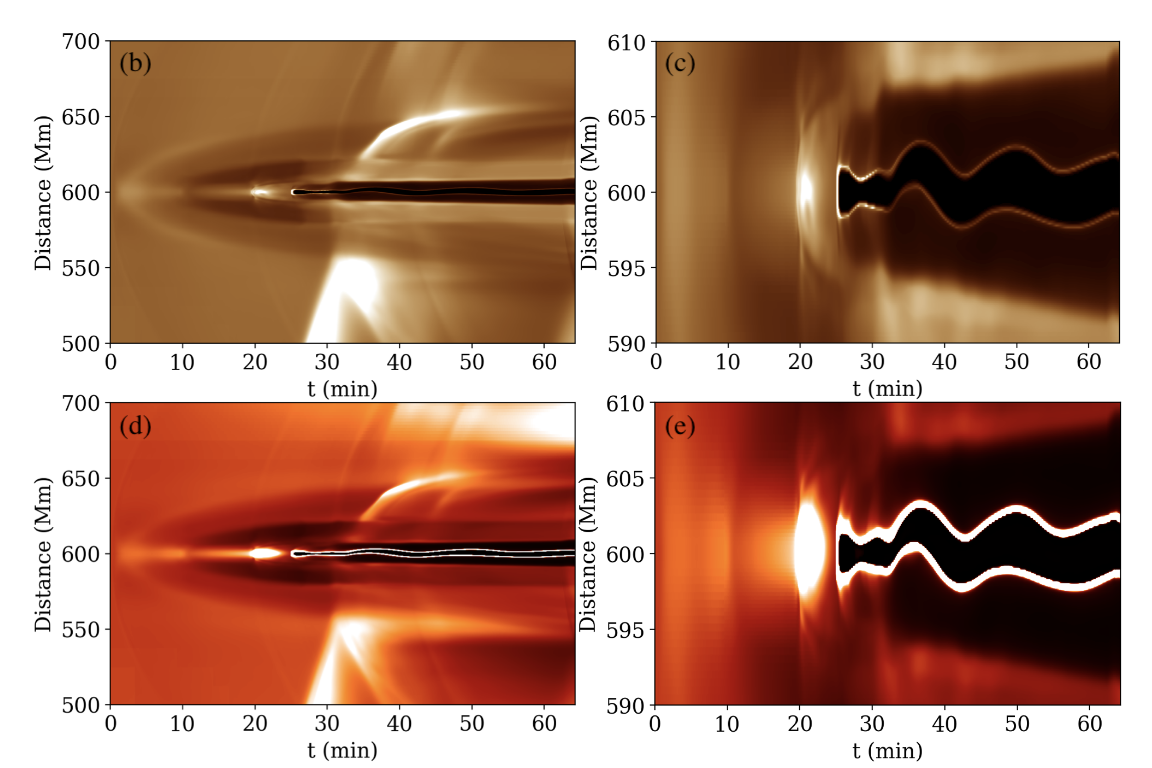}
    \caption{Panel a: Instantaneous density distribution and magnetic field lines in the entire numerical domain at $t = 25.8$ minutes. Panels b and d: Time-distance diagrams of the SDO/AIA synthetic emission in the 193 \AA\ and 304 \AA\ channels, extracted along a horizontal cut at a height of $y = 10$ Mm. Panels c and e: zoomed-in views centered on the prominence region around $x = -600$ Mm. The vertical axes indicate the distance from the eruption center. Adapted from \citet{Liakh:2025aap}.}
    \label{fig:Liakh_2025_1}
\end{figure*}

   
    Finally, in the recent study \citet{Liakh:2025aap}, several important aspects of the triggering of prominence oscillations in the flux rope prominence by the coronal waves have been demonstrated. The model combined a 2.5D catastrophe magnetic field to generate the eruption and a dipole field to form the flux rope with an artificially embedded prominence (Figure \ref{fig:Liakh_2025_1}, left). This magnetic configuration is placed into a non-adiabatic gravitationally stratified corona. This model enabled the study of the dynamics of an energetic eruption, the resulting perturbations, and the propagation of these disturbances through the non-uniformly magnetized corona upon reaching the flux rope prominence. In contrast to the previous studies \citep{Liakh:2020aap,Zurbriggen:2021solphys,Liakh:2023aap}, the coronal wave and associated energy injection into the flux rope are produced self-consistently by the eruptive event that is located at a large distance of $600$ Mm from the flux rope prominence.
   
\begin{figure*}[!h]
    \centering
    \includegraphics[width=0.9\textwidth]{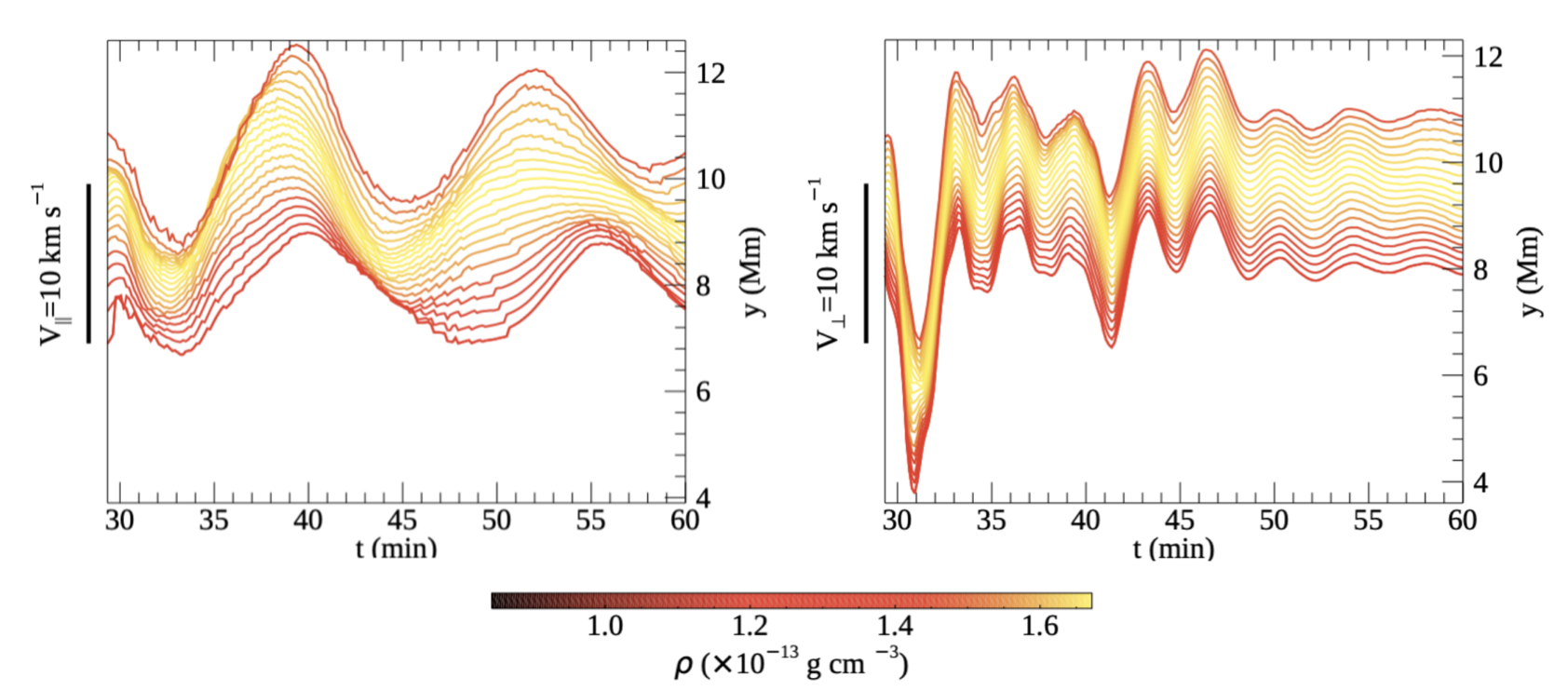}
    \caption{Temporal evolution of the longitudinal (left) and transverse (right) velocities, obtained by tracking 20 fluid elements. The right vertical axis indicates the initial height of each corresponding fluid element. The color bar represents the initial density at the location of each element. Adapted from \citet{Liakh:2025aap}.}
    \label{fig:Liakh_2025_3}
\end{figure*}

    The eruption produces a fast magnetoacoustic wave that generates a shock propagating away from the eruption site. Upon reaching the flux rope prominence, the wave produces the reflected and transmitted fronts as seen in synthetic time-distance diagrams in 193 and 304 \AA\ channels (Figure \ref{fig:Liakh_2025_1}b and d). The energetic wave interacts with the remote prominence, driving both transverse and longitudinal oscillations (Figure \ref{fig:Liakh_2025_3}). Note here that longitudinal still refers to field-aligned oscillations. These two types of oscillations are often coupled in observations and simulations \citep{Gilbert:2008apj, Luna:2016apj, Liakh:2023aap}. The period of longitudinal oscillations agrees well with the pendulum model. The period of the vertical oscillations remains constant with height, indicating that the flux rope undergoes global oscillations driven by magnetic pressure and tension forces, more so than intra-thread oscillations. Interestingly, in this experiment, both types of oscillations are attenuated significantly. Several factors have been identified that may affect the damping times. First, plasma initially contained within the flux rope is accreted onto the artificially loaded prominence. In the right panels of Figure \ref{fig:Liakh_2025_1}, a gradual darkening is visible in both channels around the prominence region, indicating that this accretion process depletes the surrounding flux rope. As we know from previous studies, such as \citet{Ruderman:2016aap}, accretion can dampen longitudinal oscillations, whereas the mechanism for transverse oscillations has not been studied before. Another important mechanism is the oscillatory reconnection that occurs below the flux rope, triggered by the passage of the wave. This oscillatory reconnection leads to the alternate formation of the horizontal and vertical current sheets. In this way, the energy can dissipate due to the increase in Ohmic heating resulting from enhanced currents.

   Simulations of prominence oscillations have confirmed that the restoring force for longitudinal oscillations is gravity projected along the magnetic field, while for transverse oscillations it is magnetic tension. Regarding damping mechanisms, in addition to non-adiabatic effects, several processes have been considered, such as mass drainage and accretion, wave leakage, and energy transfer through thread–thread interactions. Various agents have been studied as potential triggers of prominence oscillations, including repetitive flares and jets, and it has been shown how these can influence the interpretation of the observed periods. Moreover, it was demonstrated for the first time how a wave generated by an eruptive event can interact with and trigger dynamics in a distant flux rope prominence. This study, however, requires further modeling and a parametric survey.
   

 \subsection{Counterstreaming and Helical Flows}
 The counterstreaming flows are motions of the prominence threads in different directions along neighboring field lines with velocities of around $20\kms$ \citep[see, e.g.][]{Schmieder:1991aap,Zirker:1998nat,Lin:2003solphys,Wang:2018apjl}. These motions have often been connected to out-of-phase thread oscillations \citep{Lin:2003solphys} or unidirectional flows \citep{Chen:2014apj}.
 
\begin{figure*}[!h]
    \centering
    \includegraphics[width=0.7\textwidth]{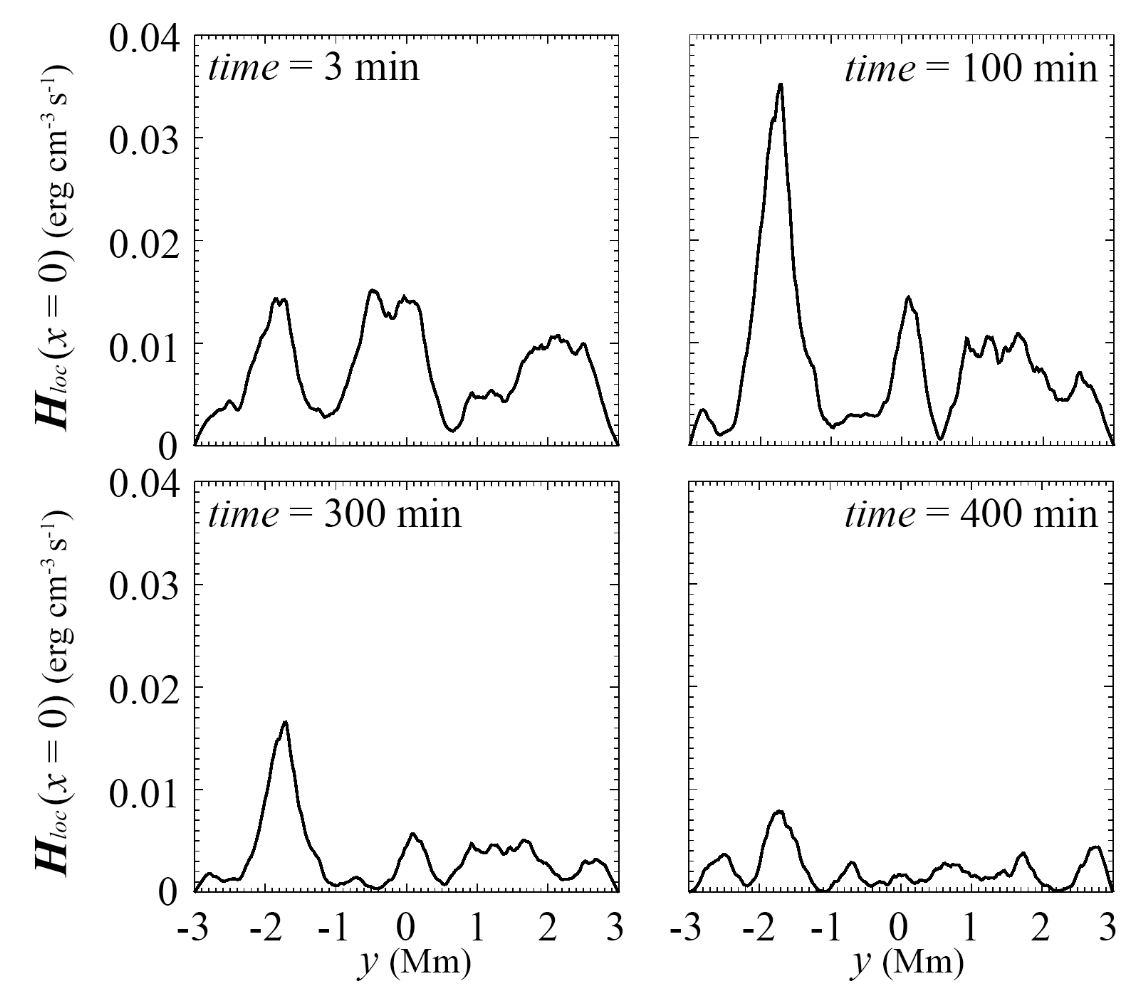}
    
    \caption{Four snapshots of the heating function distribution at the footpoints at the bottom of the numerical domain. Adapted from \citet{Zhou:2020natas}. Reproduced with permission from SNCSC.}
    \label{fig:Zhou_2020_2}
\end{figure*}

  \citet{Zhou:2020natas} performed a 2D experiment of the dipped arcade magnetic field structure located in a non-adiabatic, gravitationally stratified atmosphere, applying localized heating that is randomized in both time and space at the bottom of the numerical domain. The authors describe this approach as mimicking the turbulent heating in the solar atmosphere. The function of the randomized heating consists of two parts: temporally and spatially varying. In the horizontal direction, the heating is prescribed as a superposition of multiple wavelengths, ranging from the size of the simulation domain down to the smallest grid scale. Each wavelength enters the sum with a random phase shift. The individual heating amplitudes are scaled to a power-law spectrum of 3/4, although the authors remark that the range of 0.2\,--\,2 has little influence on the final result. In the vertical direction, the heating decreases exponentially, assuming a small scale height. The temporal evolution of the randomized heating follows $5$-minute heating episodes with slight variations. The temporal function then consists of the sum of several Gaussian functions in time. The evolution of the randomized heating at the bottom of the numerical domain is shown in Figure \ref{fig:Zhou_2020_2}. Despite intending to mimic turbulence, this distribution shows the resulting relationship to be highly structured (rather than an anticipated non-gaussian $\sim$pink noise), perhaps resembling the distribution of nanoflares more closely.

\begin{figure*}[!b]
    \centering 	
    \includegraphics[width=0.9\textwidth]{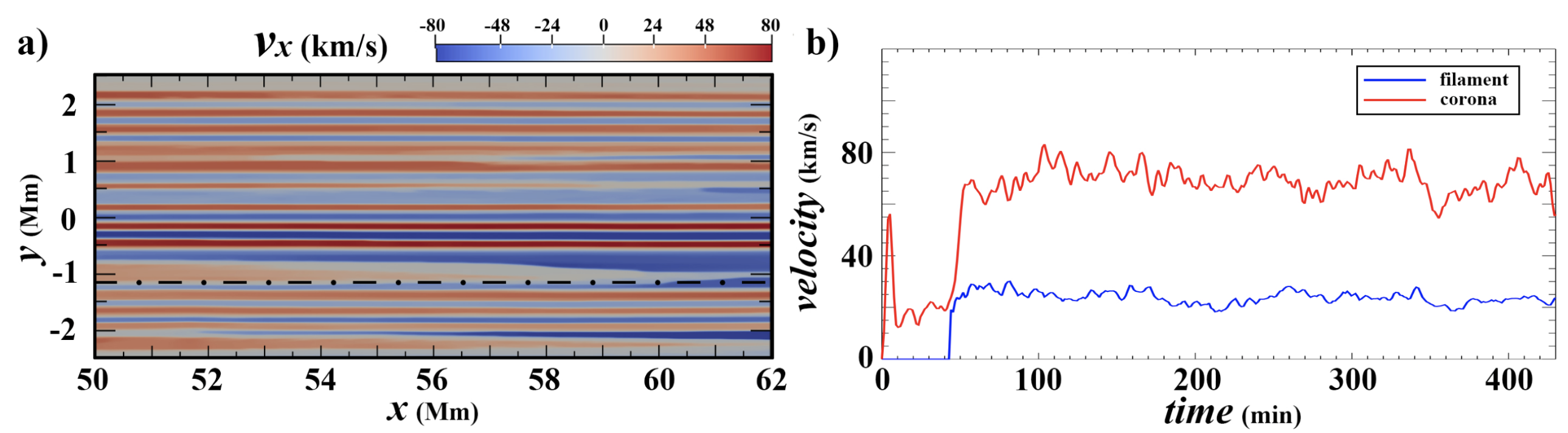}
    \caption{Left: Velocity distribution near the center of the simulation box. Right: Evolution of the averaged velocities inside the filament threads (blue line) and outside the filament threads (red line) for the materials moving in the positive x-direction. Adapted from \citet{Zhou:2020natas}. Reproduced with permission from SNCSC.} \label{fig:Zhou_2020_5}
\end{figure*}

 As a result of the randomized heating, the evaporated plasma flows along the field lines, condensing at some point due to the runaway cooling and thermal instability. In this case, the evolution of the plasma naturally reproduces the counterstreaming flows. The horizontal velocity of plasma in the numerical domain reaches high values, as shown in Figure \ref{fig:Zhou_2020_5}. The authors noted the difference in the evolution of the coronal and prominence plasma shown in the right panel. While the coronal plasma flows with a velocity of $70-80\kms$, the prominence plasma moves more slowly with a velocity of $30\kms$, similar to observational values. Using synthetic imaging in H$\alpha$, based on the method described by \citet{Heinzel:2015aap}, and in the SDO/AIA 193 \AA\ channel using the \texttt{Chianti} database as detailed in \citet{Xia:2014apj}, the authors concluded that the counterstreaming flows observed in H$\alpha$ are primarily linked to longitudinal oscillations. In the 193 \AA\ channel, counterstreaming flows are represented mainly by the unidirectional flow of hot plasma. This simulation gives a possible explanation of the observations of EUV counterstreaming and their faster flows \citep{Alexander:2013apjl} in comparison to H$\alpha$ observations.

\begin{figure*}[!h]
    \centering
    \includegraphics[width=0.9\textwidth]{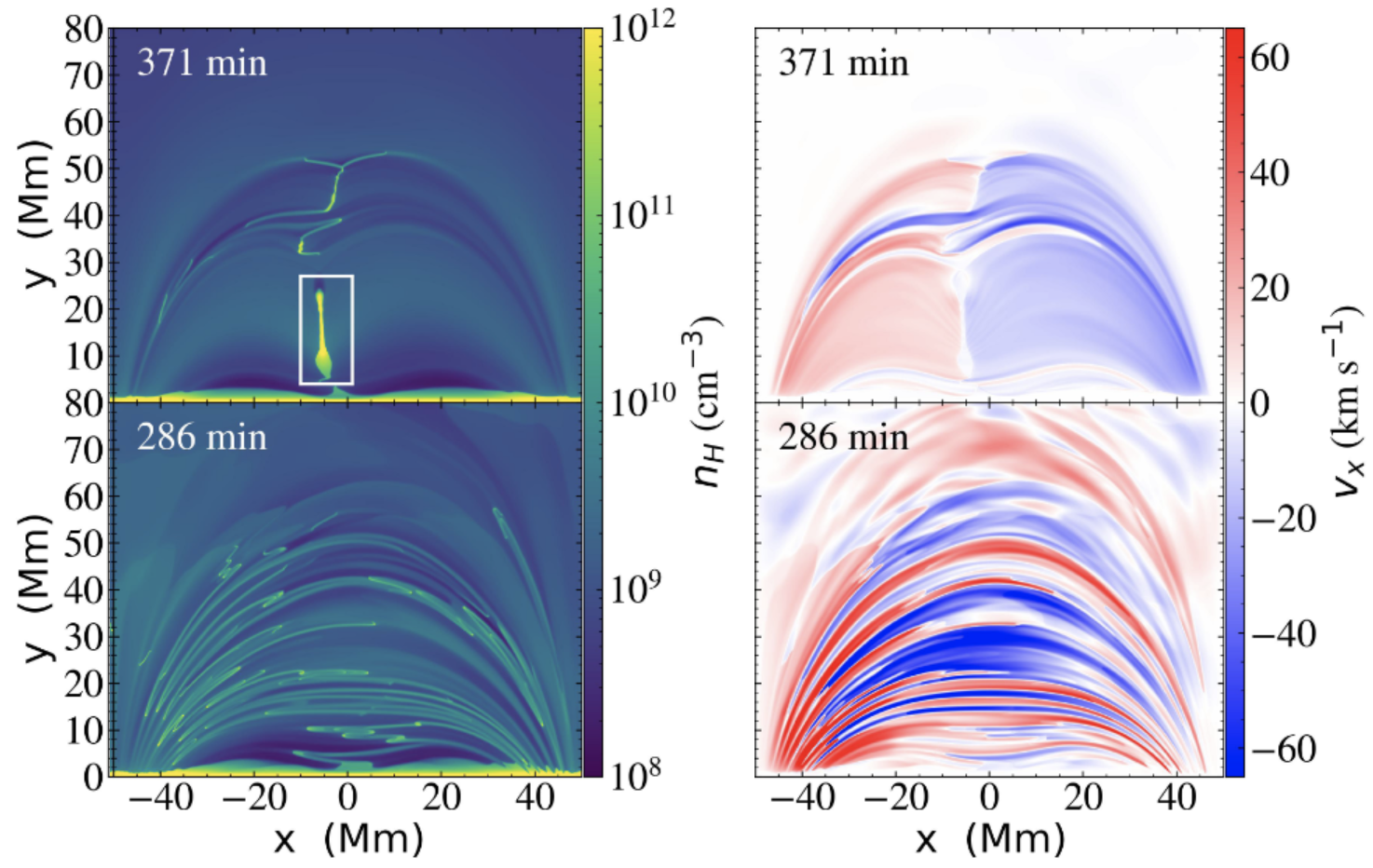}
    \caption{Left: Number density distribution in the simulation domain. The white box outlines the main prominence body in the steady heating case. Right: Distribution of the horizontal velocity component. Top panels correspond to the steady heating case; bottom panels show the stochastic heating case. Adapted from \citet{Jercic:2024aap}. } \label{fig:Jercic_2024_1}
\end{figure*}

 \citet{Jercic:2023aap} has also performed a 2D parametric study varying the height and intensity of the pulses of the randomized heating, which mimics the potentially spatially diverse positioning of the small reconnection events or nanoflares. The authors found that higher and stronger pulses lead to condensations forming faster, which in turn form longer threads. Overall, the evolution of the condensations gives rise to counterstreaming flows, some of which exhibit oscillatory behavior, consistent with the findings of \citet{Zhou:2020natas}. Building on this, \citet{Jercic:2024aap} extended the analysis by comparing the condensations formation and dynamics in a 2.5D dipped arcade configuration under steady versus stochastic heating conditions (Figure \ref{fig:Jercic_2024_1}). The steady heating produces flows of the evaporated plasma that meet at the center of the dipped arcade and form the vertically aligned condensation. In the case of stochastic heating, a variety of different thread motions occurred, including counterstreaming flows, transverse oscillations resulting from the stretching of the magnetic field lines by threads, and upflows and downflows around the footpoints.

 Another type of prominence motion is associated with the helical flows of the prominence and coronal plasma along the magnetic field lines of the flux rope or sheared arcade. They are typically observed as rotational flows in the dark coronal cavities \citep{Li:2012apjl,Panasenco:2014solphys}, which are believed to be the cross-sections of the flux rope or the sheared arcade \citep[][]{Gibson:2010apj}. The velocities of these motions usually show the range of values $5-75\kms$ \citep{Ohman:1969solphys,Liggett:1984solphys,Schmit:2009apjl,Wang:2010apjl,Li:2012apjl}. The rotational flows in magnetic flux tubes have been recently analytically studied in the context of MHD wave propagation and energy transport \citep{Skirvin:2023mnras, Skirvin:2024apj}. The main challenge in interpreting these motions from the observations is that it is difficult to reconstruct the 3D velocity field from the on-limb observations, where these flows appear to be rotations. Moreover, the 3D magnetic field structure of the prominence can be very complex. Assuming that the plasma flows along the magnetic field in the low-$\beta$ environment, it is difficult to characterize which type of motion is observed. Motions that appear to be rotation may, in reality, actually represent counterstreaming flows or oscillations. On the one hand, \citet{Gunar:2023ssr} pointed out that the multipoint observations would help to solve the projection effect problem. On the other hand, using the numerical models, we can explain the triggering mechanism of these types of motions, their velocities, and their lifetimes. Moreover, modern tools such as Lightweaver and DexRT enable accounting for departure from the local thermodynamic equilibrium (LTE) assumption and the generation of non-LTE (NLTE) synthetic spectra. In this way, we can give the basis for the interpretation of observations. We further discuss the synthetic data in Section \ref{sec:synthetic}.

\begin{figure*}
    \centering
    \includegraphics[width=0.9\textwidth]{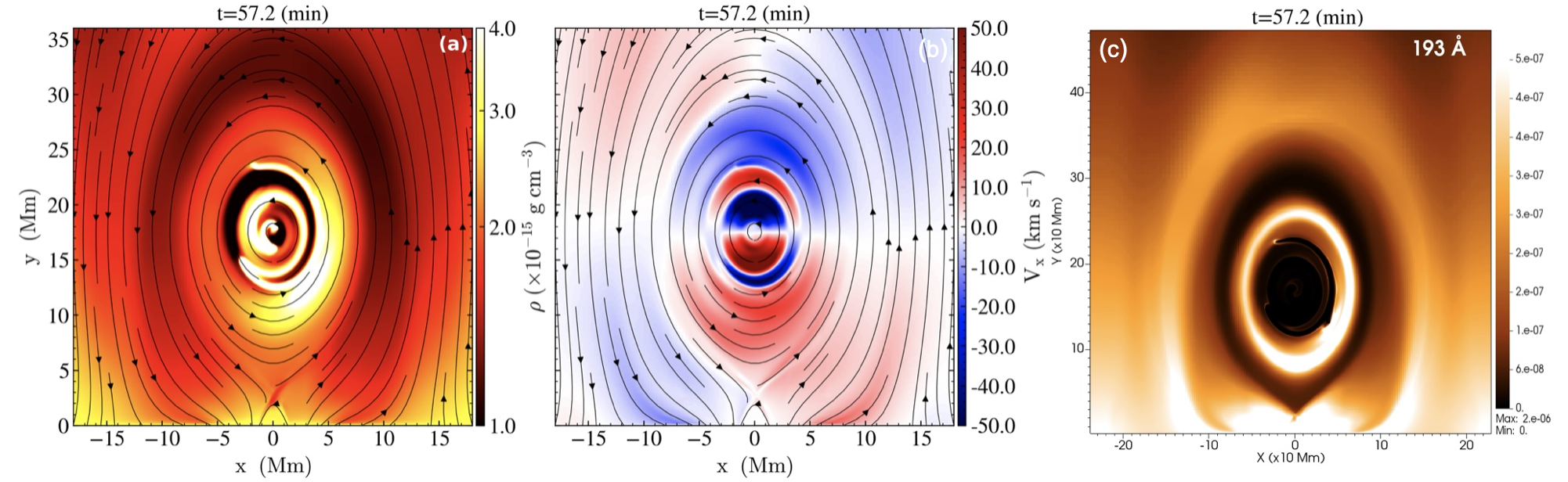}
    \caption{Distribution of the density (a), horizontal velocity (b), and synthetic image in 193 \AA\ AIA channel shown at about $57$ minutes. The black lines in panels a and b denote the magnetic field lines. Adapted from \citet{Liakh:2023apjl}.} \label{fig:Liakh_2023_1}
\end{figure*}

   The first model demonstrating the driving of flows in the coronal cavity was presented by \citet{Liakh:2023apjl}. This 2.5D simulation combines flux rope formation from a sheared arcade, achieved through footpoint motions as in \citet{Jenkins:2021aap} and \citet{Brughmans:2022aap}, with randomized heating as used by \citet{Zhou:2020natas}. The randomized heating from the lower atmosphere introduces temperature asymmetry, which initiates coherent plasma flows along the loops before magnetic reconnection occurs. Once the flux rope forms via reconnection, these flows evolve into a net rotational motion of the coronal plasma (Figure \ref{fig:Liakh_2023_1}a, b). This scenario is consistent with a suggestion of \citet{Wang:2010apjl}. As thermal instability develops within the flux rope, dense condensations form and continue the rotational motion. The central condensation sustains this rotation for over an hour, with initial speeds exceeding $60\kms$. Synthetic EUV images generated in four AIA channels confirm that the co-rotation of the coronal and prominence plasma in this way is visible in observational bandpasses. The formation of a dark coronal cavity is particularly evident in the 193 \AA\ band (Figure \ref{fig:Liakh_2023_1}c). This model successfully reproduces the observed dynamics of prominence plasma inside the coronal cavity.

 Rotational flows within the flux rope prominence have also been reported in the recent study by \citet{Li:2025arxiv}. In their simulations, an emerging dipolar field reconnects with the background magnetic arcade, forming an asymmetric flux rope inclined toward the null point. The rotational flows inside the flux rope are initiated by the formation of secondary flux ropes that subsequently merge with the primary one. In contrast, similar merging events in symmetric flux ropes do not result in rotational dynamics, as shown by \citet{Jenkins:2021aap} and \citet{Brughmans:2022aap}. These results suggest that asymmetries, either in the heating during flux rope formation or in the flux rope structure itself, create favorable conditions for the development of rotational motions.

  \subsection{Downflows and Rayleigh–Taylor Instability}
  
  The downflows of the prominence plasma are associated with mass drainage along or across the magnetic field lines under the action of gravitational force, which inevitably leads to a reduction in the total prominence mass contained within the filament channel. Observations suggest that the loss of up to $70$ percent of prominence mass can lead to the loss of equilibrium and eruption \citep[e.g.][]{Bi:2014apj,Jenkins:2018solphys}. Theoretically, the exact percentage required depends on the condition of the hosting magnetic field and how close it is to instability \citep[][]{Jenkins:2019apj,Fan:2020apj}.
  
  In the numerical experiment by \citet{Xia:2012apjl}, the initial magnetic configuration consisted of a simple magnetic arcade field. Magnetic dips formed self-consistently under the influence of gravity when condensations formed at the apex of the arcades. The authors analyzed the vertical force balance and found that the gravitational force acting on the prominence plasma was nearly perfectly counteracted by the Lorentz force from the curved magnetic field, thereby providing support to the prominence. However, as mass continued to accumulate, the prominence mass eventually reached a saturation point. When additional plasma evaporated from the chromospheric footpoints and condensed near the apex, it could no longer be stably supported and started to drain. The study also noted that stronger magnetic fields are more resistant to deformation and can delay plasma accumulation, resulting in earlier mass drainage.

  Mass drainage has also been reproduced in fully 3D configurations. For example, \citet{Xia:2014apjl} modeled the formation of a 3D flux rope through converging and shearing motions at the footpoints. The flux rope lifted denser plasma from the lower corona. Increased local radiative losses resulted in a significant pressure drop, creating a pressure gradient that drove the plasma toward the center of the flux rope. However, a portion of the lifted plasma drained along the magnetic field lines toward the lower atmosphere, reaching speeds of up to $100\kms$. In the synthetic EUV images, this drainage appeared as a prominence barb. Furthermore, \citet{Xia:2016apj} found that plasma drainage is balanced by continuous evaporation due to the heating in the chromosphere, thereby reproducing the complete prominence-corona mass cycle. 

  Another scenario of the downward motions of the plasma is associated with a relatively weak magnetic field and reconnection. In \citet{Keppens:2014apj}, the numerical setup resembles one in \citet{Xia:2012apjl}, but the quadrupolar magnetic arcades already include the dipped region in the center. The plasma condensation located at the higher field lines deforms magnetic dips, making them deeper. The evaporated plasma built up, forming the funnel prominence. At the top, where the magnetic field is shallower, the plasma starts to `spill over' and drain down. Eventually, the impulsive dynamics resulting from transient coronal rain, combined with the deepening of the magnetic field by accumulated plasma, lead to the numerical reconnection of the magnetic field and the formation of a flux rope. In this way, more field lines gradually reconnect, and the flux rope grows and transports downward with a vertical velocity of around $0.6\kms$. This evolution was also obtained by \citet{Jenkins:2021aap} in the flux rope in the weak magnetic field case ($3$ G). \citet{Jenkins:2021aap} described this process as the heavy condensation deforming the magnetic dips, which pushes the field lines in front of the condensation. This compression locally increases the current density gradients. Due to the explicit resistivity included in this experiment, magnetic reconnection occurs and forms a flux rope, which moves downward as more field lines gradually reconnect \citep[following][]{Low:2012apj}.
  
  \citet{Jercic:2024aap} also obtained the flux rope formation at the top of the magnetic arcade structure in the case of steady heating. The steady heating, as in the previous study, led to the formation of a clear vertical prominence structure (Figure \ref{fig:Jercic_2024_1}, top). At the upper part of the magnetic structure, where the field is weak, the heavy condensation deepens the magnetic field under the influence of gravity. As a result, the reconnection is driven by the heavy plasma, leading to the formation of a blob within a mini flux rope. Unlike in \citet{Keppens:2014apj}, this mini flux rope is ejected upward as a result of the reconnection process. This event more resembles the nanojets observations \citep{Antolin:2021natas}, in which the trajectory of this jet is perpendicular to the magnetic field. In the case of stochastic heating, the model reproduced the more horizontal threads, which quickly move along the magnetic field, often moving back and forth until eventually draining down. In this case, the reconnection does not occur as the fast plasma flows prevent the concentration of sufficient mass to significantly deform the threaded magnetic field. It is important to emphasise here that the magnitude of the spatial energy deposition/heating adopted in the stochastic heating case of a gaussian peaking at 0.1~erg~cm$^{-3}$~s$^{-1}$ and a scale height of 3~Mm is at the upper end of what has traditionally been considered a nanoflare. This should be explored in more detail.

  In the situation where the denser plasma is positioned above the less dense plasma, it is expected that the development of the HD Rayleigh-Taylor instability can occur, forming vertically oriented fine structures \citep{Berger:2008apj,Berger:2010apj,Hillier:2012apj1,Hillier:2012apj2}. The strong magnetic field can provide a stabilization effect for Rayleigh-Taylor instability \citep{Terradas:2016apj}. There have been works dedicated to studying the Rayleigh-Taylor instability in prominences using 2.5D and 3D modeling with the \texttt{MPI-AMRVAC}.

\begin{figure*}[!h]
    \centering
    \includegraphics[width=0.9\textwidth]{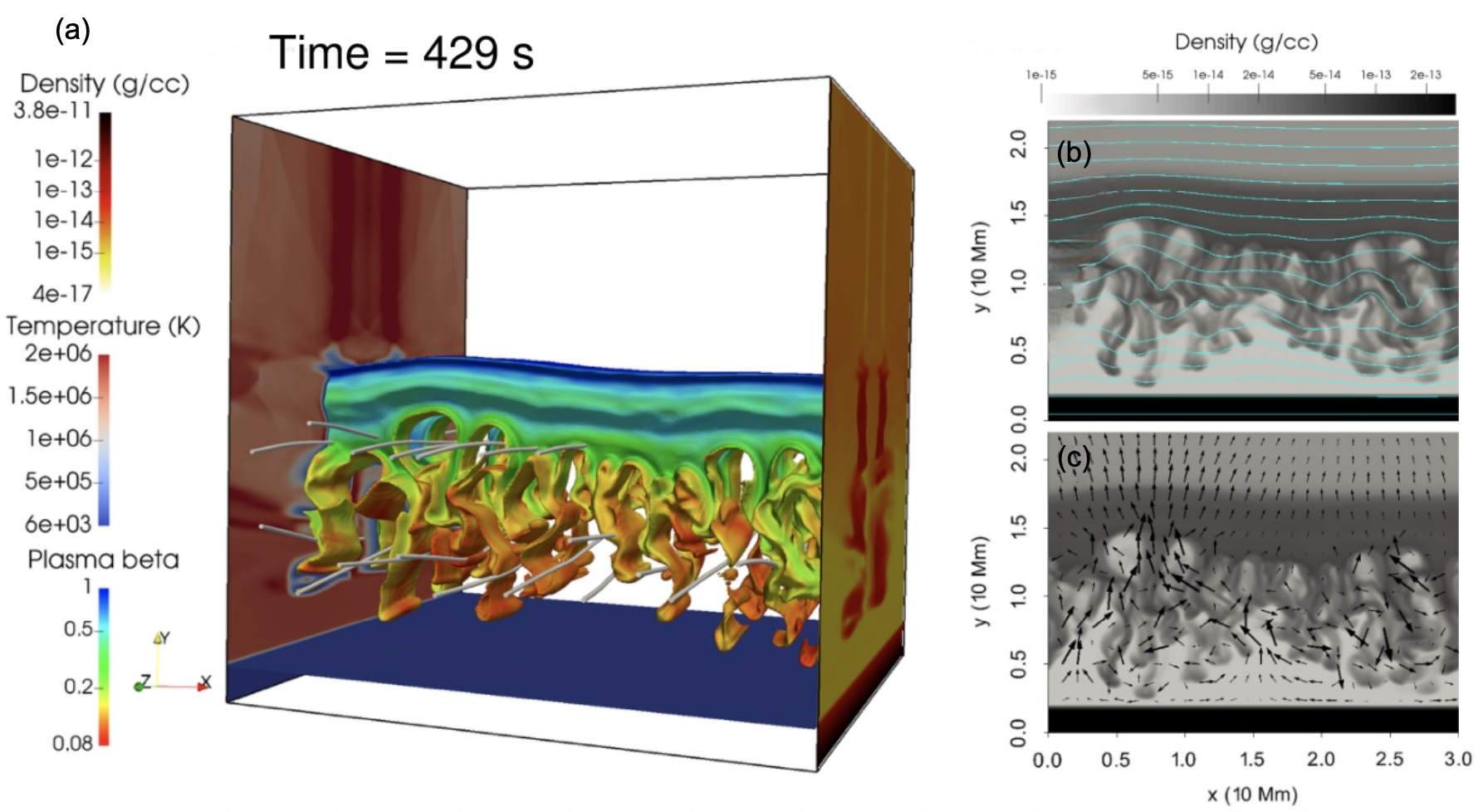}
    \caption{A 3D view of a snapshot at $429$ s, in which the isocontours are colored by local plasma-$\beta$ (a). Panels b and c show (z-)averaged density. The light blue lines denote projected magnetic field lines. The black arrows show the integrated velocity field in the plane. Adapted from \citet{Xia:2016apjl}. \textcopyright\ AAS. Reproduced with permission.} \label{fig:Xia_2016b_1}
\end{figure*}

\begin{figure*}[!b]
    \centering
    \includegraphics[width=0.9\textwidth]{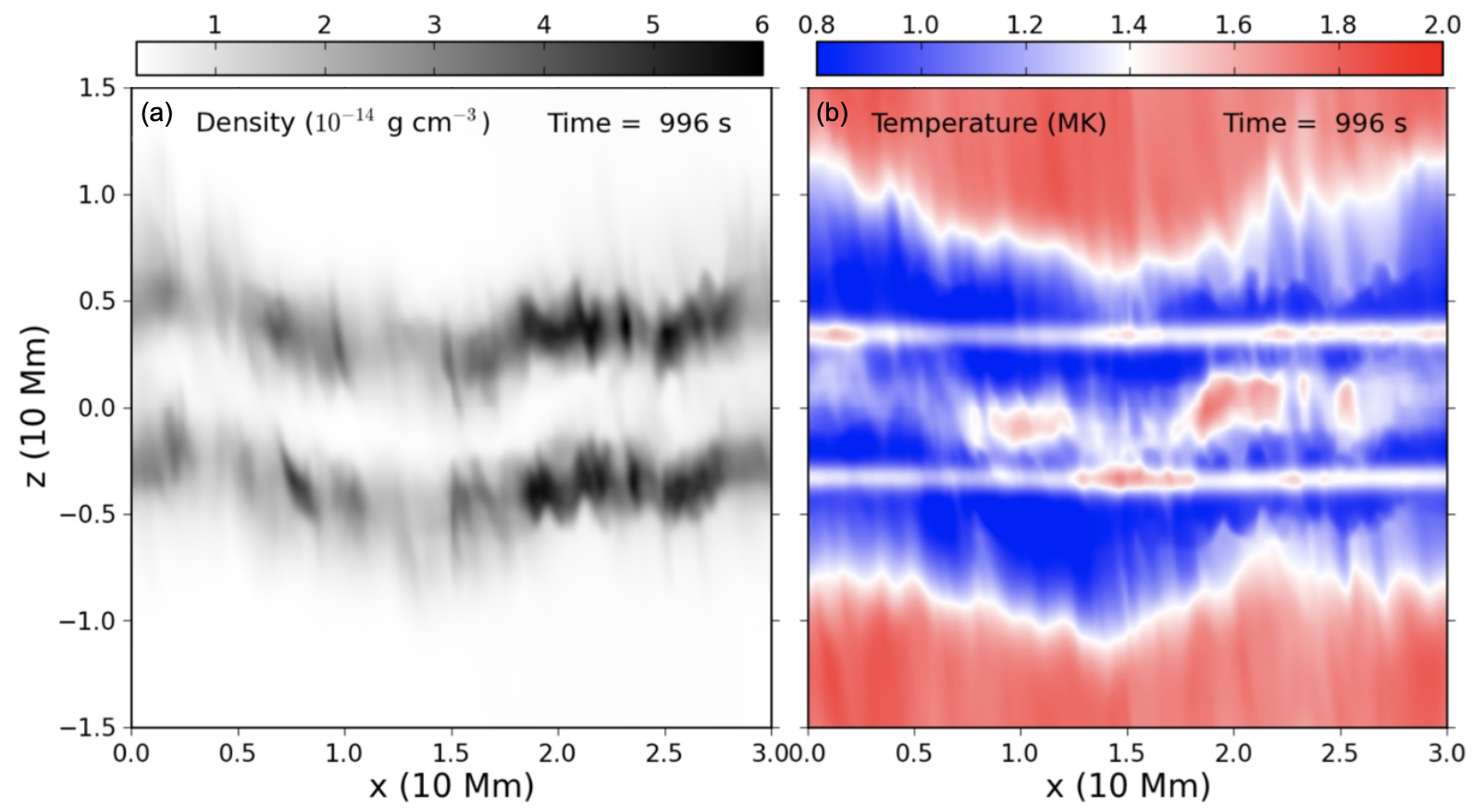}
    \caption{Top views of y-averaged density (left) and temperature (right), excluding regions with heights lower atmosphere. Adapted from \citet{Xia:2016apjl}. \textcopyright\ AAS. Reproduced with permission.} \label{fig:Xia_2016b_2}
\end{figure*}

 \citet{Keppens:2015apjl} has used a 3D experiment of the adiabatic chromosphere, transition region, and corona. The prominence is defined as a layer of higher density and lower temperature hovering in the corona. This prominence was permeated by the planar magnetic field, which has horizontal and out-of-plane (transverse) components of the magnetic field. The system is initialized in a state of vertical force balance. In the transverse direction, a pressure imbalance exists. The initial value of the transverse kinetic energy reflects the initial non-equilibrium. This kinetic energy is quickly transformed into vertical and horizontal kinetic energies, which are associated with the development of the Rayleigh-Taylor instability. In the numerical domain, the falling fingers are developed down from the prominence. At the same time, the authors observed the formation of bubbles consisting of hot coronal plasma. These bubbles show fast-rising motions. Initially, the downward motions prevail with speed exceeding $60\kms$. After some time, almost all the falling fingers reach the lower atmosphere and are deflected by the density gradient and compressed magnetic field. The authors noted that the magnetic tension of the anchored field lines, which provides a line-tying condition, can modify the evolution of the Rayleigh-Taylor instability \citep{Terradas:2015apj}. \citet{Xia:2016apjl} have extended this study, considering two sheet-like prominences with a coronal layer in between. These two prominences are permeated by a similar magnetic configuration as considered by \citet{Keppens:2015apjl}. The numerical box is shown in Figure \ref{fig:Xia_2016b_1}. Both prominence sheets showed a strongly coherent evolution, much as was discussed already in \citep[][]{Zhou:2017apj}. The authors suggested that the vertical prominence pillars seen on the limb might be due to Rayleigh-Taylor instability, and they are also the horizontal threads when seen on the disk, as shown in Figure \ref{fig:Xia_2016b_2}. Finally, \citet{Changmai:2023aap} employed a 2.5D version of the experiment by \citet{Keppens:2015apjl} in which the plasma circulation was evolved for a significantly longer time, examining the far-reaching non-linear evolution of the Rayleigh-Taylor instability as it reached a fully turbulent state, invoking comparisons against similar observational conclusions as well as providing predictions \citep[][]{Leonardis:2012apj}.

 Considering the HD Rayleigh-Taylor instability, the Atwood number, $\mathcal{A}=\frac{\rho_{+} - \rho_{-}}{\rho_{+} + \rho_{-}}$, is important when studying the fluid behavior. In the formula, $\rho_{+}$ corresponds to the denser plasma, and $\rho_{-}$ to more tenuous plasma, respectively. If the Atwood number is close to unity, the falling fingers grow larger and enter further into the tenuous layer than the lighter fluid does into the denser layer. If the Atwood number is close to zero, the falling fingers and rising plumes are expected to enter the opposite regions to a similar extent. \citet{Moschou:2015adspr} found in their 3D non-adiabatic experiment of the dipped arcade that the Atwood number is close to unity for all of their falling fingers. Indeed, condensations are frequently on the order of 100 times denser than the surrounding corona.

\begin{figure*}
    \centering
    \includegraphics[width=1.0\textwidth]{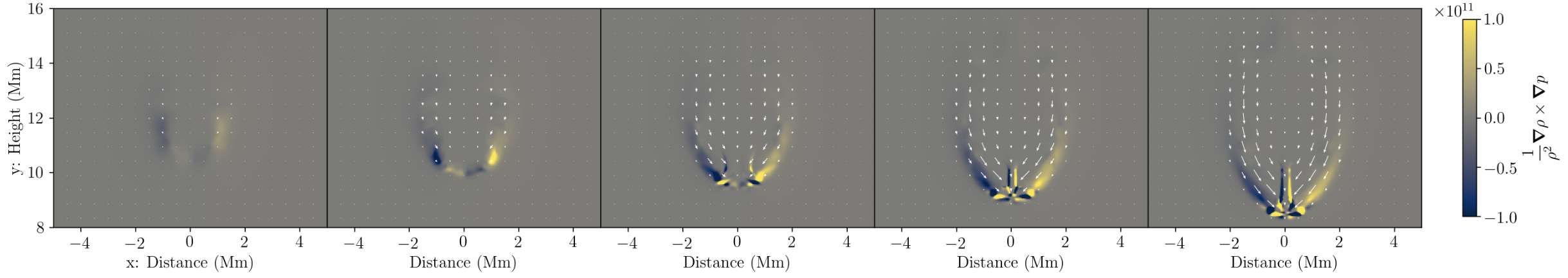}
    \caption{The temporal evolution of the baroclinicity distribution within the formed flux rope of the $3$ G magnetic field and 31 km grid resolution case. The velocity field is overplotted as white arrows. Adapted from \citet{Jenkins:2021aap}. \textcopyright\ ESO. Reproduced with permission.} \label{fig:Jenkins_2021_1}
\end{figure*}

 In addition to the Atwood number, \citet{Jenkins:2021aap} studied the baroclinicity contribution to the vorticity. Since the misalignment between the gas pressure and density gradients can produce the vortices often associated with the falling fingers of Rayleigh-Taylor instability. In their 2.5D experiment, where the magnetic flux rope axis is oriented in the invariant direction, the usual downward-falling fingers cannot be observed. However, when plasma condenses due to the thermal instability and slides down towards the magnetic dip in the cross-section of the flux rope, the authors reported on a sort of Rayleigh-Taylor instability localised to a given flux surface. The distribution of the baroclinicity shows large values in the locations of the sliding condensations on both sides with respect to the magnetic dips, showing a U-shape (Figure \ref{fig:Jenkins_2021_1}). For their magnetic configuration in the 2D setting, the Rayleigh-Taylor instability is completely suppressed, but these baroclinic signatures suggested such an evolution may be evident in a 3D domain.

\begin{figure*}[!h]
    \centering
    \includegraphics[width=0.6\textwidth]{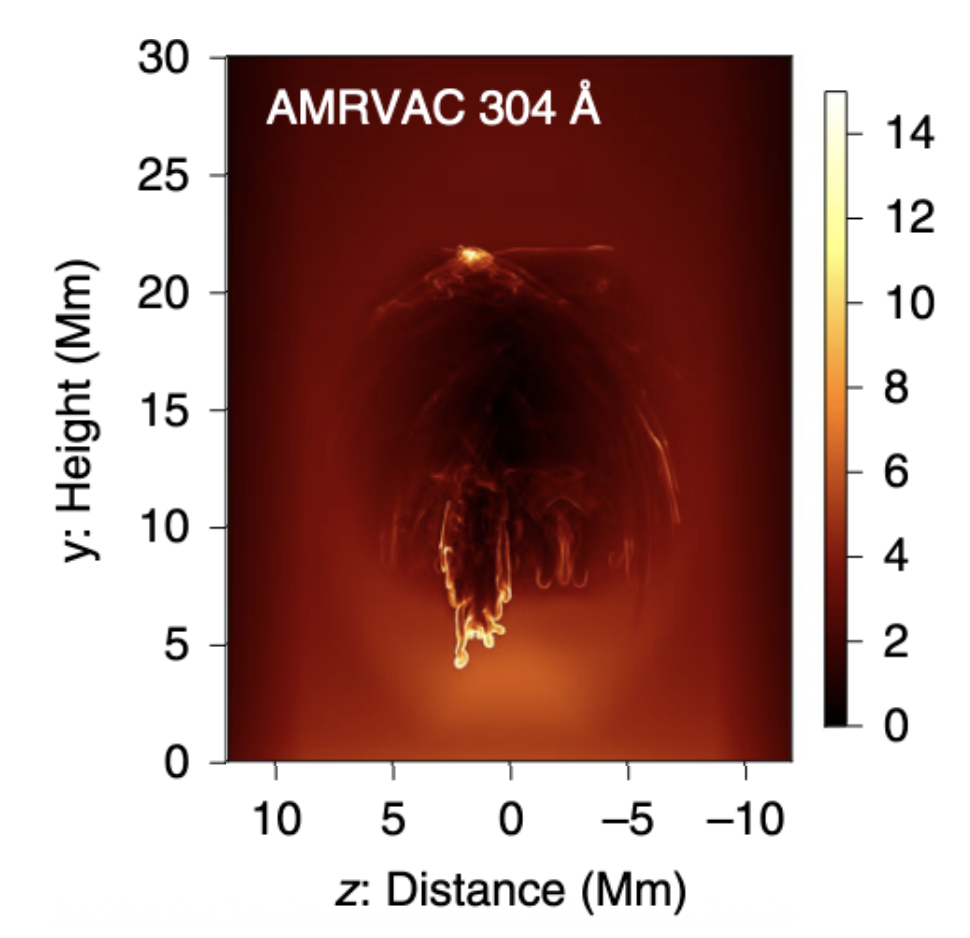}
    \caption{Synthetic image of the simulated prominence and falling fingers in SDO/AIA 304 \AA\ channel, viewed along the flux rope axis. Adapted from \citet{Jenkins:2022natas}. Reproduced with permission from SNCSC.} \label{fig:Jenkins_2022_1}
\end{figure*}

\begin{figure*}[!b]
    \centering
    \includegraphics[width=0.9\textwidth]{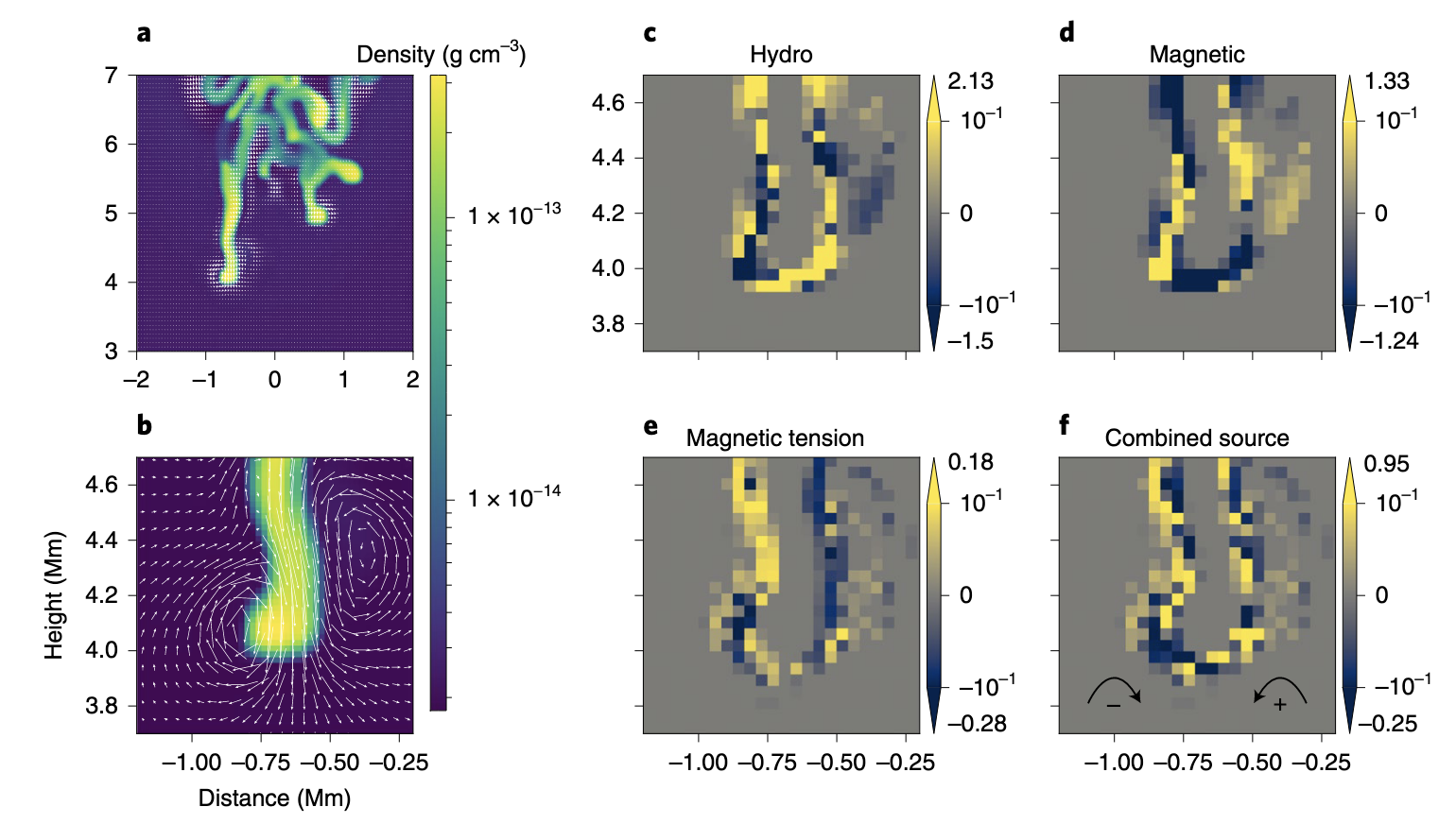}
    \caption{Panels a, b: Full-resolution density (color scale) and velocity (white arrows) for a cut through the falling finger visible at the height of approximately 4 Mm in Figure \ref{fig:Jenkins_2022_1}. Panels c–f: Half-resolution contributions to the evolving vorticity, units $s^{-2}$. Color maps are saturated to $\pm 0.1$, with maximum and minimum values indicated by the ends of the color bars. Adapted from \citet{Jenkins:2022natas}. Reproduced with permission from SNCSC.} \label{fig:Jenkins_2022_2}
\end{figure*}

  \citet{Jenkins:2022natas} studied in more detail the Rayleigh-Taylor instability in the 3D magnetic flux ropes in the non-adiabatic corona. In their work, the Atwood number is very close to unity as anticipated, and from the overall evolution, it is clear that the falling fingers dominate over rising bubbles (Figure \ref{fig:Jenkins_2022_1}). The authors also investigated the possibility of vortex cell formation due to shearing flows at the periphery of the falling fingers. They looked into the velocity field around one of the falling fingers, shown in Figure \ref{fig:Jenkins_2022_2}, finding the presence of the clockwise and counterclockwise vortexes at the left and right flanks of the finger, respectively. However, the typical Kelvin-Helmholtz swirls do not develop due to a combination of very large density contrasts of the falling finger and the suppression of it by magnetic forces. The authors considered the decomposed MHD baroclinitic contributions \citep{Shelyag:2011aap}, as well as the contribution from magnetic tension \citep{Canivete:2020aap}, shown here in Figure \ref{fig:Jenkins_2022_2}. They found that the HD baroclinicity dominates over other terms almost everywhere around the falling fingers. Magnetic tension provides the restoring force to suppress the vortices, having the opposite sign to the velocity field at both flanks. This reinforces the theory that these dynamics within prominences are driven by the gravitational interchange instability, the generalisation of the classical Rayleigh-Taylor initial condition \citep[][]{Goedbloed:2019book}. Crucially, these features were shown to have length scales that matched exactly those found in observations, suggesting that numerical resolutions might finally be reaching that necessary to fully resolve a physically meaningful relation.
 
\begin{figure*}
    \centering
    \includegraphics[width=0.95\textwidth]{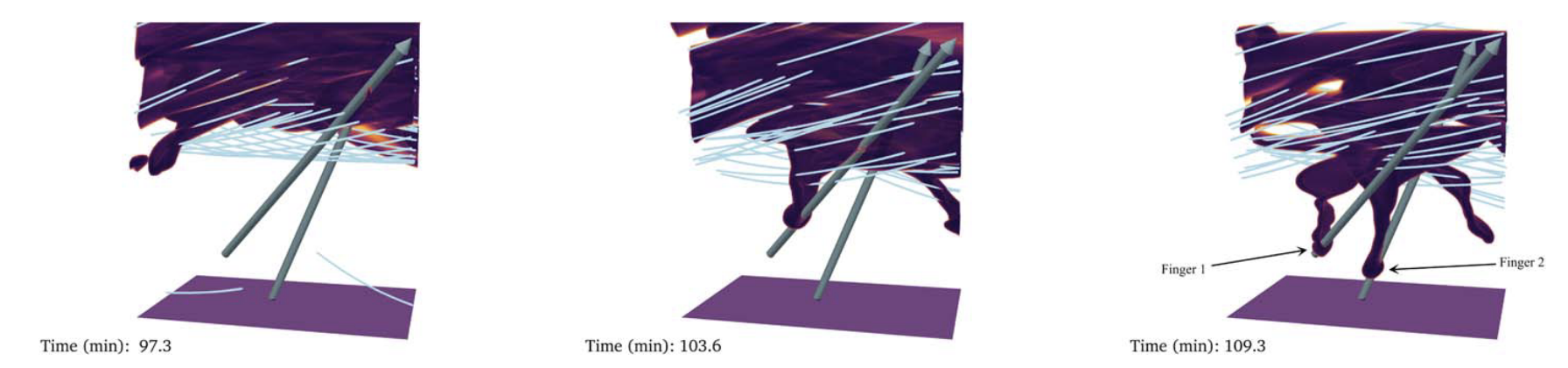}
    \includegraphics[width=0.8\textwidth]{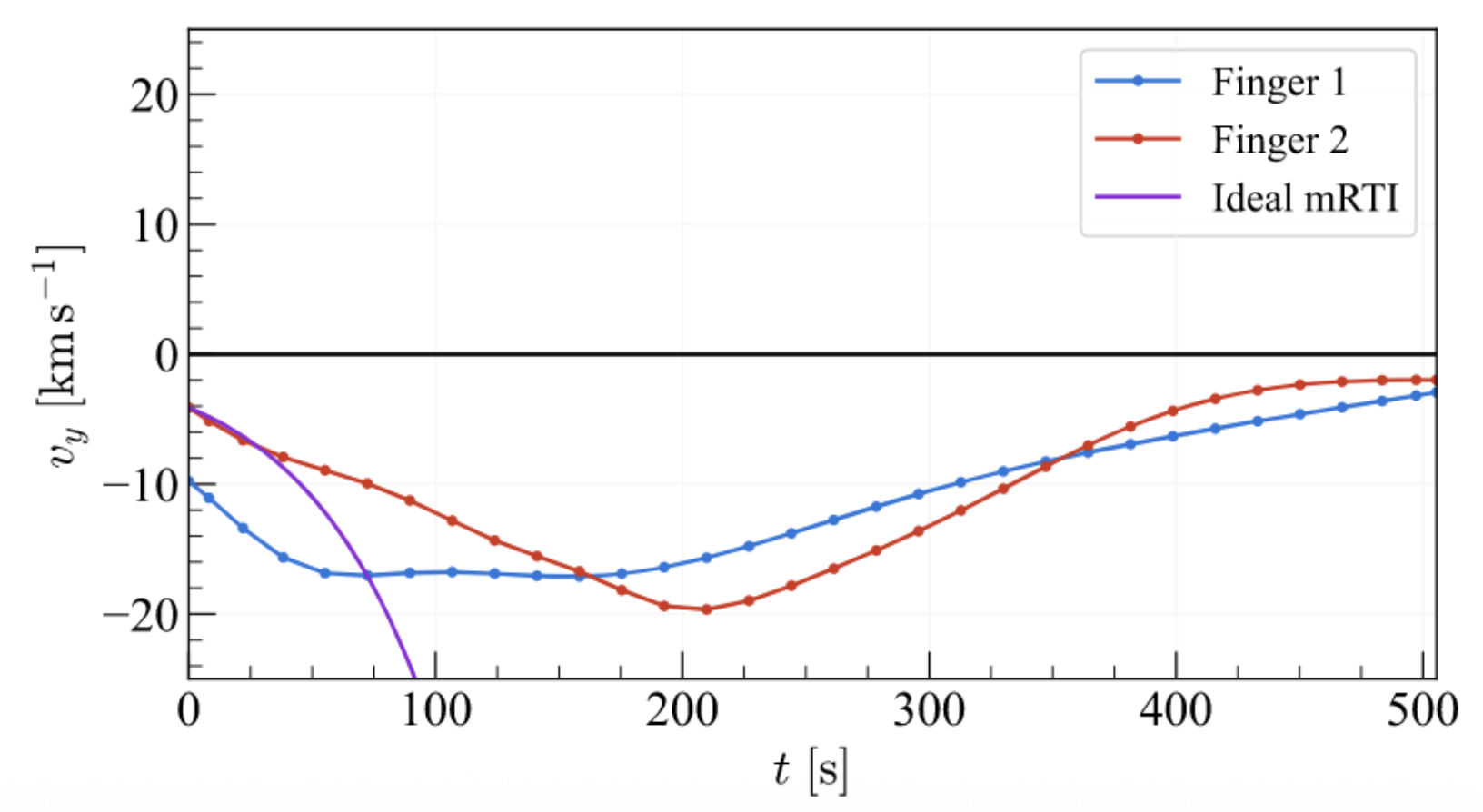}
    \caption{Top: The evolution of the magnetic Rayleigh-Taylor instability. The solid lines represent magnetic field lines, and the volume rendering displays those cells whose temperature, $T$, in MK lies below the cutoff value of 0.1 MK. The two arrows represent rays along which two falling fingers are studied, with Fingers 1 and 2 as indicated on the right. Bottom: The evolution of the vertical velocity $V_y$ is provided with the analytical solution of the linear magnetic Rayleigh-Taylor instability in purple along the rays shown in the top panel as a function of time. Blue-colored curves refer to the dynamics of Finger 1, and red-colored curves to the dynamics of Finger 2. The black solid line signifies the zero level. Adapted from \citet{Donne:2024apj}.} \label{fig:Donne_2024_3}
\end{figure*}
 

 Finally, \citet{Donne:2024apj} obtained the same falling fingers from the spine of the flux rope following the prominence formation according to the levitation-condensation scenario. These three large falling fingers show individual dynamics, as shown by the temporal evolution in Figure \ref{fig:Donne_2024_3}. Finger 1 starts with the speed $10\kms$, accelerating to $17\kms$ before decelerating. Finger 2 enters with a lower speed, $2\kms$, but accelerates significantly to $20\kms$ before decelerating to almost zero velocity. This propagation of the finger is stabilized by the magnetic tension, which counteracts the gravitational force. The analysis in the bottom panel of Figure \ref{fig:Donne_2024_3} shows that Finger 2 closely follows the analytical linear magnetic Rayleigh–Taylor solution during the first 30 seconds. Beyond this point, nonlinear effects such as the magnetic braking become significant and the evolution diverges from the analytical prediction \citet{PopescuBraileanu:2021aap}.


 To conclude, numerical modeling of prominence dynamics has significantly advanced our understanding of prominence evolution. Simulations have successfully reproduced observed behaviours, including oscillations, downflows, and rotational motions. These models demonstrated the importance of processes such as thermal instability, magnetic reconnection, and the Rayleigh–Taylor instability in shaping the evolution of prominences. By using synthetic images, many observed properties of the prominences have been reproduced.

      \section{Modeling of Coronal Rain}\label{sec:rain} 
    
 During our discussion on prominence modeling, we focused on the downflows, which in many cases can also be interpreted as coronal rain. In this section, however, we focus on models where condensations at no point accumulate in magnetic dips. The study of coronal rain is important for several reasons, among others: a) the coronal mass cycle can be investigated by following the exchange of mass between the chromosphere and corona; b) formation of the coronal rain is often attributed to the thermal non-equilibrium and thermal instability, which are fundamentally connected to the nature of coronal heating; c) the dynamic condensations trace the magnetic field of the solar corona, something we cannot yet routinely measure with direct methods; d) post-flare coronal rain provides information on the energy budget from the solar flares and evolution in the post-flare phase.
 
  \begin{figure*}
 	\includegraphics[width=0.9\textwidth]{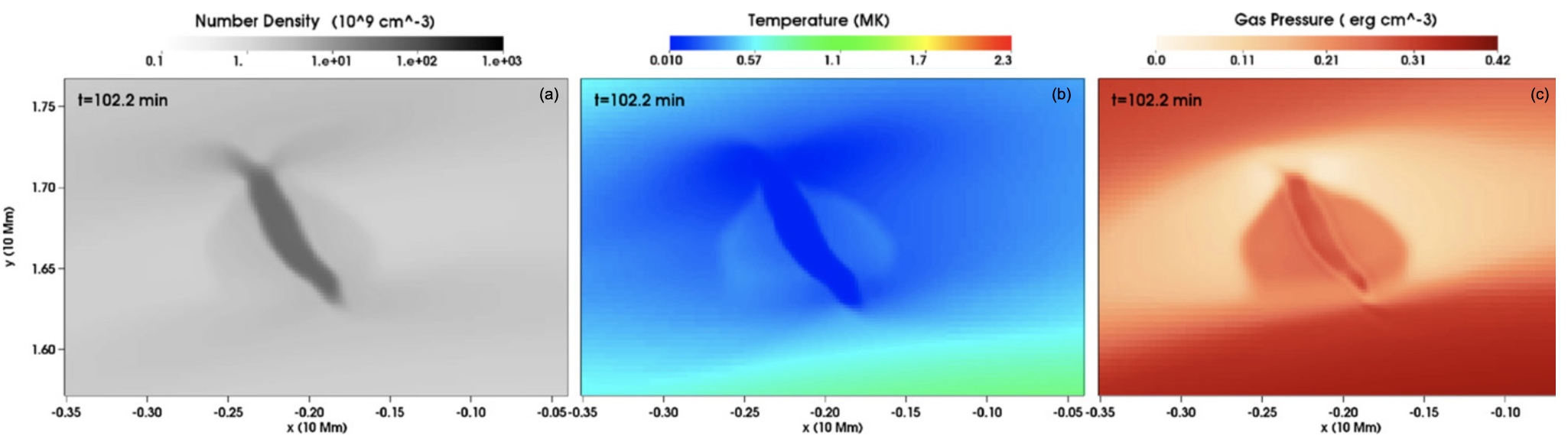}
 	\caption{At $ 102.2$ minutes, the number density (left column), temperature (middle column), and gas pressure (right column) distributions in a zoomed area of the condensation and rebound shock formation. Adapted from \citet{Fang:2015apj}. \textcopyright\ AAS. Reproduced with permission.} \label{fig:Fang_2015_3}
 \end{figure*}
 
   \begin{figure*}[!h]
 	\includegraphics[width=0.9\textwidth]{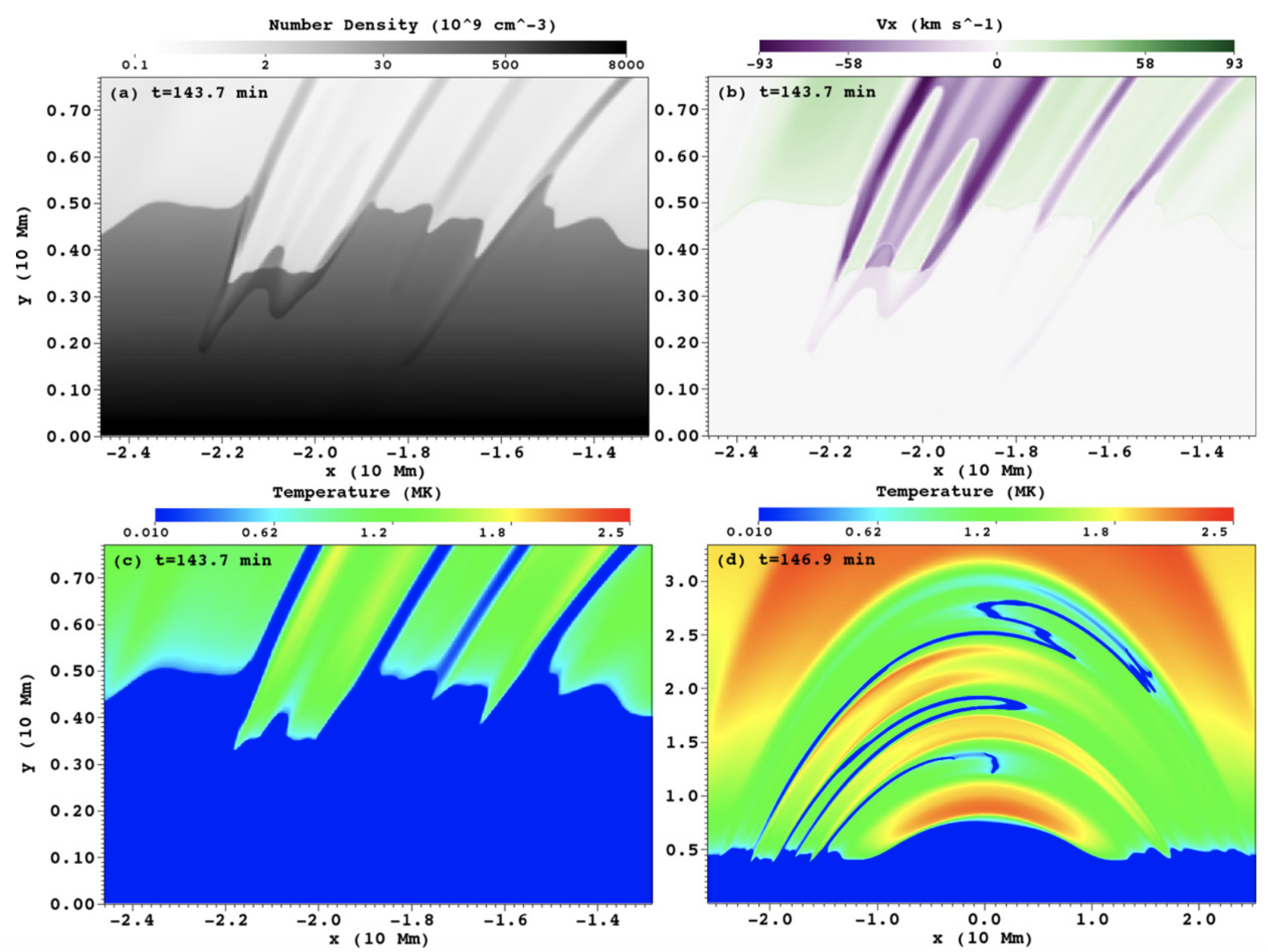}
 	\caption{Panel (a)-(c): The number density, the horizontal velocity component, and the temperature distributions at $143.7\mins$. These three panels show the downflows and upflows in the chromosphere and transition region. Panel (d) shows a larger area, providing a temperature map at $ 146.9\min$, which reveals the global distribution of coronal rain. Adapted from \citet{Fang:2015apj}. \textcopyright\ AAS. Reproduced with permission.} \label{fig:Fang_2015_6}
 \end{figure*}
 
 \begin{figure*}[!h]
 	\includegraphics[width=0.9\textwidth]{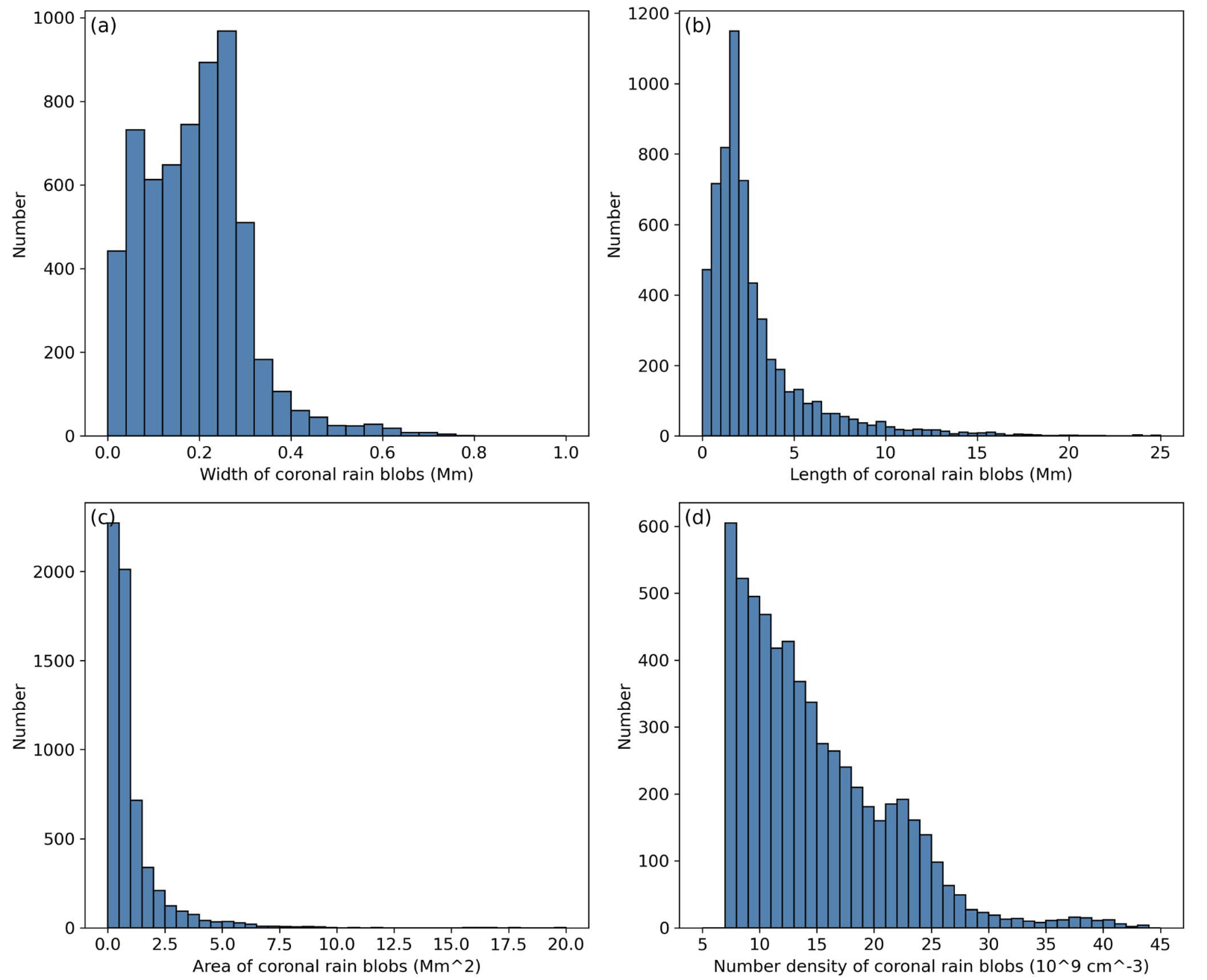}
 	\caption{Distribution of width, length, area, and mean number density of coronal rain structures. Adapted from \citet{Li:2022apj}.} \label{fig:Li_2022_4}
 \end{figure*}

\begin{figure*}[!b]
	\includegraphics[width=0.9\textwidth]{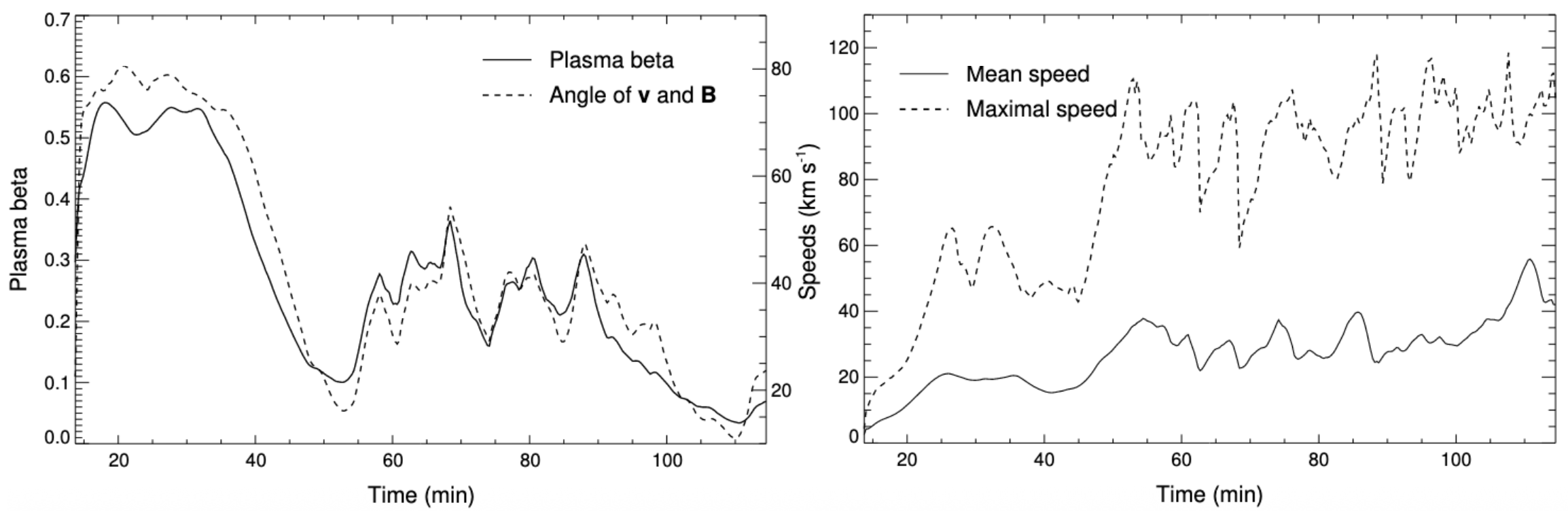}
	\caption{Left: Time evolution of coronal rain mean plasma-$\beta$ (solid line) and the angle between local velocity and magnetic field vector (dashed line). Right: Instantaneous mean speed (solid line) and maximal speed (dashed line) of coronal rain blobs. Adapted from \citet{Xia:2017aap}. \textcopyright\ ESO. Reproduced with permission.} \label{fig:Xia_2017_4}
\end{figure*}

  \begin{figure*}[!h]
	\includegraphics[width=0.9\textwidth]{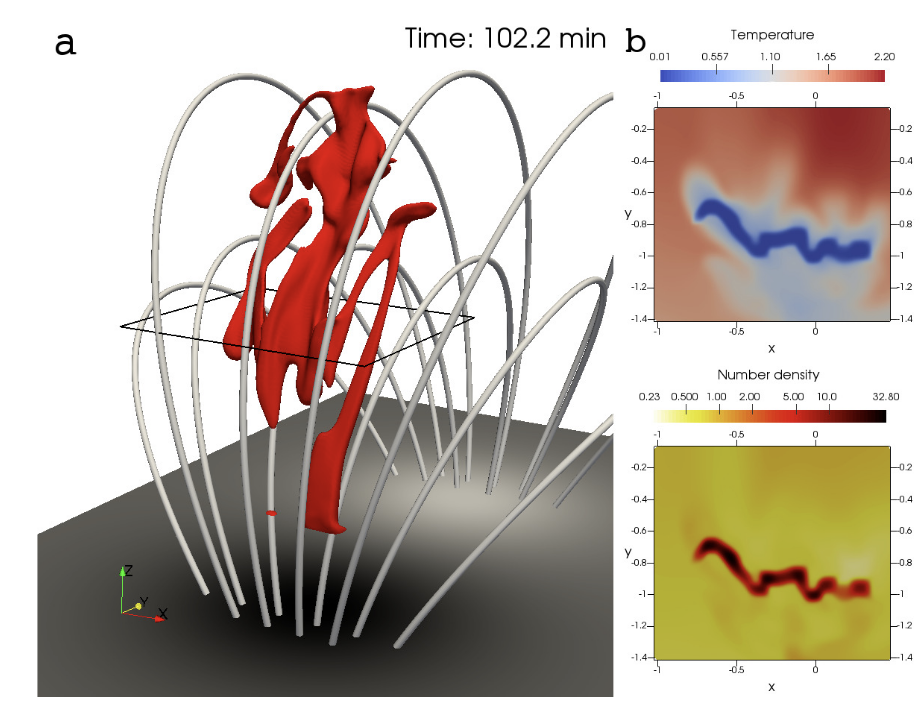}
	\caption{Coronal rain at $102.2\mins$. Panel a shows a 3D view with density isosurfaces and selected magnetic field lines. Panel b shows a plane slice across the rain blobs, indicated by the black frame in Panel a, showing the temperature map and number density distribution. Adapted from \citet{Xia:2017aap}. \textcopyright\ ESO. Reproduced with permission.} \label{fig:Xia_2017_7}
\end{figure*}

\begin{figure*}[!b]
	\includegraphics[width=0.9\textwidth]{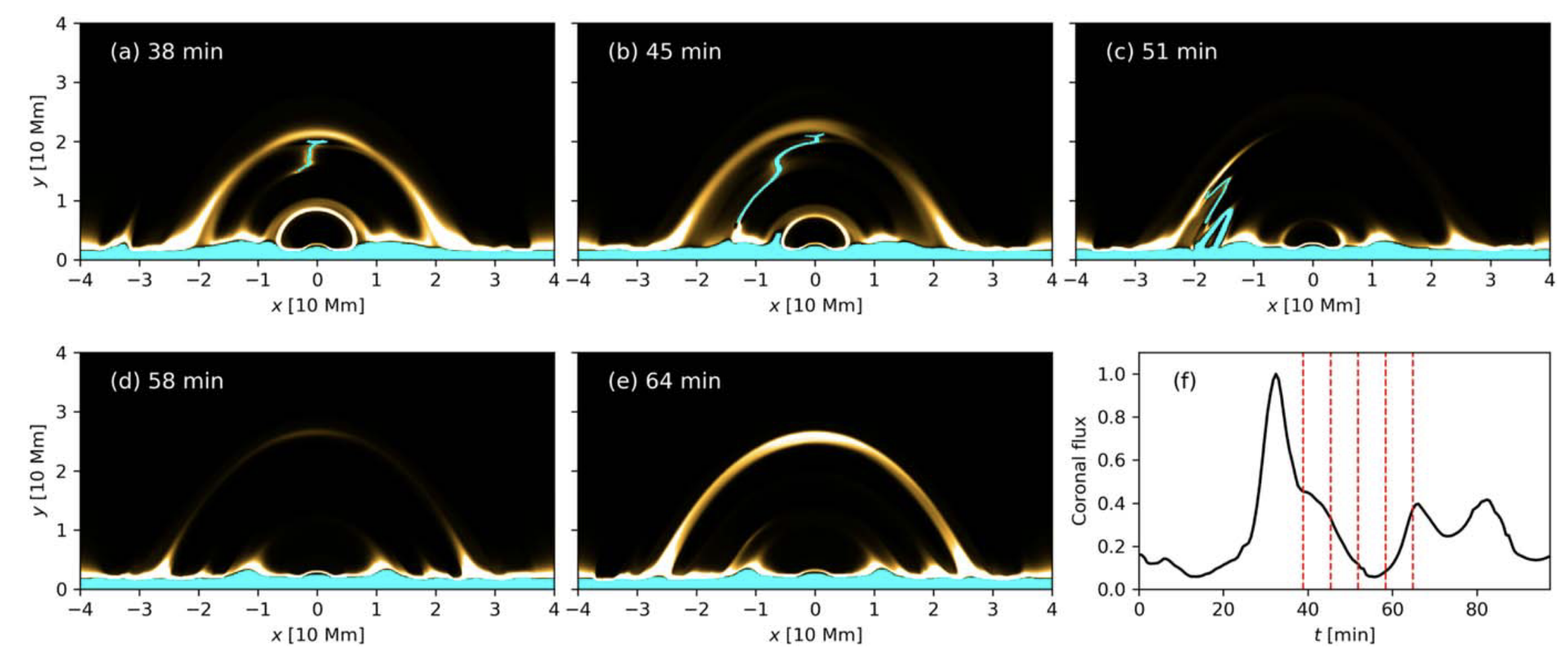}
	\caption{Time evolution of synthetic EUV 171 \AA\ data. Panels a-e show the synthetic image of the post-flare loops. The regions in cyan have temperatures lower than 0.1 MK. (f): Time evolution of the integral EUV 171 \AA\ flux from a region higher than $5$ Mm. Red vertical dashed lines in panel f correspond to the times of panels a–e. Adapted from \citet{Ruan:2021apjl}. \textcopyright\ AAS. Reproduced with permission.} \label{fig:Ruan_2021_3}
\end{figure*}

  We begin with \citet{Fang:2015apj}, who adopted a 2.5D model to study this phenomenon. The initial atmosphere is a gravitationally stratified chromosphere, transition region, and corona. The energy equation included thermal conduction, thin radiative losses, and background heating. The magnetic field in this model is defined by the 2.5D sheared arcade. After the system reached an equilibrium state, localized heating was prescribed at the footpoints of the magnetic arcade. The evaporated plasma rises along the magnetic field lines. When the heating is insufficient to balance the radiative losses, a local thermal instability develops, leading to a rapid decrease in temperature. This results in the formation of condensations. The region where thermal instability develops involves a drop in pressure local to the unstable region, leading to siphon flows in both directions away from the condensing region and along the magnetic field. The formation of the condensation thus leads to rebound shocks that propagate away from the condensation region (Figure \ref{fig:Fang_2015_3}). Similar to the studies of prominence threads, such as \citet{Claes:2020aap}, \citet{Hermans:2021aap}, and \citet{Jenkins:2021aap}, \citet{Fang:2015apj} obtained a faster growth rate of condensation in the direction perpendicular to the magnetic field. After formation, this condensation drains to the lower atmosphere along the magnetic arcade, finishing the mass cycle. When authors followed the falling blob in the lower atmosphere, they found that it produced the rebound upflow. Hence, many of the footpoints of the condensation-loaded field lines show the presence of simultaneous upflows as observed by \citet{Tripathi:2009apj} and downflows as observed by \citet{Antolin:2010apj} (Figure \ref{fig:Fang_2015_6}b). The authors continued their simulation long enough that a second event of localised heating led to further mass-cycles. 
  
  Similar to previous work, \citet{Li:2022apj} performed a 2.5D experiment of randomly heated magnetic arcades, using the same turbulence-inspired yet nanoflare structured heating as described by \citet{Zhou:2020natas}. Additionally, \citet{Li:2023apjl} considered a combination of the flux emergence with the 2.5D randomly heated arcades. In contrast to \citet{Fang:2015apj}, \citet{Li:2022apj} obtained the formation of the blobs throughout the time of the numerical experiment. Moreover, they found the periodicity of the mass cycle in this model was $83\mins$. The formation of the blobs is associated with asymmetric rebound shocks due to the different speeds of the siphon flows on both sides of the blob. Due to the repetitive cycle of the condensations formation, the authors obtained the statistics of the properties of the coronal rain shown in Figure \ref{fig:Li_2022_4}. Thus, the widths of the coronal condensations are typically less than $800$ km and their lengths vary in the range of a few hundred kilometers to $25$ Mm. The blobs move downward with a mean velocity of around $ 22.4\kms$. The properties of the coronal rain are in agreement with observations \citep{Muller:2005aap,Antolin:2012apj}. Typically, the acceleration of the blobs is below the free-fall gravitational acceleration and has already been shown to be caused by a gas pressure gradient that forms ahead of the plasma falling along the magnetic flux tube \citep{Martinez:2020aap,Adrover:2021aap}. The blobs tend to continue to grow after their initial formation as enhanced radiative losses drive further thermal instability in their transition regions, located in the heads and tails as the blobs fall under gravity. As explained before, the temperature of the transition region between condensation and corona falls into the maximum value of the radiative cooling curves used in the \texttt{MPI-AMRVAC} code, which creates favorable conditions for yet more material condensing under the thermal instability.
 
  
  An earlier study by \citet{Xia:2017aap} considered coronal rain in 3D numerical simulations of the magnetic arcades nested into a non-adiabatic gravitationally stratified atmosphere, including chromosphere, transition region, and corona. They were able to reproduce condensation at the top of the magnetic arcade and rebound shocks in a 3D configuration. The further evolution of the condensation in 3D differs significantly from the 2.5D study by \citet{Fang:2015apj}. The initial long condensation formed at the loop top starts to fragment due to the development of the Rayleigh-Taylor instability \citep[much as already described in][]{Moschou:2015adspr}. These falling blobs move in the direction of gravity with velocity angle $80\degree$ with respect to magnetic field (Figure \ref{fig:Xia_2017_4}, left) until reaching the region of the strong magnetic field and low plasma-$\beta$ at the lower heights and then drain along these field lines to the footpoints as shown in the left panel of Figure \ref{fig:Xia_2017_4}. The authors also demonstrated the presence of strong density inhomogeneities within the coronal rain clumps resembling observed multi-stranded coronal rain \citep{Antolin:2015apj2}, as shown in Figure \ref{fig:Xia_2017_7}. The horizontal extension of fragmentation was about 10 Mm, while the thickness was around 1 Mm, as shown in the central panels of Figure \ref{fig:Xia_2017_7}.
  
 Post-flare coronal rain is yet another form of coronal condensation. Observations show that these condensations typically form near the loop tops of post-flare arcades \citep{Scullion:2016apj}, much the same as the \textit{ordinary} coronal rain studied above. The \texttt{MPI-AMRVAC} code has recently been applied to multidimensional studies of solar flares \citep[see, e.g.,][]{Ruan:2023apj, Druett:2023solphys, Druett:2024aap, Ruan:2024apj}. In this review, we do not address the preceding flare modeling in detail and refer the reader to the cited works for specific details. Instead, we focus on the post-flare evolution of the plasma. 
 
 \citet{Ruan:2021apjl} focused their study on the evolution of the solar flare from pre-flare, through the impulsive phase, into its gradual phase until post-flare coronal rain forms. The mechanism of the formation of the post-flare coronal rain can be summarized as follows: the magnetic energy from the flare is converted into heat as a result of the reconnection during the impulsive phase. Using the scaling laws of \citet{Yokoyama:1998apjl} and characteristics in the numerical experiment, such as magnetic field strength, coronal number density, and half-length of the post-flare loop, \citet{Ruan:2021apjl} reproduced the actual temperature in the experiment, which suggests that the flare temperature can be well predicted from these properties.
 
 The energy loss at the beginning of the gradual phase is defined by thermal conduction, therefore, the heat is conducted away from the reconnection region down to the footpoints of the post-flare loops. The heating leads to the evaporation of the chromospheric plasma, which gradually fills the post-flare loops. Around this time, the radiative losses become dominant due to the concentration of plasma at higher heights \citep{Cargill:2004apj}. Thermal instability develops, causing a drop in temperature and gas pressure, which ultimately leads to the condensation of evaporated plasma and the depletion of the post-flare loops. In the synthetic 171 \AA\ images, this appears to be the dark post-flare loops phenomenon remarked upon in observations \citep{Song:2016apj}. First, this darkening happens when initially hot loops start to cool down due to thermal instability and formation of condensation (Figure \ref{fig:Ruan_2021_3}a). Second, after the condensation forms, the loops recover their temperature; however, the dark appearance can be explained by plasma depletion, as the plasma drains back to the lower atmosphere (Figure \ref{fig:Ruan_2021_3}b-d). The second injection of cold plasma from the lower atmosphere into the low-pressure region occurs due to the same siphoning effect as for \citet{Donne:2024apj}. 
 


    \section{Post-processing Models as Synthetic Observations }\label{sec:synthetic}

    At the time of writing, \texttt{MPI-AMRVAC} solves a set of equations and source terms as outlined in Section~\ref{sec:methods} but does not account for radiation transport aside from the radiative losses sink term in the energy equation.
    For now, all endeavours to compare results obtained using \texttt{MPI-AMRVAC} with their motivating observations have therefore been carried out as a post-processing step.
    For example, it is now relatively commonplace to synthesise \texttt{MPI-AMRVAC} simulations for qualitative and quantitative comparison against EUV observations taken by the SDO/AIA.
    This we refer to as `approximate forward modelling', where strong approximations are made as to the behaviour of radiation throughout a simulation domain, often reducing the calculation to precomputed look-up tables that map the local density, pressure, and temperature to some intensity with limited or no considerations for the back processing of the radiation field on itself by the local plasma conditions.
    This is most applicable in the optically thin regime, as radiation interacts very little between the source and the observer by definition.
    In the optically thick regime, we are required to consider the explicit propagation of radiation. In this case, local volumes are influenced by radiation from regions potentially several megameters away, primarily from the solar surface but also from the inter-condensation radiation.
    These efforts are relatively new, with application of these methods to \texttt{MPI-AMRVAC} output only beginning within the last few years, but are already being widely adopted.
    This we refer to as `NLTE modelling'.
    In what follows, we will summarise these methods and some specific conclusions that have come as a direct consequence of these synthetic observation representations.
    
    \subsection{Approximate Forward Modeling}

    Beginning with the radiative transfer equation as described in \citet{Rybicki:1986book},
    \begin{equation}
        I_\lambda(\tau_\lambda) = I(0)e^{-\tau_\lambda} + \int_0^{\tau_\lambda} e^{-(\tau_\lambda - \tau'_\lambda)} S_\lambda(\tau'_\lambda) d\tau'_\lambda \label{eq:radiation_transport}
    \end{equation}
    where $I_\lambda(\tau_\lambda)$ is the emergent intensity of light, of wavelength $\lambda$, measured at a point in space. 
    Beginning with an intensity $I_\lambda (0)$ (read background intensity, can be zero) and passing through a medium with total optical thickness $\tau_\lambda$, the source function $S_\lambda$ measures the addition/removal of intensity to the lightbeam as a consequence of local conditions within the $\tau'_\lambda$ subvolume. 
    In a non- or weakly scattering medium (where the cross-sectional scattering term $\sigma J \approx 0$), $S_\lambda$ is defined as $S\equiv j_\lambda / \alpha_\lambda$ where $j_\lambda$ and $\alpha_\lambda$ are the emissivity and absorption coefficients, respectively.

\begin{figure}
    \centering
    \includegraphics[width=0.45\linewidth]{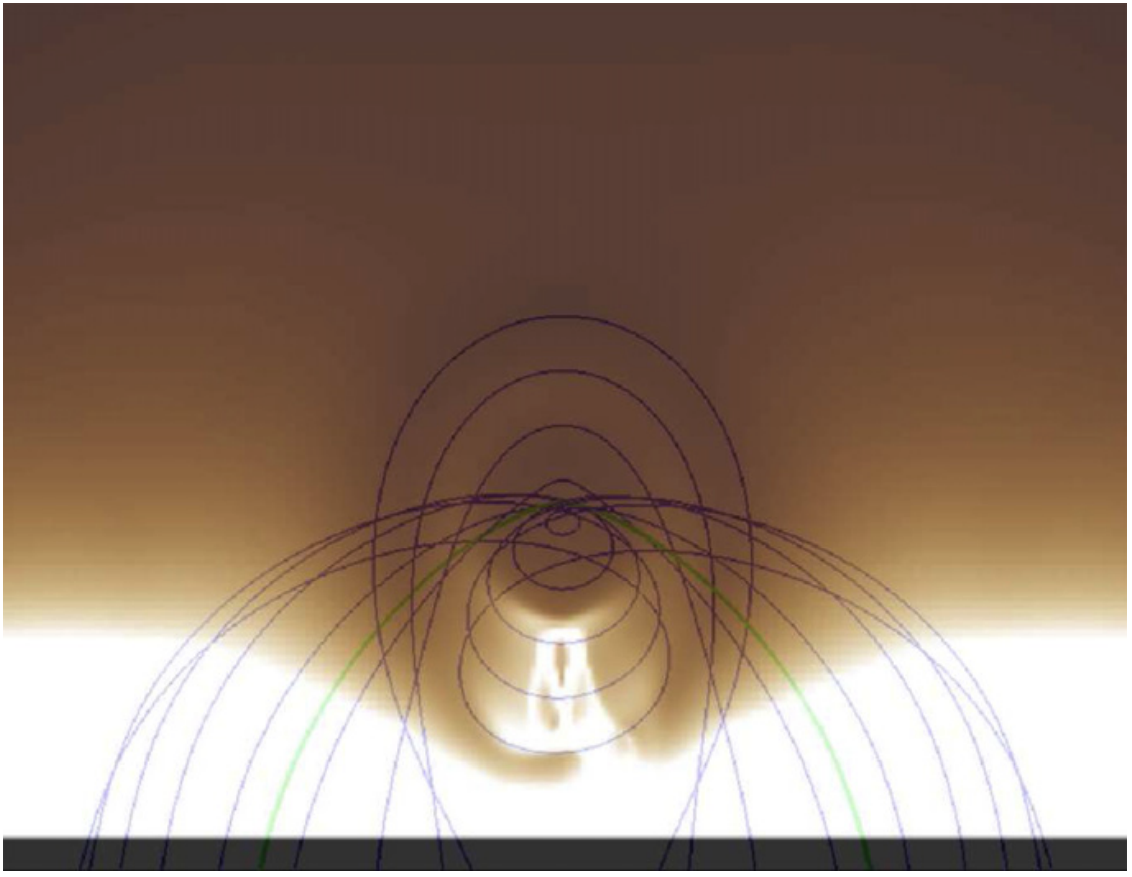}
    \caption{EUV synthesis of a prominence condensation formed in situ within a stable 3D magnetic flux rope. The 193~\AA\ synthesis for the model shown in Figure~\ref{fig:Xia_2014_1} viewed along the axis of the flux rope. Adapted from \citet{Xia:2014apj}. \textcopyright\ AAS. Reproduced with permission.}
    \label{fig:Xia_2014b_2}
\end{figure}

    \citet{Xia:2014apjl} was the first to present the appearance of prominence condensations formed via the thermal instability within a stable 3D magnetic flux rope configuration.
    Herein, the authors considered only the emissivity quantity, $j_\lambda(\tau') = \frac{A_\mathrm{b}}{4\pi} n^2_e(\tau')G_\lambda(n_\mathrm{e}(\tau'), T(\tau'))$, where $A_\mathrm{b}$ is the abundance of the relevant atomic species, $n_\mathrm{e}$ the electron number density, and $G_\lambda$ the contribution function for the relevant spectral window provided by the \texttt{Chianti} database \citep[][]{Landi:2013apj}.
    Since this quantity is purely local, the resulting synthesis images are constructed as a line of sight (LOS) integration and assume that the plasma condition is optically thin, the result shown here in Figure~\ref{fig:Xia_2014b_2}.
    As the absorption coefficient is dropped in this representation, non emissive prominence material simply appears less bright than the surrounding brighter loops, with the PCTR appearing bright due to the intermediate temperatures present there.
    The limited resolution owed to the computing capacity available at the time imposed a strong restriction on the field-aligned pressure scale height and, in turn, the recoverable scales of the prominence fine structure \citep[see also][]{Kaneko:2015apj}.
    Hence, no fine structure is present in either the base simulation or the resulting syntheses.
    The authors thus refer to the formed monolithic sheet as a macroscopic condensation, but nevertheless recover the EUV horns commonly remarked upon in observations, tracing the inside curvature of the associated coronal cavity.

\begin{figure}[!b]
    \centering
    \includegraphics[width=1.0\linewidth]{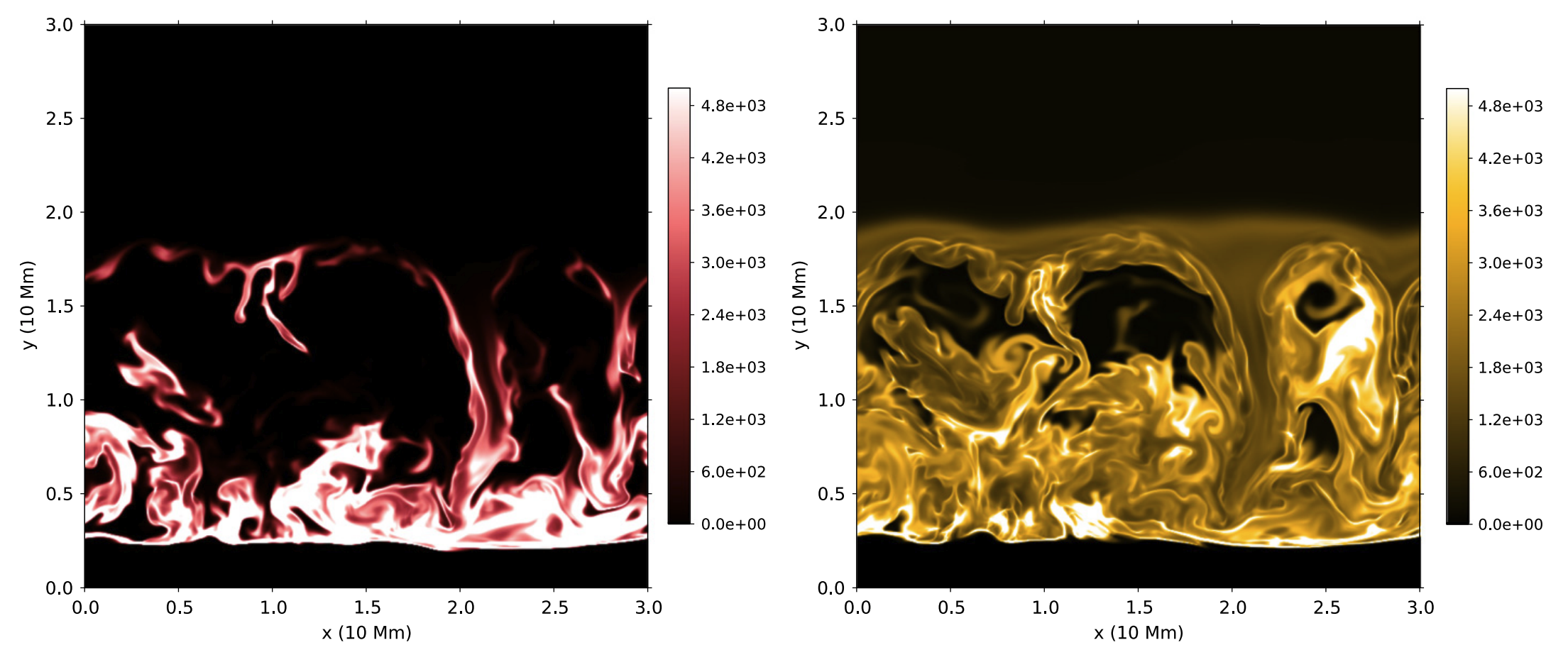}
    \caption{Appearance of Rayleigh-Taylor and Kelvin-Helmholtz unstable plasma dynamics within a solar prominence in the 304 and 171~\AA\ passbands of SDO/AIA. Adapted from \citet{Keppens:2015apjl}. \textcopyright\ AAS. Reproduced with permission.}
    \label{fig:Keppens_2015_1}
\end{figure}

    Shortly thereafter, \citet{Keppens:2015apjl} removed this scale height limitation by restricting the extent of their 3D domain to the dips of the magnetic topology, here assumed horizontal, responsible for the stable suspension of prominence plasma. 
    Owed to the unsheared orientation of the magnetic field in the initial condition, the loaded prominence material interface deformed in accordance with the HD Rayleigh-Taylor instability, leading to the formation of the characteristic fingers and bubbles, yielding intricate intensity variations when synthesised in the EUV passbands as shown in Figure~\ref{fig:Keppens_2015_1}.
    $\alpha_\lambda$ is not considered here either.
    Of particular interest, the appearance of secondary instabilities reminiscent of the Kelvin-Helmholtz instability is also clear within their syntheses.
    Their highly resolved case study lies below the resolution of equivalent observations, however, leaving direct comparisons to follow-up studies.
    Nevertheless, it is not yet clear whether such simulations are yet recovering the fundamental length scales associated with the condensations, or whether they are still limited even on cell scales in terms of their properties.
    One should caution the reader from overinterpreting the appearance of the dynamics in the 304~\AA\ passband as the underlying optically thin photoionisation approximation is not physically correct for the primary photon donor He~{\sc ii} to this passband.


\begin{figure}[!b]
    \centering
    \includegraphics[width=1.0\linewidth]{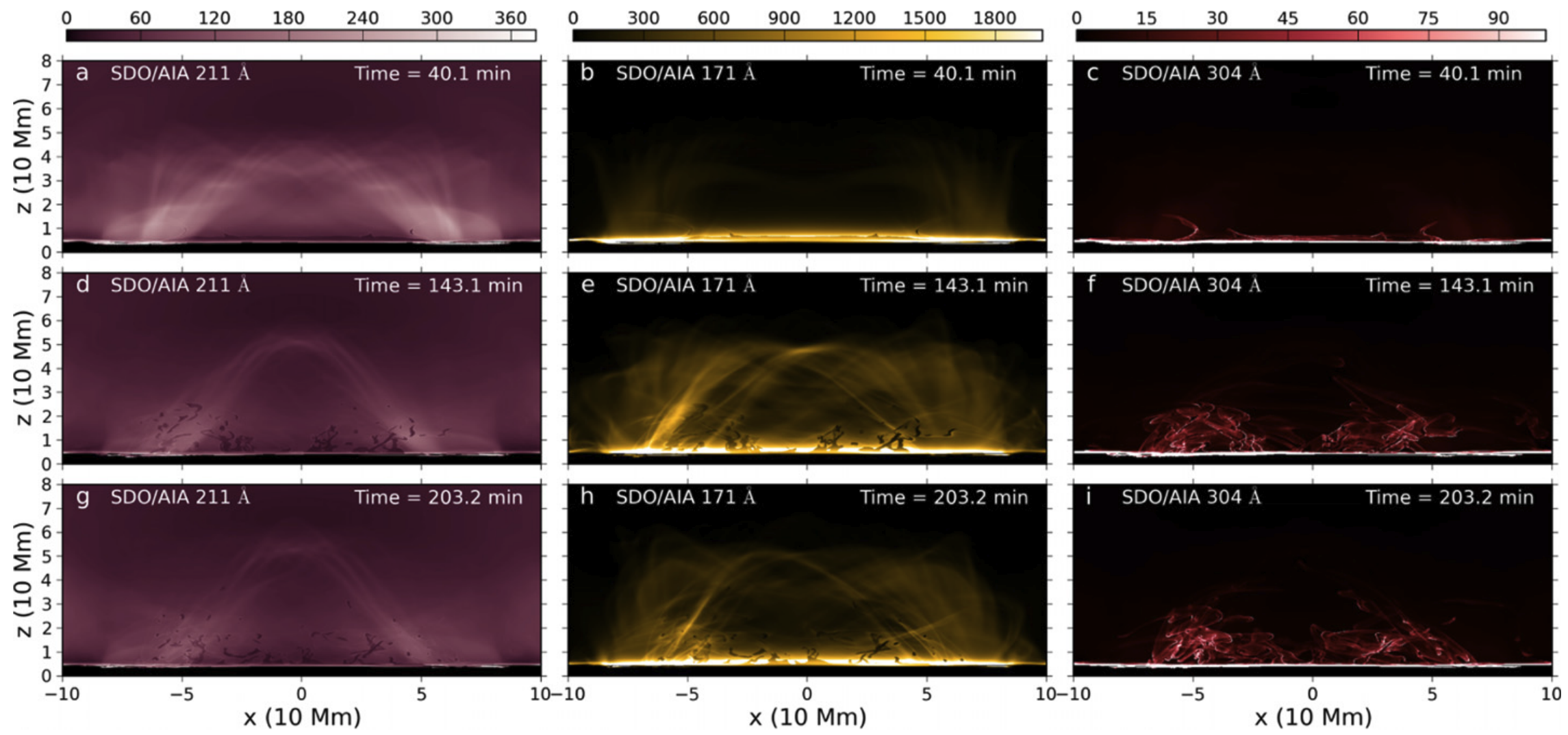}
    \caption{Synthetic representations of plasma circulation dynamics as if observed with the 211, 171, and 304~\AA\ passbands of SDO/AIA. The approximate treatment of $\alpha_\lambda$ yields a stronger intensity contrast for the discrete condensations in comparison to previous works. Adapted from \citet{Xia:2016apj}. \textcopyright\ AAS. Reproduced with permission.}
    \label{fig:Xia_2016_2}
\end{figure}
    
    In the seminal work of \citet{Xia:2016apj}, as detailed in Section~\ref{sec:plasma}, they used a similar method to their earlier work in \citet{Xia:2014apjl} to synthesise condensations formed \textit{ab-inito} via the evaporation-condensation mechanism. 
    Now at a much higher resolution, the authors considered any column mass that exceeded 2$\times$10$^{10}$~cm$^{-3}$ to adhere to approximately optically thick behaviour, imitating the role of $\alpha_\lambda$, and giving hints as to the spatial distribution of cool plasma along a LOS and across the synthesised field of view. 
    This work represents the first highly successful qualitative comparison between prominences synthesised in EUV and corresponding observations, shown in Figure~\ref{fig:Xia_2016_2}. 
    The same authors then apply this pseudo-$\alpha_\lambda$ approach in \citet{Xia:2016apjl} to a twin-layer solar prominence, recovering yet-finer structure aided by the more optically thick compliant method. 
    Intriguingly, the authors present density and temperature-integrated representations of their domain, which yield striations reminiscent of the characteristic threaded appearance of solar filaments and are shown to be the projection of the Rayleigh-Taylor instability fingers visible from the side, but no synthetic counterparts are offered.

\begin{figure}
    \centering
    \includegraphics[width=0.9\linewidth]{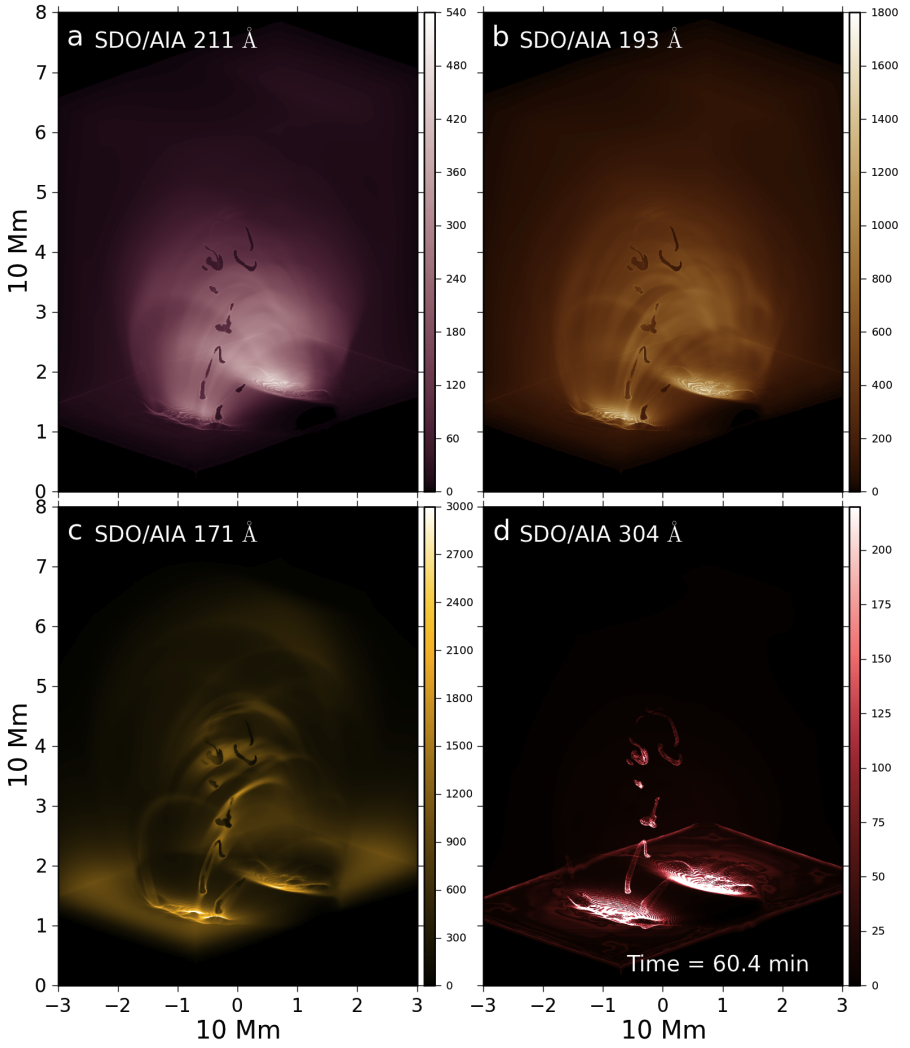}
    \caption{211, 193, 171, and 304~\AA\ synthetic representations of coronal rain formation within a constantly heated coronal bipole configuration. Adapted from \citet{Xia:2017aap}. \textcopyright\ ESO. Reproduced with permission.}
    \label{fig:Xia_2017_1}
\end{figure}

    In \citet{Xia:2017aap}, the authors present the EUV synthesis of coronal rain condensations induced by constant footpoint heating at chromospheric heights.
    The corresponding syntheses describe brightenings at these footpoints in all examined spectral lines, and once the condensations form, the pseudo-$\alpha_\lambda$ approach yields the discrete absorption features seen in Figure~\ref{fig:Xia_2017_1} that propagate towards the bottom boundary and fragment in the process.
    Since discrete brightenings are observed throughout the low solar atmosphere, be that the so-called `moss' or the finite-scale nano flares, such constant heating and broad brightenings represent a first order approximation to the evaporation-condensation mechanism.
    The authors draw attention to the fact that the current configuration lies in a parameter space perhaps outside that of the Sun on account of the formed condensations being larger than those observed, despite having a cell resolution of 78~km. 
    This strongly suggests that the constant heating is nonphysical since smaller, discrete perturbations will disrupt the instability pathway and create smaller condensations.

\begin{figure}
    \centering
    \includegraphics[width=0.5\linewidth]{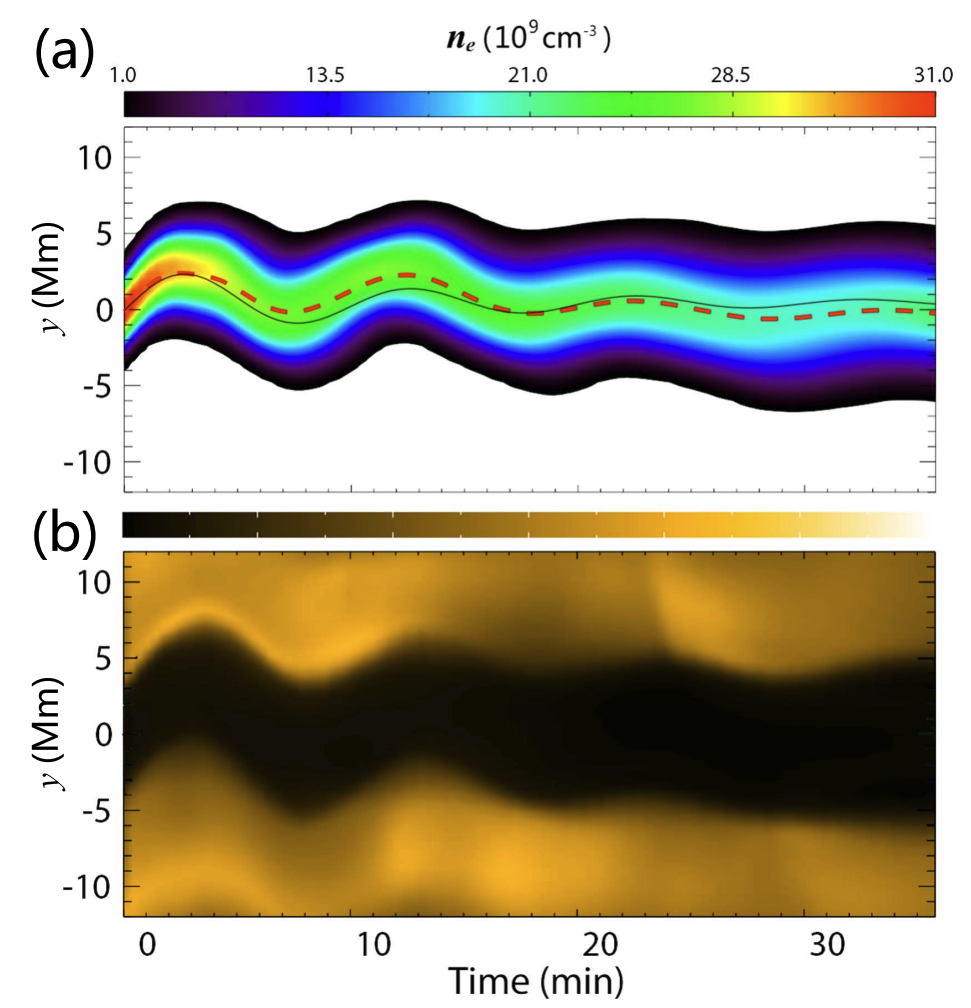}
    \caption{Primitive electron number density in an oscillating prominence spine and equivalent EUV synthesis demonstrating that the results are largely $n_e$ independent. Adapted from \citet{Zhou:2018apj}. \textcopyright\ AAS. Reproduced with permission.}
    \label{fig:Zhou_2018_4}
\end{figure}

    The EUV syntheses of prominence oscillations explored by \citet{Zhou:2018apj} are constructed using a temperature-only version of $G_\lambda$ as it can be shown that $G_\lambda$ depends only weakly on $n_\mathrm{e}$. 
    Even with such a simplification, the authors go on to demonstrate that time-distance cuts reproduce oscillatory frequencies that match observations.
    This is because such bulk oscillations consider the entire prominence extent, and finer structure was unable to form as a consequence of the lower resolution of 300+~km employed, as was previously the case for \citet{Xia:2014apjl}.
    Hence, the syntheses are directly proportional to the primitive temperature via the contribution function mapping, shown in Figure~\ref{fig:Zhou_2018_4}, meaning the oscillations in the simulation understandably match well those in the syntheses.
    Identical methods have also been applied to coronal rain studies, in which similar synthetic features are present and complementary conclusions are drawn \citep[e.g.][]{Li:2022apj, Li:2023apjl, Liakh:2023apjl}.
    Of particular interest is the recent work of \citet{Li:2025arxiv} where, although already present within the primitive variables, the accompanying EUV syntheses describe a host of secondary instabilities readily observable within the EUV passbands of SDO/AIA and are ripe for direct comparisons, albeit at lower resolutions.


\begin{figure}
    \centering
    \includegraphics[width=1.0\linewidth]{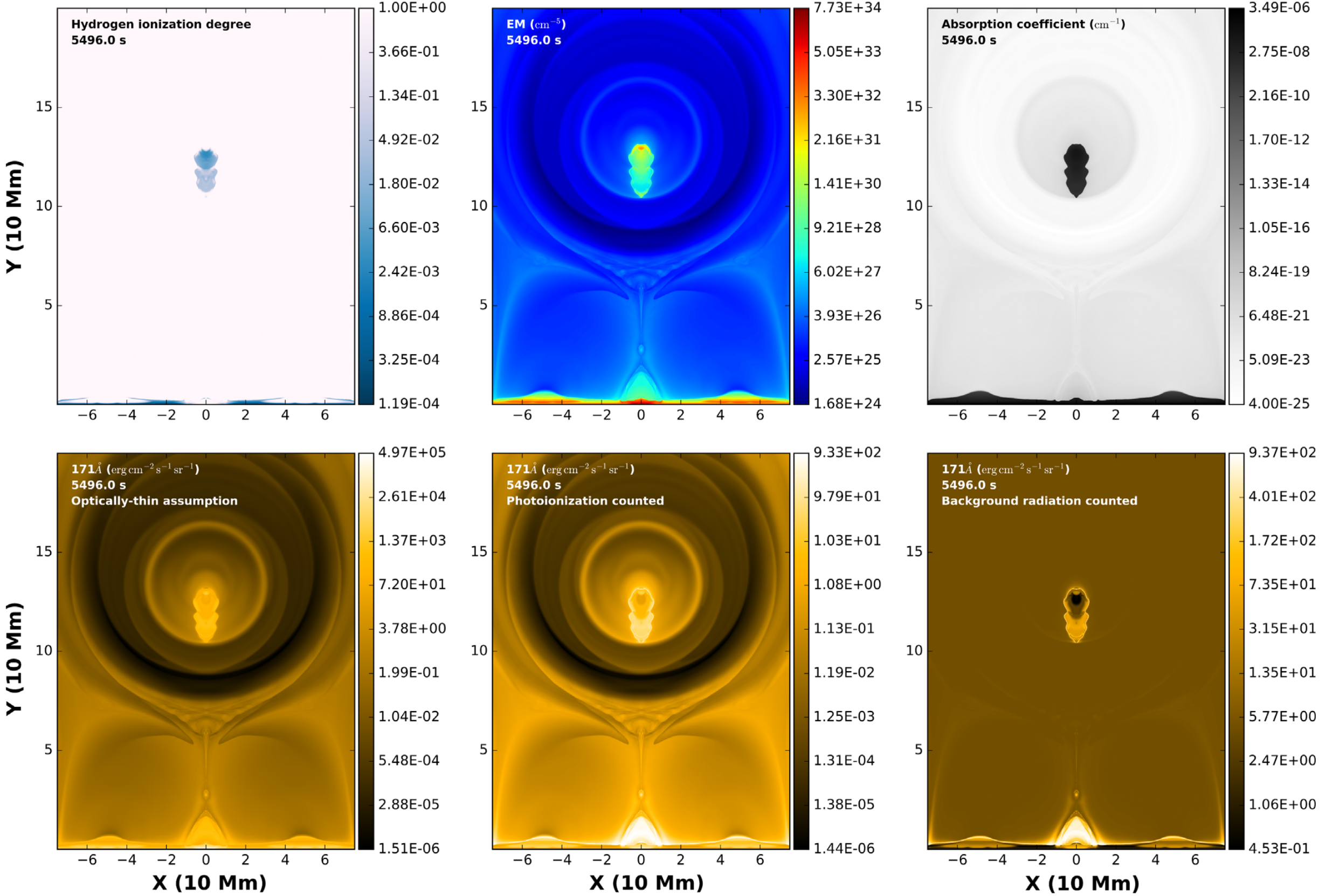}
    \caption{Application of different assumptions in the synthesis calculation process. (a) Derived ionisation degree under the LTE assumption, (b) emission measure as $\int n_e^2 dl$, (c) the EUV absorption coefficient $\alpha_\lambda$ as in the text, (d) the optically thin assumption of $j_\lambda$ only, (e) additionally considering the photon removal of $\alpha_\lambda$, and (f) homogenisation of background intensity $I_\lambda(0)$ to emphasise prominence/filament absorption. Adapted from \citet{Zhao:2019apj}. \textcopyright\ AAS. Reproduced with permission.}
    \label{fig:Zhao_2019_1}
\end{figure}

    \citet{Zhao:2019apj} were the first to additionally consider the EUV absorption coefficient $\alpha_\lambda$ following heritage established by \citet{vanDoorsselaere:2016fass}.
    Absorption in the EUV iron lines observed by SDO/AIA is provided in part by the photoionisation of H~{\sc i}, He~{\sc i}, and He~{\sc ii} atoms by the Fe photons, leading to a decrease in their intensity in those locations.
    \texttt{MPI-AMRVAC} does not explicitly consider those populations and their evolutions, requiring an approximation that is consistent with the physics.
    The authors consider LTE conditions where the ionisation state is set by the temperature in the Saha equation, and the relative populations of the H~{\sc i}, He~{\sc i}, and He~{\sc ii} elements are then found through an iterative process maintaining the fixed total population.
    It is important to emphasise here that the LTE approximation applies only to $\alpha_\lambda$, not the contribution functions within $j_\lambda$, which assume the standard `coronal approximation'.
    The absorption is therefore approximated as $\alpha_\lambda(\tau') = (n_\mathrm{HI}(\tau') + n_\mathrm{HeI}(\tau')) + \sum_s w_\mathrm{s}(\tau')A_\mathrm{b,s}\sigma_\mathrm{s}$ with the weights $w_\mathrm{s}$ given by $w_\mathrm{HI}=1-n_\mathrm{HII}/n_\mathrm{HI},~w_\mathrm{HeI}=1-(n_\mathrm{HeII}-n_\mathrm{HeIII})/n_\mathrm{HeI}$, and $w_\mathrm{HeII}=n_\mathrm{HeIII}/n_\mathrm{HeI}$, and the photoionisation cross-section $\sigma_\mathrm{s}$ provided from experiments.
    Now with an active photon loss in the integration, the authors show how this significantly modifies the resulting syntheses at each stage in the process, reproduced here in Figure~\ref{fig:Zhao_2019_1}, resulting in stronger contrast between the background and the filament or prominence projection.
    Naturally, this leads to the clearer appearance of finestructuring since prominence material actively removes intensity rather than passively contributing zero, alongside the PCTR being enhanced in several of the considered spectral lines in accordance with observations.

\begin{figure}
    \centering
    \includegraphics[width=0.95\linewidth]{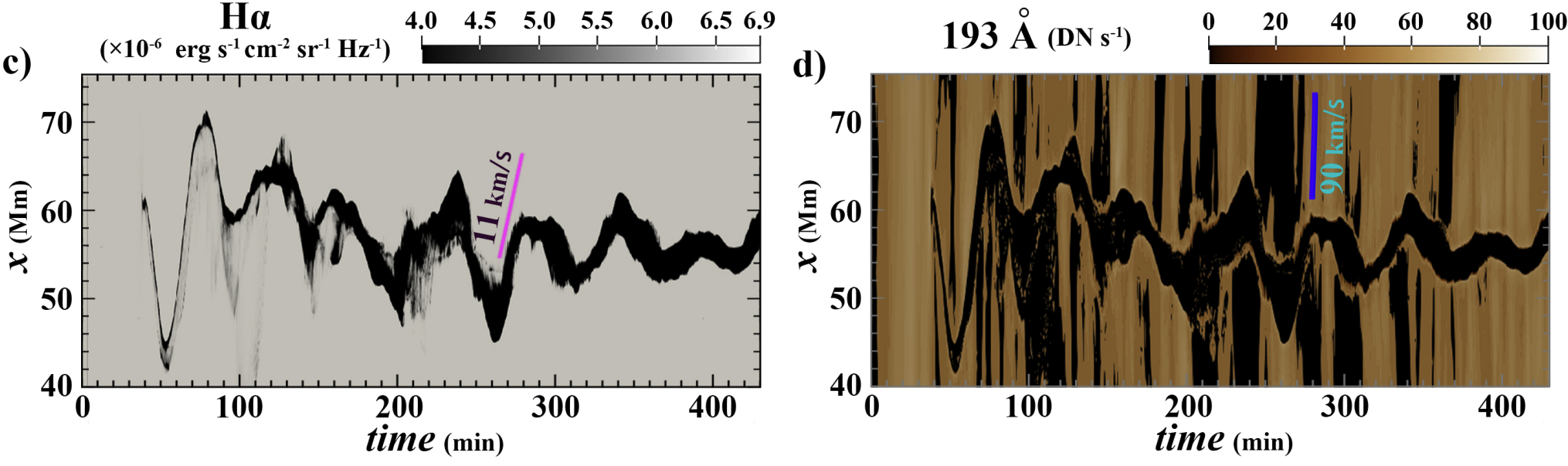}
    \caption{Appearance of prominence thread oscillations according to (c) hydrogen H$\alpha$ linecore and (d) EUV 193~\AA\ synthesis. Adapted from \citet{Zhou:2020natas}. Reproduced with permission from SNCSC.}
    \label{fig:Zhou_2020_1}
\end{figure}

    Under the assumption of a constant source function along the LOS, one can reduce Equation~\ref{eq:radiation_transport} to,
    \begin{equation}
        I_\lambda(\tau_\lambda) = I(0)e^{-\tau_\lambda} +  S_\lambda(1 - e^{-\tau_\lambda}), \label{eq:simple_radiation_transport}
    \end{equation}
    and following \citet{Heinzel:2015aap}, one can approximate the emission from the more optically thick hydrogen H$\alpha$, a line commonly employed in observations to study filaments and prominences.
    \citet{Zhou:2020natas} did exactly this for their study on the counterstreaming flows induced within self-consistently formed condensations suspended along the magnetic field of the sheared arcade type.
    Therein, they are the first study based on \texttt{MPI-AMRVAC} results that we present here to include both the EUV and hydrogen H$\alpha$ syntheses (Figure \ref{fig:Zhou_2020_1}).
    This is also carried out under the LTE approximation for the EUV synthesis, albeit employing a different underlying population approximation, and considers the H and He photoionisation cross sections found in \citet{Anzer2005:apj}.
    For hydrogen H$\alpha$, $\alpha_\lambda$ is approximated using empirical relationships between the electron number density $n_\mathrm{e}$ and the number density of the second level of Hydrogen $n_\mathrm{2}$ as $\alpha_\lambda(\tau') = \frac{\pi e^2}{m_\mathrm{e}c}f_{23}(\tau')n_\mathrm{2}\phi(\lambda,\tau')$ \citep[see][for details]{Heinzel:2015aap}.
    Under their geometrical configuration, they were able to show the threaded appearance of the condensations in both EUV and hydrogen H$\alpha$, stating that the majority of the fine structure therein was exactly aligned with or only slightly deviating from the magnetic field orientation.
    In conclusion, the need for a fully 3D domain was stressed by the authors so as not to require ad hoc approximations for integration length and permit a more self-consistent superposition of multiple discrete condensations with differing properties.

\begin{figure}
    \centering
    \includegraphics[width=1.0\linewidth]{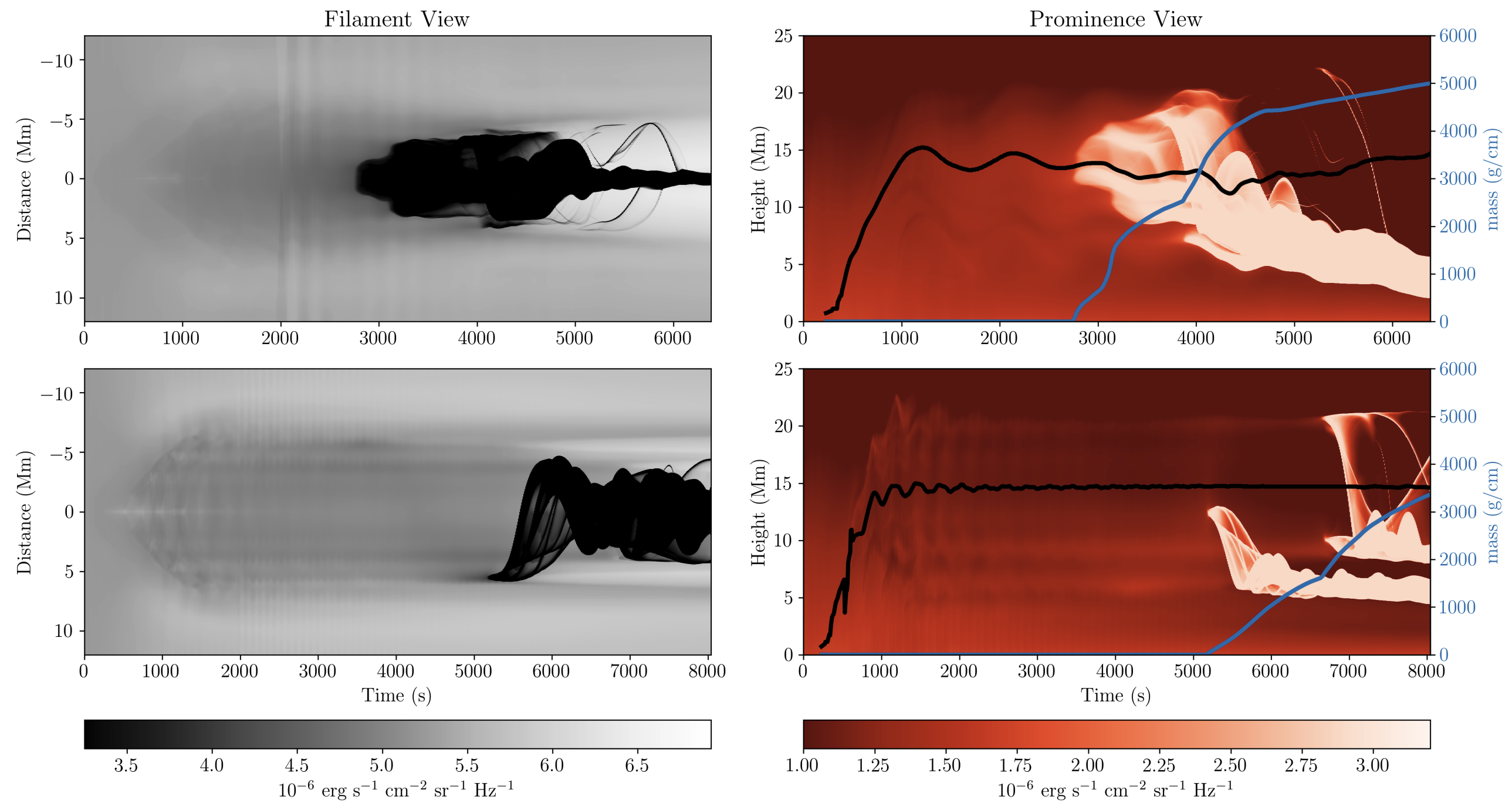}
    \caption{Filament and prominence projection hydrogen H$\alpha$ time-distance syntheses for a 2.5D \textit{ab-initio} prominence condensation formation simulation in which discrete condensations can be tracked to trace the magnetic field topology before collecting in concave-up magnetic dips. The black line traces the flux rope center, the blue line the total mass within the condensations. Adapted from \citet{Jenkins:2021aap}. \textcopyright\ ESO. Reproduced with permission.}
    \label{fig:Jenkins_2021_2}
\end{figure}

    In \citet{Jenkins:2021aap}, the authors used the above method to concentrate on the hydrogen H$\alpha$ appearance of discrete condensations in a very high resolution simulation of \textit{ab-initio} condensation formation within a 2.5D flux rope that resolved down to scales of $\approx$~5.7~km.
    In the underlying simulation, condensations formed not only in situ within the magnetic dips but throughout the flux rope, fell due to gravity, and collected in the concave-up portions of the magnetic topology.
    The accompanying syntheses suggest that these small structures would be visible within observations, tracing the magnetic field topology throughout their formation and evolution and giving direct hints as to the curvature of the host field, shown here in Figure~\ref{fig:Jenkins_2021_2}.
    This could have important consequences for the modeling of eruptive structures as it will provide an estimate of the magnetic helicity, as already tentatively assumed within observations, and the energy budget derived from estimates to the magnetic twist.
    Until just recently, this study hypothesized the existence of structures on the finest scales, far below the instrument resolution of all modern state-of-the-art telescopes.
    Excitingly, however, \citet{Schmidt:2025Natas} have just observationally confirmed the accuracy of the numerical model and accompanied syntheses by resolving scales down to less than 10~km, an impressive feat.

\begin{figure}
    \centering
    \includegraphics[trim = 0 -45 0 0, clip, width=0.49\linewidth]{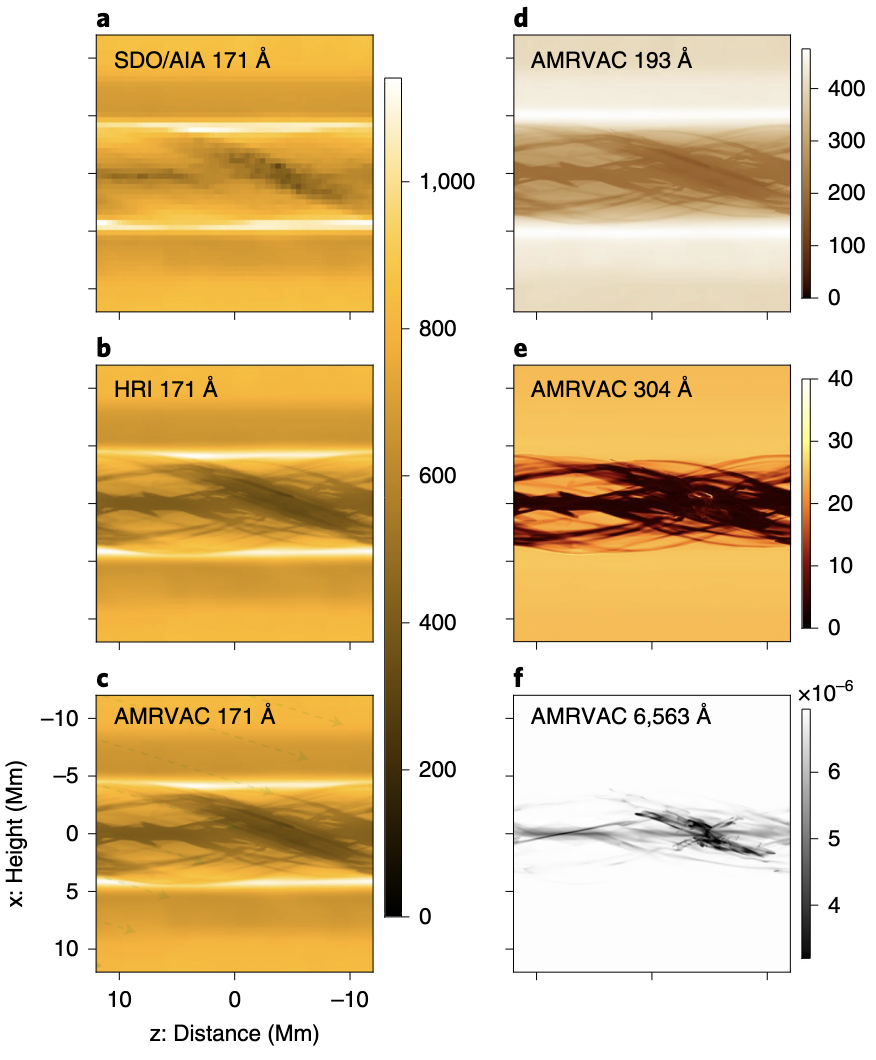}
    \includegraphics[trim = 450 0 0 0, clip, width=0.49\linewidth]{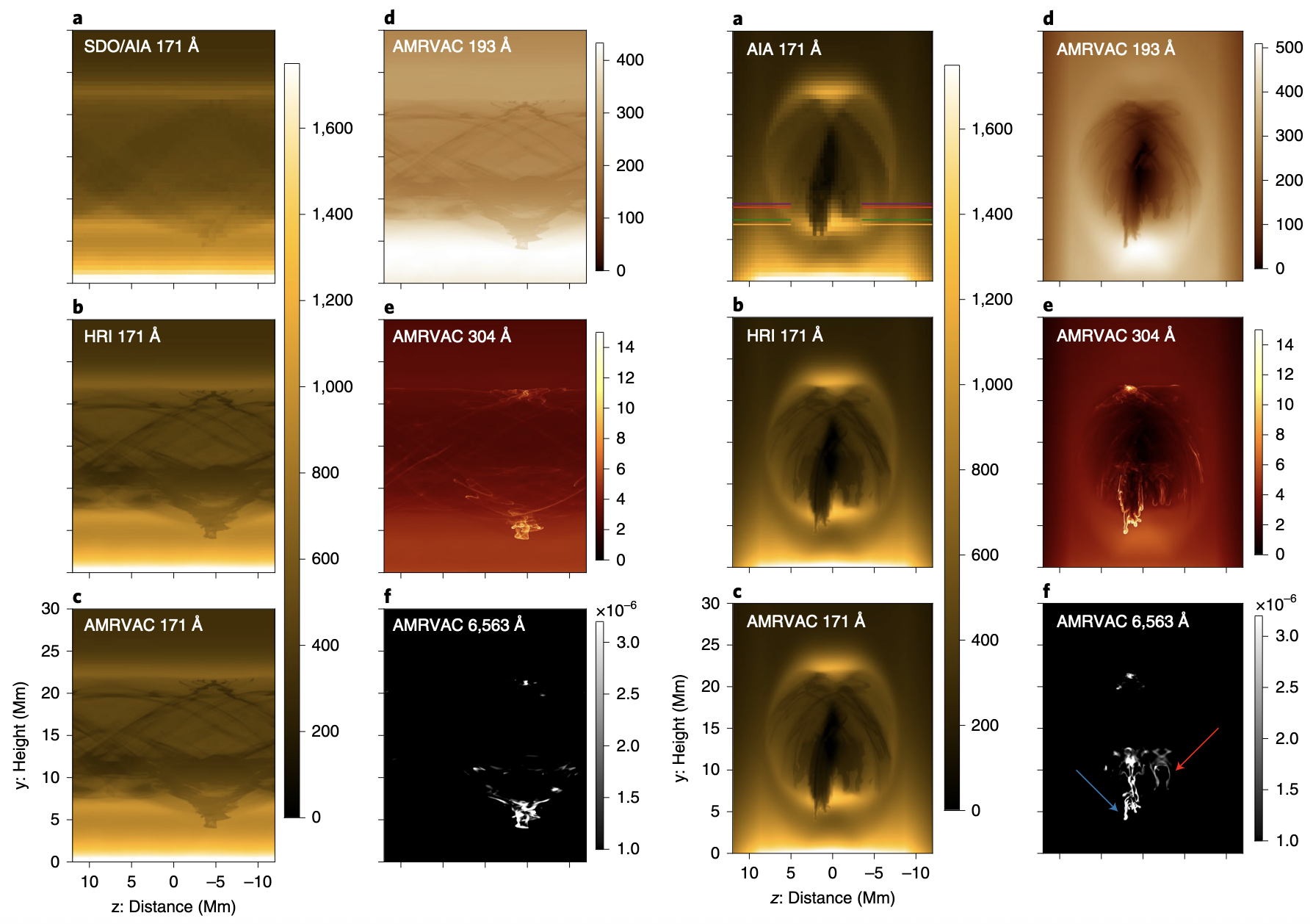}
    \caption{Multi instrument resolution EUV and hydrogen H$\alpha$ syntheses for both the filament (left six panels) and prominence (right six panels) projections of discrete condensations undergoing the gravitational interchange instability within a static flux rope magnetic topology, simultaneously reproducing the filament and prominence fine structure as observed in the real solar atmosphere. Adapted from \citet{Jenkins:2022natas}. Reproduced with permission from SNCSC.}
    \label{fig:Jenkins_2022_3}
\end{figure}

    These authors went on in \citet{Jenkins:2022natas} to consider the appearance and behaviour of condensations in light of the `prominence paradox', which describes the disconnect between the appearance of solar prominences and filaments despite them being identical phenomena.
    Although a fundamental physics question in itself, the projection effect has important radiation implications since the identical underlying structure has to appear differently depending on the viewing angle.
    These authors adopted the same approaches as \citet{Zhao:2019apj} and \citet{Zhou:2020natas}, considering LTE populations for the material absorption, and produced the first synthetic evidence of the Rayleigh-Taylor fingers, propagating under the interchange instability, being responsible for both the vertical prominence striations and the horizontal threads of filaments.
    The results reproduced here in Figure~\ref{fig:Jenkins_2022_3} therefore extend the previous work of \citet{Xia:2016apjl} as presented in Figure~\ref{fig:Xia_2016b_2}.
    Further study is required as this case example succeeded in creating only a single condensation event, whereas prominences are known to have high mass turnover rates that fuel multiple simultaneous condensations throughout the magnetic structure \citep[][]{Kaneko:2018apj}.

\begin{figure}
    \centering
    \includegraphics[width=0.9\linewidth]{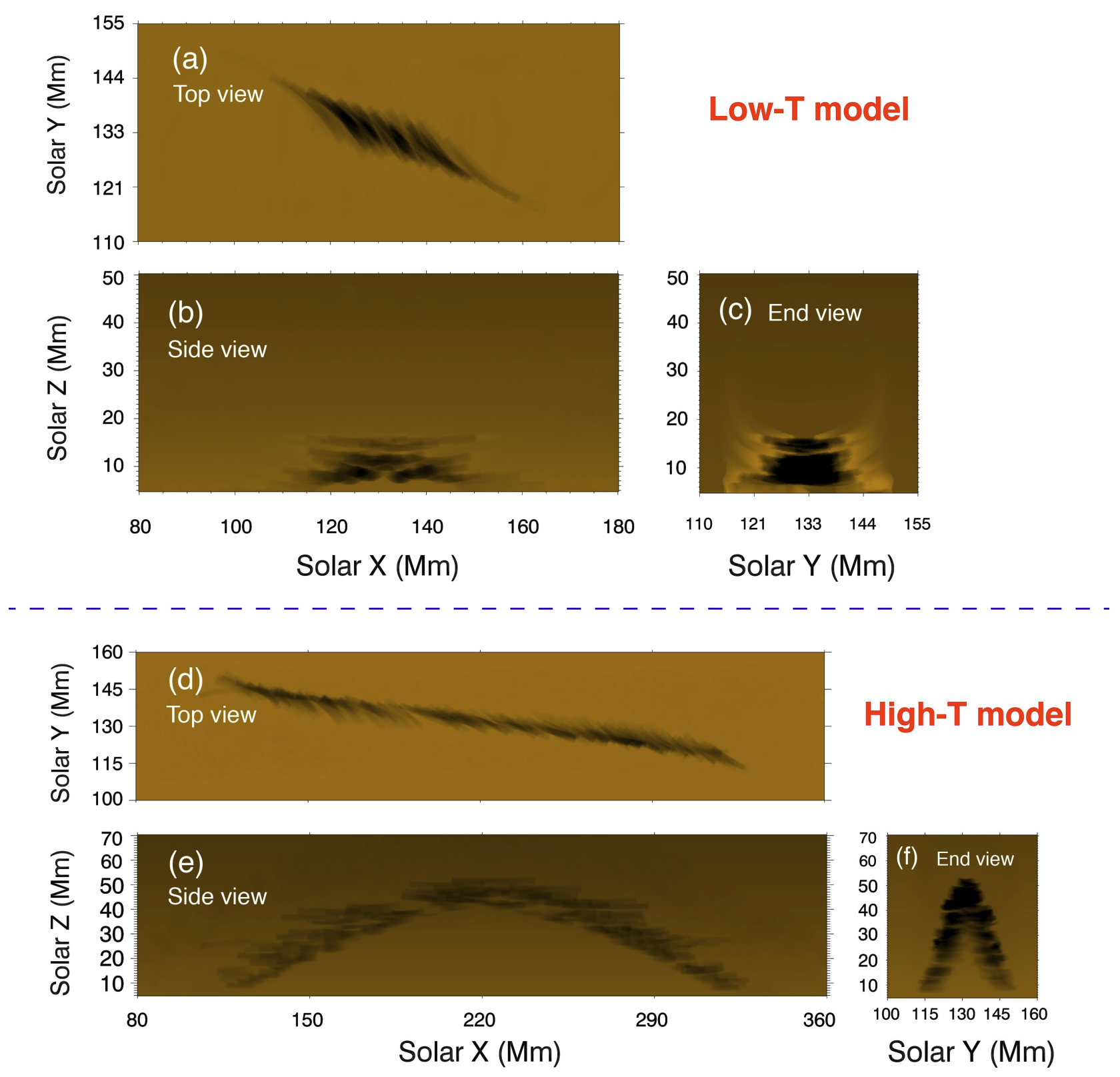}
    \caption{Difference in appearance of EUV synthesised HD threads computed along magnetic field lines extracted from magnetic flux ropes with low (Low-T) and high (High-T) twist. Adapted from \citet{Guo:2022aap}.}
    \label{fig:Guo_2022_3}
\end{figure}

    Shortly thereafter, \citet{Guo:2022aap} considered an ensemble of 1D simulations whose magnetic configurations were traced from and later interpolated back to a 3D flux rope equilibrium.
    Their EUV syntheses, similar to \citep[][]{Zhou:2020natas}, demonstrated how threads with finite angles to the polarity-inversion line could naturally form as a consequence of the 3D structure having twist with a fixed turn number.
    This aids our understanding of real prominences within the solar atmosphere as, under an assumed flux rope topology, we can more confidently extract estimations to field line connectivity, radius of curvature, and hence total enclosed magnetic flux based on filament thread structures alone \citep[][]{Kucera:2022apj}.
    However, as the underlying simulations were HD 1D with the implicit assumption of low plasma-$\beta$ and zero magnetic field deformation as outlined in Section~\ref{subsec:hydrodynamics}, this is a self-fulfilling conclusion as nothing other than field-aligned structure formation is permitted, see Figure~\ref{fig:Guo_2022_3}.
    \citet{Zhou:2025apj} performed similar simulations now in multi-dimensions, where the full MHD equations are modified to only consider field-aligned evolutions.
    Termed fixed-field HD (ffHD), this approximation significantly increases the computational efficiency but continues to prohibit field-perpendicular evolutions.
    Nevertheless, these authors demonstrate that synthetic representations succeed well in reproducing the global appearance of prominences.
    Overall, this class of models opens up the study of ideal instabilities in large nontrivial magnetic topologies, previously overcomplicated by nonlinear developments and interactions with the magnetic field.

    Finally, the eruptive work of \citet{Liakh:2023apjl} considered the appearance of the global propagating waves commonly referred to as EIT waves, owed to their first detection using the EIT instrument on board SOHO \citep[][]{Thompson:1998georl}.
    Once again, the temperature-dependent contribution functions directly imprint the propagation of both the fast and slow magnetoacoustic waves associated with the eruption of the primary flux rope.
    Once these waves impact the nearby low-lying flux rope hosting a newly formed prominence condensation, the syntheses previously presented in Figure~\ref{fig:Liakh_2025_1} capture the subsequent oscillations and echo the results found by \citet{Zhou:2018apj}, in particular capturing both the incident-reflected wave and the internal conversion of the fast-to-slow mode that in turn propagates around the flux rope and causes significant perturbation to the prominence material hosted there.

    \subsection{NLTE Statistical Equilibrium Modeling}        

    So far, we have largely discussed the common EUV synthesis approaches that do not consider the spectral appearance of these discrete condensations, whether prominences, filaments, or coronal rain.
    Indeed, since the passband filters available on SDO/AIA are broadband, these represent a wavelength integration of sorts.
    To explore how these simulations appear in spectra, one requires modeling of the radiation field and the different population states of whichever species and transition is responsible for the spectral line.
    The basis of the H$\alpha$ synthesis method of the previous section is exactly this, calculated for a selection of prominence slab models, wherein the resulting relationship between the primitive atmosphere and the condition of the converged $n=3$ and $n=2$ levels within the Hydrogen atom is approximately linear \citep[][]{Heinzel:2015aap}.
    Nevertheless, to self-consistently ascertain the statistical equilibrium of the radiative environment and the discrete level populations of a set of multi-level atoms residing within an atmosphere taken from an MHD simulation, one requires NLTE approaches.
    As the focus of this review is to collect and present the relevant studies that have been carried out using \texttt{MPI-AMRVAC} simulations, it is far beyond the scope to embark on a thorough introduction to spectral synthesis methodology, the theory, and the very mature background associated.
    A complete theoretical overview is available in the recent review of \citet{Leenaarts:2020lrsp} and the references therein, as well as a description of the specifics implemented for prominence, filament, and coronal rain studies by \citet{Heinzel:2015assl}, and so we aim instead to guide the reader through the main recent achievements.
    
\begin{figure}
    \centering
    \includegraphics[width=1.0\linewidth]{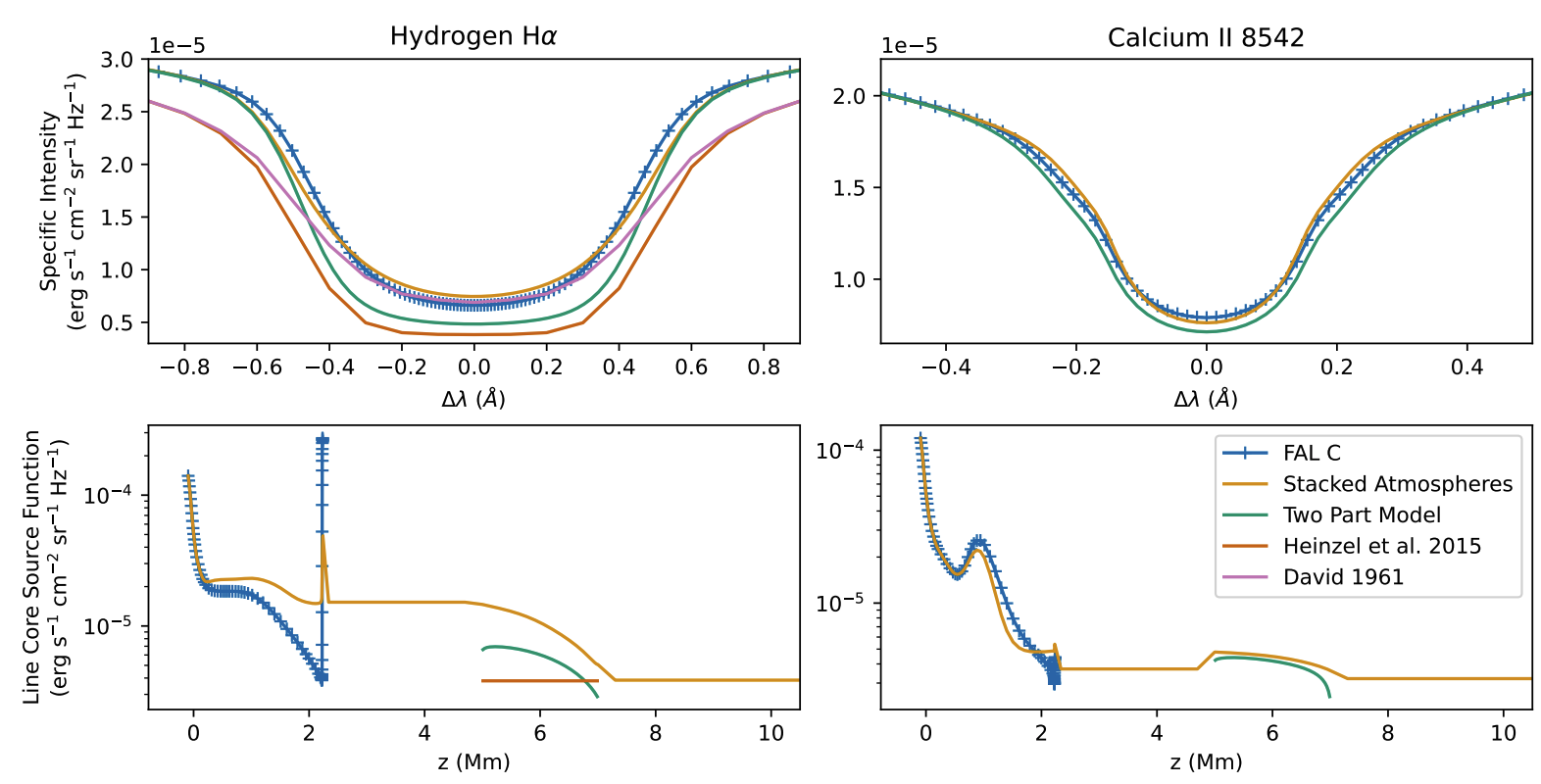}
    \caption{Difference in emergent intensity for the hydrogen H$\alpha$ and calcium~{\sc ii}~IR caused by atmospheric construction method. The stacked atmosphere approach within the one-and-a-half-dimensional (1.5D) geometry leads to radiation trapping that enhances the source function throughout both the lower and discrete condensation atmospheres. Adapted from \citet{Jenkins:2023aap}.}
    \label{fig:Jenkins_2023_2}
\end{figure}
    
    The first application of NLTE synthesis to \texttt{MPI-AMRVAC} simulations of discrete condensations was performed by \citet{Jenkins:2023aap}, where the authors made use of the recently developed \textit{Lightweaver} framework \citep[][]{Osborne:2021apj}.
    \textit{Lightweaver} adopts similar numerical machinery to the heritage RH radiative transfer code but in the form of a modular Python framework, and has been extensively tested against RH, RADYN, and SNAPI \citep[][]{Pereira:2015aap, Allred:2015apj, Milic:2018aap}. 
    To provide performance, it incorporates a C++ backend with hand-tuned parallelisation and vectorisation for modern architectures.
    As a proof of concept, these authors extracted 1D columns from a 3D simulation of prominence condensation formation and demonstrated a successful application in both the prominence and filament 1.5D geometries, where the extracted atmosphere is considered an infinite plane parallel. 
    For the filament projection, it was shown that modified boundary conditions were paramount so as to account for the radiation trapping that occurred between a chromosphere and filament atmosphere contained within a single stratification, otherwise yielding the filament atmosphere in positive contrast against the background illumination from the solar disk.
    A summary of this is presented in Figure~\ref{fig:Jenkins_2023_2}.
    This was not the first example of radiative transfer applied in such a geometry, but was the first time radiation was computed through such a highly structured, self-consistent stratification, with the need for special considerations only previously mentioned in passing in a footnote of \citet{Paletou:1993aap}.
    By precomputing the FAL-C atmosphere as the boundary illumination, the scattered radiation from the prominence is unable to modify the chromospheric source function, remedying the apparent enhanced emission therein.
    This birthed the \textit{Promweaver} package that serves as a prominence-inspired wrapper to the core \textit{Lightweaver} functionality, valid for any 1D discrete and vertically\,--\,horizontally suspended condensation, including coronal rain - although this is yet to be explicitly explored.

\begin{figure}
    \centering
    \includegraphics[width=0.6\linewidth]{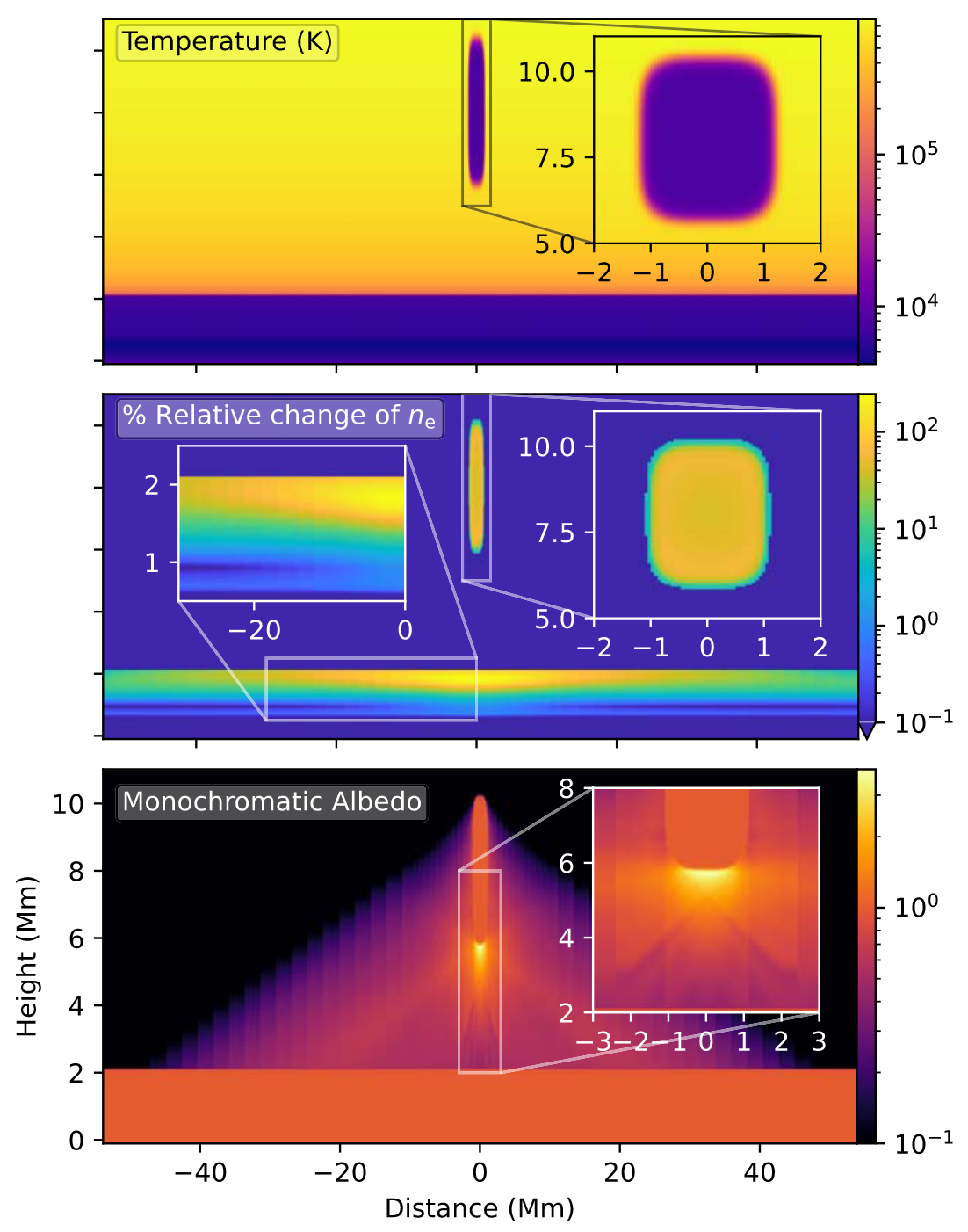}
    \caption{Converged properties of 2.5D Toy model prominence atmosphere. Upper panel is the primitive temperature distribution with a cutout of the suspended condensation, middle panel is the converged electron number density distribution demonstrating the enhanced population in the underlying chromosphere, and lower panel is the angle-integrated monochromatic albedo for the Lyman head. Adapted from \citet{Jenkins:2024apjl}.}
    \label{fig:Jenkins_2024_1}
\end{figure}

    The source function pumping found in the study of \citet{Jenkins:2023aap}, albeit as an unwanted side effect of the geometry, demonstrated that a very real radiative connection existed between prominences and their underlying chromospheres, mediated in an as yet unquantified manner by the geometry.
    In follow up work, \citet{Jenkins:2024apjl} considered a 2D domain containing a model chromosphere and a toy prominence, presenting estimates to the prominence albedo even with the additional dimension along which impinging radiation could escape without reflecting, summarised in Figure~\ref{fig:Jenkins_2024_1}.
    A critical consideration herein was the quadrature order adopted in the convergence, with the common Carlson sets (A2\,--\,A8) being wholly inadequate, and even high-order Gauss-Trapezoidal (G-T) schemes proving unsatisfactory, leading to strong radiation anisotropies from the numerical scheme alone.
    The authors advocated for a uniform quadrature, such as that of HEALPix, so as to evenly redistribute the prominence-reflected radiation but highlighted the numerical cost that such schemes imposed \citep[yet still notably far less than the underperforming GT schemes that considered 4\,--\,8x more rays][]{Gorski:2005apj}.
    They showed that not only was this non negligible and influenced the population equilibria within the chromosphere, but also demonstrated that enhanced populations in the underside of the suspended prominence shared spectral characteristics with the `Bright-rim' phenomena commonly remarked upon in observations of prominences positioned towards the limb.
    This case study served as another proof of concept matching observations; yet further statistical studies using magnetohydrostatic or fully MHD solutions are required to ascertain the extent of the valid parameter space.

    The temporal evolution of the statistical equilibrium approach adopted in \textit{Lightweaver} and associated \textit{Promweaver}, be that for toy or full MHD simulations with \texttt{MPI-AMRVAC}, is not considered.
    Although \textit{Lightweaver} contains the necessary machinery, this additional step continues to present strong numerical limitations for non trivial atmospheres and requires more focused study.
    As it is, each computed snapshot is unaware of preceding evolutions and the potentially important radiative relaxation times known to influence very dynamic atmospheres such as shocks \citep[cf.][]{Heinzel:2015aap}.
    Nevertheless, several exploratory studies have been carried out in time, as isolated solutions, to ascertain a zeroth-order approximation for the kind of spectral features we can anticipate from dynamic evolutions within prominence material.

\begin{figure}
    \centering
    \includegraphics[width=1.0\linewidth]{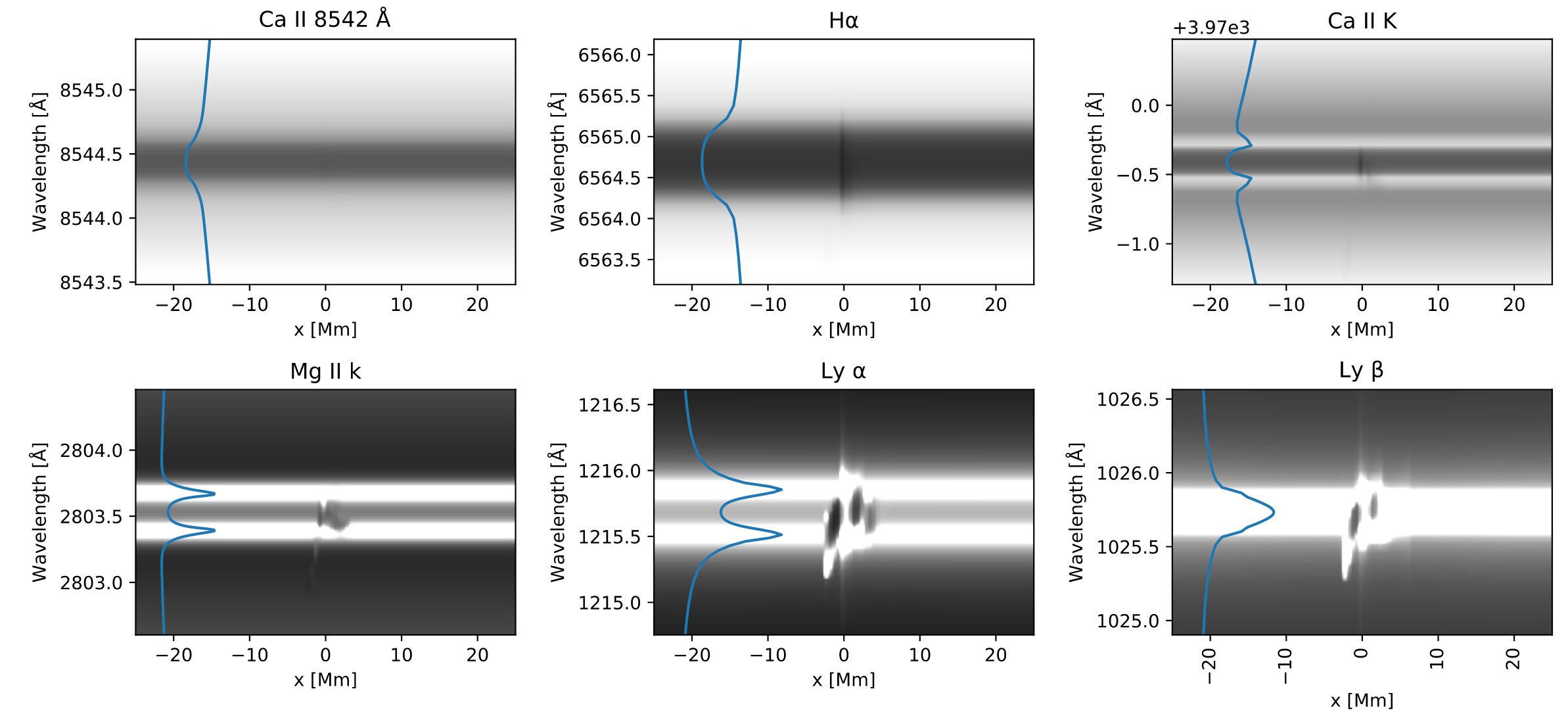}
    \caption{Spectral signatures of rotating plasma within solar filaments, the on-disk projection of solar prominences. In particular, for the more optically thick lines, a diagonal asymmetric pattern is present on either side of $x=0$, showing how one portion of the flux rope's plasma is moving away from the observer and the other towards. Blue profiles represent the quiet FAL-C illumination. Adapted from \citet{Pietrow:2024aap}}
    \label{fig:Pietrow_2024_2}
\end{figure}
    
    In a lower energy example, \citet{Pietrow:2024aap} applied \textit{Promweaver} to the rotating prominence cavity simulations of \citet{Liakh:2023apjl} to demonstrate the spatial and temporal markers of a coherent cavity rotation for comparison against the long debated `solar tornado' phenomenon.
    These authors demonstrated that velocity-skewed profiles were very much present within the synthetic spectra for a range of spectral lines, in particular those that are very optically thick, as can be seen in Figure~\ref{fig:Pietrow_2024_2}. 
    This lends credence once again to previously dismissed claims that prominence material may not only fundamentally rotate within its host magnetic field, but that these evolutions may be observable \citep[][]{Gunar:2023ssr}.

\begin{figure}
    \centering
    \includegraphics[width=0.7\linewidth]{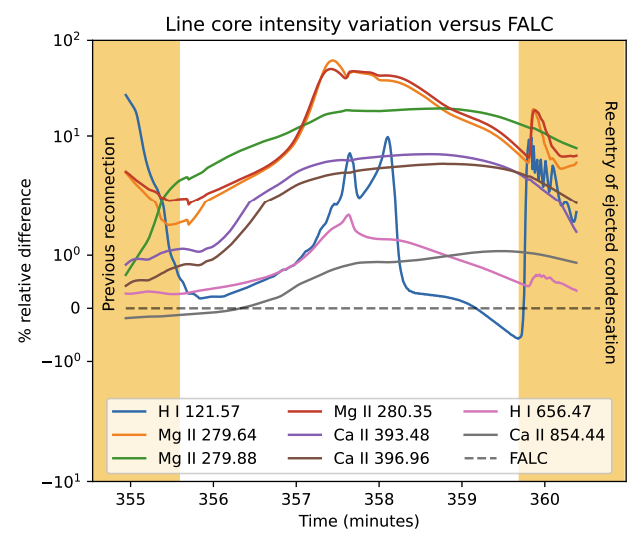}
    \caption{Response of several commonly observed linecores to the reconnection event within a prominence atmosphere, sharing similar impulsive rise and decay phases as general solar flare phenomena. The reconnection event does not occur in isolation, and so the vertical yellow bands identify the pre- and post-reconnection regions to be excluded. Adapted from \citet{Jercic:2024aap}.}
    \label{fig:Jercic_2024_2}
\end{figure}

    A higher energy example was presented by \citet{Jercic:2024aap}, in which a discrete reconnection event occurred within a monolithic slab prominence simulated with \texttt{MPI-AMRVAC}.
    In synthesising the emerging spectra with \textit{Lightweaver}, the authors were able to show that such an evolution may imprint a strong signature on the emergent spectra in the form of a burst of emission not too dissimilar to that of a small flare, reproduced in Figure~\ref{fig:Jercic_2024_2}.
    Nevertheless, such evolution is a classical example of a shocked atmosphere and further study should be targeted here, including a comparison of treatments in multidimensions to ensure the underlying physics is being accurately captured in the synthetic spectra.
    Furthermore, it has been proposed that radiation exiting extended 1D atmospheres that contain strong source function enhancements could incorrectly thermalise as a consequence of the reduced geometry and escape probability (Heinzel, P., private communication).
    This should be explicitly explored in future works.

\begin{figure}
    \centering
    \includegraphics[width=0.9\linewidth]{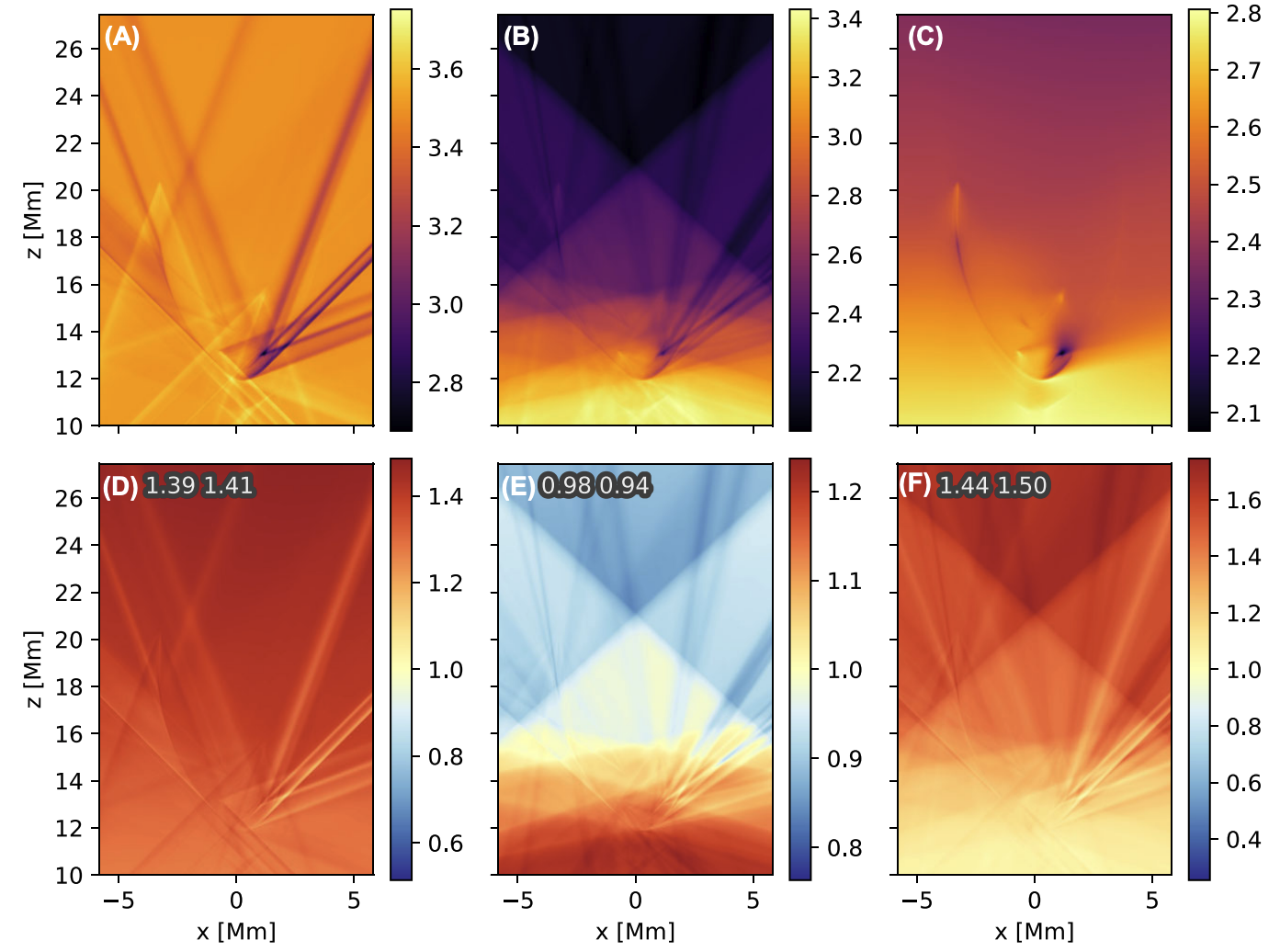}
    \caption{Comparison between the converged average radiation field $J$ solutions for the same underlying atmosphere but different discrete ray quadrature approximations. Top row are solutions considering (a) A4 quadrature set of \citet{Carlson:1963apny}, (b) 10 ray/octant set of \citet{Stepan:2020aap}, and radiance cascades. The bottom row is the A/C, B/C, and A/C, respectively. The influence of the discrete ray approximations is visible as enhanced `beaming'. Adapted from \citet{Osborne:2025rasti}.}
    \label{fig:Osborne_2025_4}
\end{figure}

    The latest state-of-the-art in application of radiative transfer to the problem of discrete condensations within the solar atmosphere is \textit{DexRT}, a new radiative transfer code that adopts the `radiance cascade' ray propagation machinery to address the `penumbra criteria' of finite features (emitting or occluding), the underlying limitation faced by the classical short-characteristics solvers in this field.
    The exact method is developed by A. Sannikov (in preparation)\footnote{A preprint is available here: \href{https://github.com/Raikiri/RadianceCascadesPaper}{https://github.com/Raikiri/RadianceCascadesPaper}} for computer graphics rendering.
    The extension of this to solar radiative transfer has been completed by \citet{Osborne:2025rasti} following unsatisfactory convergence between solutions adopting different order discrete ray quadratures in even simplified multidimensional atmospheres \citep[][]{Jenkins:2024apjl}.
    These authors have demonstrated the applicability of this new approach to \texttt{MPI-AMRVAC} solar modeling using the highly structured condensations of \citet{Jenkins:2021aap}, yielding smooth radiation fields where short characteristics could not, summarised in Figure~\ref{fig:Osborne_2025_4}.
    The 1D and 2D solutions computed with \textit{Lightweaver}/\textit{Promweaver} and \textit{DexRT} largely agree on the global structure of the spectra, as anticipated, but differ significantly in the details for the more optically-thick spectral lines where accurately representing the multidimensional radiation field is critical.
    Work is progressing on extending this technical demonstration to 3D, to include partial frequency redistribution and additional scattering tensors.
    
     
     
     
      \section{Prospects of Future Studies}\label{sec:future} 

Recent advances in numerical modeling have highlighted the need for increasingly realistic simulations to capture the complexity of coronal condensations. 

\begin{itemize}

\item\uline{Formation of prominences and coronal rain in 3D models:} 
A particularly active area is the formation and evolution of prominences and coronal rain in three dimensions. Hybrid models that combine aspects of both prominence and coronal rain formation are necessary, as they allow us to understand the full mass cycle between the solar chromosphere and corona.

\item\uline{Prominence dynamics in 3D models:}
 Another line of work focuses on prominence dynamics driven by coronal waves. This first requires detailed parametric studies in 2D to understand how energy from eruptive events is deposited in distant prominences. In 3D, it becomes essential to investigate how a line-tied magnetic field condition influences the induced prominence motions. Moreover, it is important to vary not only the distance to the wave source but also the angle between the wavefront and the flux rope axis.
Another major direction is the modeling of 3D prominence eruptions that include realistic prominence plasma, enabling studies of how processes such as mass drainage, reconnection, oscillations, or rotations may affect the pre-eruptive evolution and lead to the loss of equilibrium and the onset of eruption.

\item\uline{Synthetic observations:} 
To accurately interpret observations, synthetic images and spectra should be used in combination with 3D models to account for projection effects. This is particularly relevant in the era of high-resolution observations. It is valuable to compare multiple cold plasma lines such as H$\alpha$, H$\beta$, Ca II, He I D3, and Mg II, observed by space-based instruments like Hinode/Solar Optical Telescope (HINODE/SOT) and the Interface Region Imaging Spectrograph (IRIS), as well as ground-based telescopes, including the SST, GREGOR, and the Daniel K. Inouye Solar Telescope (DKIST). These cold lines should be studied alongside coronal emission channels such as those from SDO/AIA or Solar Orbiter.

\item\uline{Modeling of partially-ionized coronal condensations:} 
Two-fluid studies of coronal condensations represent an important direction. It is known that the velocities of neutral and charged particles can decouple, which can affect many aspects of the dynamics of coronal condensations. For instance, \citet{Jercic:2025apj} studied 1D prominence formation and growth in a two-fluid setting and revealed the decoupling of neutral and ion motions in the prominence–corona transition region. \citet{Popescu:2025aap} investigated coronal condensations in a 3D null-point configuration (neglecting gravity but including ionization–recombination processes) and demonstrated how the temperature drop is accompanied by the recombination of charged particles. The neutrals are slowed down by recombination and decouple in velocity at the edges of the condensation. Two-fluid effects are crucial for prominence plasma in many aspects, including the dynamics, condensation formation, and the development of Rayleigh–Taylor instability, among others. The ultimate future modelling is solving 3D MHD equations coupled with radiative transfer to self-consistently form partially-ionized, NLTE prominence and coronal rain plasma and study their structures and dynamics.
    
\end{itemize}

%

%

%

%
 \begin{acks}
 	We acknowledge the careful reading of the manuscript by an anonymous reviewer.
 \end{acks}

 \begin{authorcontribution}
V. L. wrote Sections \ref{sec:prom_field_plasma} -- \ref{sec:rain}, \ref{sec:future} and abstract.

J. J. wrote Sections \ref{sec:intro} -- \ref{sec:methods}, \ref{sec:synthetic}. 

 All authors reviewed the manuscript.
 \end{authorcontribution}
 \begin{fundinginformation}
 V. L. acknowledges funding from the European Union’s Horizon Europe research and innovation programme under the Marie Skłodowska-Curie grant agreement No. 101126636. J. J. acknowledges support through the European Space Agency (ESA) Research Fellowship Program in Space Science. We acknowledge support by the European Research Council (ERC) Advanced Grant PROMINENT from the ERC under the European Union’s Horizon 2020 research and innovation programme (grant agreement No. 833251 PROMINENT ERC-ADG 2018)
 \end{fundinginformation}
%
%
%
 \begin{ethics}
 \begin{conflict}
The authors declare that they have no conflicts of interest.
 \end{conflict}
 \end{ethics}

%
%
 \bibliographystyle{spr-mp-sola}
 \bibliography{export-bibtex.bib}  
%
%
%
%

\end{document}